\documentclass[%
reprint, onecolumn,
 amsmath,amssymb,
 aps,
 longbibliography
]{revtex4-1}

\usepackage{graphicx}
\usepackage{subfig,enumerate}
\usepackage[mathscr]{euscript}
\usepackage{upgreek}
\usepackage{cancel}
\usepackage{dcolumn}
\usepackage{bm}
\usepackage{hyperref}
\usepackage{xcolor}

\begin{document}

\newcommand{\Pe}{\textrm{Pe}}
\newcommand{\pard}[2]{\frac{\partial #1}{\partial #2}}
\newcommand{\ord}[2]{\frac{d #1}{d #2}}

\title{Steady translation of a chemically active viscous drop along a rigid wall}
\author{Nikhil Desai}
 \email{n.desai.2@bham.ac.uk}
\affiliation{School of Mathematics, University of Birmingham, Edgbaston, Birmingham, B15 2TT, United Kingdom}
\author{S\'ebastien Michelin}
\affiliation{LadHyX, CNRS – Ecole Polytechnique, Institut Polytechnique de Paris, 91120 Palaiseau, France}


\begin{abstract}
Chemically active drops release a chemical solute, gradients in whose concentration cause interfacial flows that advect the emitted solute. This non-linear coupling between solute distribution and fluid flow causes the drop to swim via an advective instability, if the system's Peclet number Pe--ratio of advective-to-diffusive solute transport--is larger than a finite threshold. Theoretical analyses often model active drops as diffusiophoresis-driven active particles, primarily to simplify the implementation of boundary conditions at the drop surface. However, physical insights about drop motion that rely on this simplification risk having limited applicability to fluid drops, especially when the drop is interacting chemo-hydrodynamically with its environment. Towards this, we study here the translation of a chemically active viscous drop parallel to a rigid wall, where it is confined by an external force. We consider a propulsion mechanism where the drop swims due to interfacial Marangoni stresses and diffusiophoretic flows, and compare our results to those of the ``isotropic active particle'' model, to examine the validity of the latter. Our analysis shows that, for the same confining force, a less viscous drop swims closer to the wall than a more viscous drop. This results in faster swimming, due to stronger concentration gradients driving the drop's motion. It is also shown that for moderate Pe, the swimming speed of the drop depends weakly on its mobility ratio (relative strength of diffusiophoretic and Marangoni flows). This weak dependence provides a quantitative justification for studying diffusiophoresis-driven active particles to infer the motion of Marangoni-stress-driven active drops.
\end{abstract}

\maketitle

\section{Introduction}\label{intro}

Chemically active drops are one of the simplest realizations of a microscopic swimmer. These surfactant-stabilized drops convert chemical energy into motion through physicochemical interactions that occur at the interface between the drop and its surrounding fluid~\citep{Herminghaus2014, Maass2016}. Depending on the properties of their constituents, active drops can mimic biological microlocomotion such as chemotaxis~\citep{Jin2017}, predator-prey interactions~\citep{Meredith2020}, and upstream swimming~\citep{Dey2022}. Thus, they can be used to identify the passive mechanisms behind biologically complex behavior. The ease of fabrication of these drops, as compared to other chemically-driven microswimmers like Janus particles~\citep{Ebbens2016}, also makes them an attractive tool to tackle challenging microfluidic tasks like reagent transport, sorting and analysis~\citep{Izri2014, Ban2014, Li2018}. This variety of fundamental and practical uses continues to motivate experimental and theoretical studies on understanding and controlling the motion of chemically active drops~\citep{Birrer2022, Dwivedi2022, Michelin2023}.

Active drops use a combination of interfacial chemical reactions and fluid flows to swim persistently in a surrounding fluid: (i) the chemical reaction is effectively the drop exchanging some solute with its environment, a property termed as the drop's activity; (ii) the interfacial fluid flow results from gradients in the concentration of this chemical solute close to the drop surface, due to a second property of the drop called its mobility. Some examples of activity include interfacial reactions that change the chemical properties of the surfactant molecules residing on the drop's interface with the suspending fluid~\citep{Thutupalli2011}, or solubilization processes that cause the drop to slowly dissolve in a surfactant-saturated medium~\citep{Herminghaus2014}. Classical examples of mobility would be Marangoni flows driven by gradients in surface tension resulting from inhomogeneous surfactant coverage, or diffusiophoretic flows caused by gradients in the chemical potential of a solute~\citep{Anderson1989}. In this regard, active drops are similar to Janus microswimmers: the latter also swim due to combined effects of surface chemical reactions that produce a solute whose gradients trigger fluid flows~\citep{Moran2017}. However, unlike inherently asymmetric Janus particles, active drops are physicochemically symmetric and cannot rely on built-in design to break symmetry and swim. Symmetry-breaking in active drops occurs because of a two-way, nonlinear coupling between chemical solute distribution and fluid flow around the drop: a gradient in the concentration of the exchanged solute causes interfacial flows which transport (advect) solute along the initial gradient. The original gradient is thus amplified and a positive feedback loop is established between solute gradients and gradient-driven fluid flow, each amplifying the other. This leads to symmetry-breaking and self-propulsion of the drop. Importantly, self-propulsion occurs only if solute advection--which reinforces the original concentration gradient--is sufficiently stronger than the homogenizing effects of solute diffusion. More formally, an active drop swims only if the characteristic Péclet number (Pe) of the system is larger than a critical value, say $\Pe_\textrm{c}$~\citep{Michelin2013, Izri2014}.

The advective instability described above allows active drops to swim even under confinement, near a rigid wall~\citep{Desai2021, Desai2022, Gou2025}, within a cylindrical tube~\citep{Picella2021, Brandao2024} or inside a rectangular micro-channel~\citep{Guchhait2025, Singh2025}. The confinement can result due to an external force (e.g., gravity) causing the drop to sediment near a surface~\citep{Moerman2017, Chen2025}, although such a force is not always necessary. For example, active drops can swim along the `sidewalls' of strongly confining rectangular micro-channels (i.e., when the channel height is only slightly greater than the drop diameter)~\citep{Singh2025}. Interestingly, confinement promotes the motion of the active drop, both in terms of a reduction in the critical Péclet number for self-propulsion $\Pe_\textrm{c}$, as well as an increase in the steady swimming speed of the drop. In the presence of a single wall, the drop can destabilize--to eventually swim along the wall--from an asymmetric state where it hovers close to the wall under the balancing influences of an external force and a hydrodynamic repulsion from the wall. The latter is caused by an accumulation of the solute emitted by the drop, in the gap between the drop and the wall. The enhancement in the drop's swimming speed near a rigid wall occurs because wall proximity causes strong concentration gradients at the bottom pole of the drop (i.e., the point on the drop surface that is closest to the wall), which also experiences maximum hydrodynamic traction.

While previous studies have improved our understanding of the mechanism of enhanced propulsion near a single wall~\citep{Desai2021, Desai2022}, they suffer from an important modeling assumption in that the \emph{liquid drop} is approximated as a \emph{rigid particle} that generates a phoretic slip velocity along its surface in response to solute concentration gradients. This is done mainly to avoid the technical challenge associated with implementation of the tangential velocity and stress boundary conditions at the surface of the drop, in a geometry complicated by the presence of a wall. However, this modeling simplicity comes at the cost of limited applicability of the physical insights gained by it, to more general cases. For a phoretic particle, the influence of the phoretic slip on propulsion is localized around the bottom pole as a direct consequence of the particle's rigidity. As such, the validity of this rigidity-based argument to predict an enhancement in the near-wall swimming speed of fluid drops is not immediately obvious, particularly for drops with low viscosities. Also, the effects of the drop's key physicochemical properties on its near-wall propulsion are not yet understood; for example, its viscosity ratio with respect to (w.r.t.) the suspending fluid, and its mobility ratio (the relative strength of interfacial flows caused by diffusiophoresis as compared to those caused by Marangoni stresses). These effects remain elusive in an analysis where the drop is modeled as a rigid particle.

Therefore, we consider here the motion of a chemically active viscous drop along a rigid wall, which is driven by a general interfacial mobility mechanism wherein gradients in the concentration of an emitted solute lead to Marangoni stresses, as well as a slip in the tangential fluid velocity across the drop surface~\citep{Anderson1989, Izri2014}. Beyond simply describing the influence of the drop's viscosity and mobility on its near-wall motion, our central objective is to interrogate the validity of the minimal ``isotropic phoretic particle model'', and provide a rigorous justification, or lack thereof, for considering rigid active particles as suitable models for fluid active drops.

The structure of the paper is as follows. Section~\ref{mathMod} introduces the mathematical model: governing equations, boundary conditions and the important dimensionless parameters of the problem. Section~\ref{solMeth} describes the solution methodology, while Section~\ref{solnTypes} discusses the types of solutions that result, depending on specific parameter values. In Section~\ref{effect_lambda}, we discuss the effect of a finite viscosity ratio between the drop and the surrounding fluid, on the drop's swimming speed and emphasize the unique role played by a confining wall by comparing our results against the unbounded swimming case. In Section~\ref{effect_mc}, our understanding of the near-wall motion of chemically active drops is generalized by considering the motion of drops that swim due to combined Marangoni and diffusiophoretic effects. Section~\ref{flowFields} discusses the nature of flow inside and around the drop, and Section~\ref{discu} recapitulates our findings and offers perspectives for future investigation.

\section{Mathematical model}\label{mathMod}

\subsection{Problem description, governing equations and boundary conditions}\label{GDEs_BCs}
We consider the motion of a chemically active spherical drop of radius $R$, that is swimming parallel to a rigid wall. The drop's deformation is negligible since surface tension dominates hydrodynamic stresses in most experimental situations. Formally, the capillary number of the system, which quantifies the deformation~\citep{Leal2007}, is negligibly small, $\textrm{Ca} \ll 1$ (see Appendix~\ref{param_values}). The drop is maintained close to the wall by a vertical external force $\mathbf{F}^\textrm{e}~=~-W'\mathbf{e}_z$. Typically, active drops are $O(10)$~$\mu$m in diameter and swim at speeds of $O(10)$~$\mu$m/s, meaning that the Reynolds number is sufficiently small for the fluid's inertia to be negligible and its motion is described by the continuity and the Stokes equations,
\begin{subequations}\label{stokes_oi}
\begin{align}
    \nabla \cdot \mathbf{u}^\textrm{o,i} &= 0, \\
    -\nabla p^\textrm{o,i} + \eta_\textrm{o,i}\nabla^2 \mathbf{u}^\textrm{o,i} &= \mathbf{0},
\end{align}
\end{subequations}
where, $\mathbf{u}$ is the fluid's velocity, $p$ its pressure and $\eta$ its viscosity; `o' and `i' denote, respectively, the quantities outside and inside the drop.

In experiments, a chemically active drop slowly solubilizes (dissolves) in a surfactant-saturated suspending fluid. Our focus here is on time scales much smaller than the solubilization time scale, so the radius of the drop is assumed to be constant; for example, a typical active drop that solubilizes completely over $\approx2$~hours still swims persistently with a uniform speed for $\approx6$~minutes before its radius shrinks by more than 7\%~\citep{Hokmabad2021}. As the drop solubilizes, empty surfactant micelles in the outer fluid interact with the drop and capture some of the drop's fluid to form filled surfactant micelles, whose bulk concentration is $C$~\citep{Maass2016}. The micellar transport is described by an advection-diffusion equation,
\begin{equation}\label{advDiff_C}
    \pard{C}{t} + \mathbf{u}^\textrm{o} \cdot \nabla C = D \nabla^2C,
\end{equation}
where $D$ is the molecular diffusivity of the filled micelles.

\subsubsection{Boundary conditions on drop surface}\label{BCs_drop_surf}
The formation of filled micelles is assumed to occur isotropically over the drop surface, and can be modelled as an emission of a solute at a constant rate $\mathcal{A} >0$,
\begin{equation}\label{eqn_act}
    \left. -D \mathbf{n} \cdot \nabla C \right|_\mathcal{S} = \mathcal{A},
\end{equation}
where $\mathbf{n}$ is the outward pointing normal and $\mathcal{S}$ denotes the surface of the drop.

Additionally, due to the active drop's interfacial mobility, solute concentration gradients along the drop's surface result in interfacial flows that transport the solute around the drop and possibly cause the drop to move/swim. In general, there exist two mechanisms of mobility, i.e., gradients in filled micelle concentration along the surface of the drop can lead to fluid flow in two ways~\citep{Anderson1989}. First is Marangoni flow resulting from gradients in surface tension between the drop and the surrounding fluid. This surface tension is larger in regions of low surfactant coverage, or equivalently in regions of high bulk concentration of filled micelles. The surface tension variation is modelled as,
\begin{equation}\label{gamma_func_C}
    \gamma = \gamma_0 + \mathcal{K} \left( C - C_0 \right),
\end{equation}
where $\gamma_0$ is the surface tension for a reference concentration of the solute $C_0$ and $\mathcal{K}>0$ is the sensitivity of the surface tension to the concentration of solute. When combined with the tangential stress boundary condition~\citep{Leal2007}, eqn.~\eqref{gamma_func_C} relates the tangential stress discontinuity at the drop's surface to the surface gradient in micelle concentration,
\begin{equation}\label{tau_BC_dim}
    \left. \mathbf{n} \cdot \left( \boldsymbol{\tau}^\textrm{o} - \boldsymbol{\tau}^\textrm{i} \right) \cdot \left( \boldsymbol{\delta} - \mathbf{nn} \right) \right|_\mathcal{S} = - \mathcal{K} \left( \boldsymbol{\delta} - \mathbf{nn} \right) \cdot \nabla C,
\end{equation}
where $\boldsymbol{\tau}$ is the deviatoric stress tensor and $\boldsymbol{\delta}$ is the identity tensor. The second mobility mechanism in our model is a diffusiophoretic slip velocity across the interface, that develops in response to micellar concentration gradients~\citep{Anderson1989}; this leads to a discontinuity in the tangential velocity at the interface,
\begin{equation}\label{u_BC_dim}
    \left. \left( \mathbf{u}^\textrm{o} - \mathbf{u}^\textrm{i} \right) \cdot \left( \boldsymbol{\delta} - \mathbf{nn} \right) \right|_\mathcal{S} = \mathcal{M} \left( \boldsymbol{\delta} - \mathbf{nn} \right) \cdot \nabla C,
\end{equation}
where $\mathcal{M} >0$ is called the phoretic mobility. This latter mobility mechanism is particularly relevant for drops that are highly viscous or drops whose interfaces are densely packed with a monolayer of surfactant molecules. It is important to note here that we have considered a minimal model of solute transfer and interaction, whose physical interpretation dictates the choice of signs $\mathcal{A},\mathcal{K}>0$. In reality, active drop systems undergo a multitude of complex solute exchange mechanisms, each dictating a physically consistent choice for the signs of the activity $\mathcal{A}$ and the surface tension sensitivity $\mathcal{K}$. For example, one could also consider the solute to be surfactant monomers that are adsorbed onto the interface, to model a molecular pathway of solubilization~\citep{Herminghaus2014}. In that case, one would have $\mathcal{A}<0$ to model surfactant consumption, and also $\mathcal{K}<0$, since higher interfacial surfactant concentration is associated with lower surface tension. Mathematically, as long as the condition $\mathcal{AK}>0$ (and also, $\mathcal{AM}>0$) is satisfied, the choice of individual signs of $\left( \mathcal{A}, \mathcal{K}, \mathcal{M} \right)$ does not affect our analysis. Importantly, this criterion guarantees that the drop is anti-chemotactic, i.e., it is repelled by its own wake, which is characterized by a surplus of the solute emitted by the drop (when $\mathcal{A},\mathcal{K}>0$) or a deficit of the solute consumed by it (when $\mathcal{A},\mathcal{K}<0$). This anti-chemotactic feature is widely observed in experiments and is the central reason behind the drop being able to swim.

While the flow tangential to the drop surface follows eqn.~\eqref{u_BC_dim}, the flow normal to the drop surface satisfies the impermeability condition,
\begin{subequations}
\begin{align}
    \left. \mathbf{u}^\textrm{o} \cdot \mathbf{n} \right|_\mathcal{S} &= \mathbf{U} \cdot \mathbf{n}, \\
    \left. \mathbf{u}^\textrm{i} \cdot \mathbf{n} \right|_\mathcal{S} &= \mathbf{U} \cdot \mathbf{n},
\end{align}
\end{subequations}
where, $\mathbf{U}$ is the drop's (unknown) velocity. The present work considers only steady wall-parallel swimming so, without loss of generality, a coordinate system is chosen such that the drop swims along $\mathbf{e}_x$, i.e., $\mathbf{U}=U_x\mathbf{e}_x$ (see Fig.~\ref{bisp_coords}).

\subsubsection{Boundary conditions on the wall, and far from the drop}
The wall, denoted $\mathcal{W}$, is rigid and non-slipping, so the fluid velocity vanishes there. It is also impermeable to the solute:
\begin{subequations}
\begin{align}\label{uo_C_W}
    \left. \mathbf{u}^\textrm{o} \right|_\mathcal{W} &= \mathbf{0}, \\
    \left. -D \mathbf{n} \cdot \nabla C \right|_\mathcal{W} &= 0.
\end{align}
\end{subequations}
Far from the drop, the fluid velocity decays to zero and the solute concentration relaxes to a reference value,
\begin{subequations}
\begin{align}\label{uo_C_inf}
    \left. \mathbf{u}^\textrm{o} \right|_{r \to \infty} &= \mathbf{0}, \\
    \left. C \right|_{r \to \infty} &= C_0.
\end{align}
\end{subequations}

\subsubsection{Force-free condition}\label{force_free_cond}
In addition to the flow $\left( \mathbf{u}^\textrm{o}, \mathbf{u}^\textrm{i} \right)$ and the solute concentration $C$, the steady-state swimming speed of the drop $U_x$ and the drop-to-wall separation $h$ (see Fig.~\ref{bisp_coords_a}), also need to be calculated. These are obtained from the force-free condition on the drop: since the drop is inertialess, the total force on the drop must vanish,
\begin{equation}\label{force_free}
    \int_\mathcal{S} { \mathbf{n} \cdot \boldsymbol{\sigma}^\textrm{o}\;dS } + \mathbf{F}^\textrm{e} = \mathbf{0}.
\end{equation}
Physically, gradients in solute concentration drive interfacial flows that transport the emitted solute via advection, and maintain a polarity in solute concentration at the drop surface. The horizontal polarity--reflected in the $\mathbf{e}_x$ component of eqn.~\eqref{force_free}--powers the drop motion parallel to the wall. The vertical component of eqn.~\eqref{force_free} should yield the drop's vertical velocity $U_z$, which determines the instantaneous rate of change of the drop-to-wall separation, $U_z = dh(t)/dt$. However, steady propulsion parallel to the wall requires $U_z = 0$ (otherwise, the separation will change, resulting in an unsteady problem). In this case, the vertical component of eqn.~\eqref{force_free} is used to obtain the steady-swimming separation $h$. This is explained in Section~\ref{solMeth}.

\subsection{Reference scales and dimensionless parameters}\label{dl_pars}
We define a velocity scale,
\begin{equation}\label{u_c}
    u_\textrm{ref} = \frac{\mathcal{A}}{D} \frac{R \mathcal{K} + 3 \eta_\textrm{i} \mathcal{M}}{2 \eta_\textrm{o} + 3 \eta_\textrm{i}},
\end{equation}
which denotes the characteristic speed of an isolated inactive drop that is moving--due to both diffusiophoresis and Marangoni stresses--in a solute concentration gradient of magnitude $\mathcal{A}/D$. Then, the length, velocity, time and solute concentration are non-dimensionalized by $R$ (drop radius), $u_\textrm{ref}$, $R/u_\textrm{ref}$ and $c_\textrm{ref}=\mathcal{A}R/D$, respectively. The fluid pressure inside and outside the drop is non-dimensionalized by $\eta_\textrm{i}u_\textrm{ref}/R$ and $\eta_\textrm{o}u_\textrm{ref}/R$, respectively. In what follows, our results are reported in dimensionless form based on the above reference scales. The system is described by four dimensionless parameters:
\begin{enumerate}[(\roman*)]
    \item a P\'eclet number $\Pe = u_\textrm{ref}R/D$, denoting the relative strengths of convective to diffusive solute transport,
    \item a dimensionless external force $W=W'/\left( \eta_\textrm{o} u_\textrm{ref} R \right)$,
    \item a viscosity ratio $\lambda = \eta_\textrm{i}/\eta_\textrm{o}$, quantifying the viscosity of the drop's inner fluid relative to the fluid outside the drop, and,
    \item a mobility ratio $\tilde{m} = \eta_\textrm{o} \mathcal{M}/\left( R \mathcal{K} \right)$, quantifying the relative influence of diffusiophoresis and Marangoni stresses on the drop's self-propulsion.
\end{enumerate}
Typical values of all these dimensionless parameters can be estimated from experimental work on active drops, and this has been done in Appendix~\ref{param_values}.

\section{Solution methodology: an iterative approach}\label{solMeth}

\begin{figure}[t]
\begin{center}
\subfloat[]{\label{bisp_coords_a}\includegraphics[width=5.5cm]{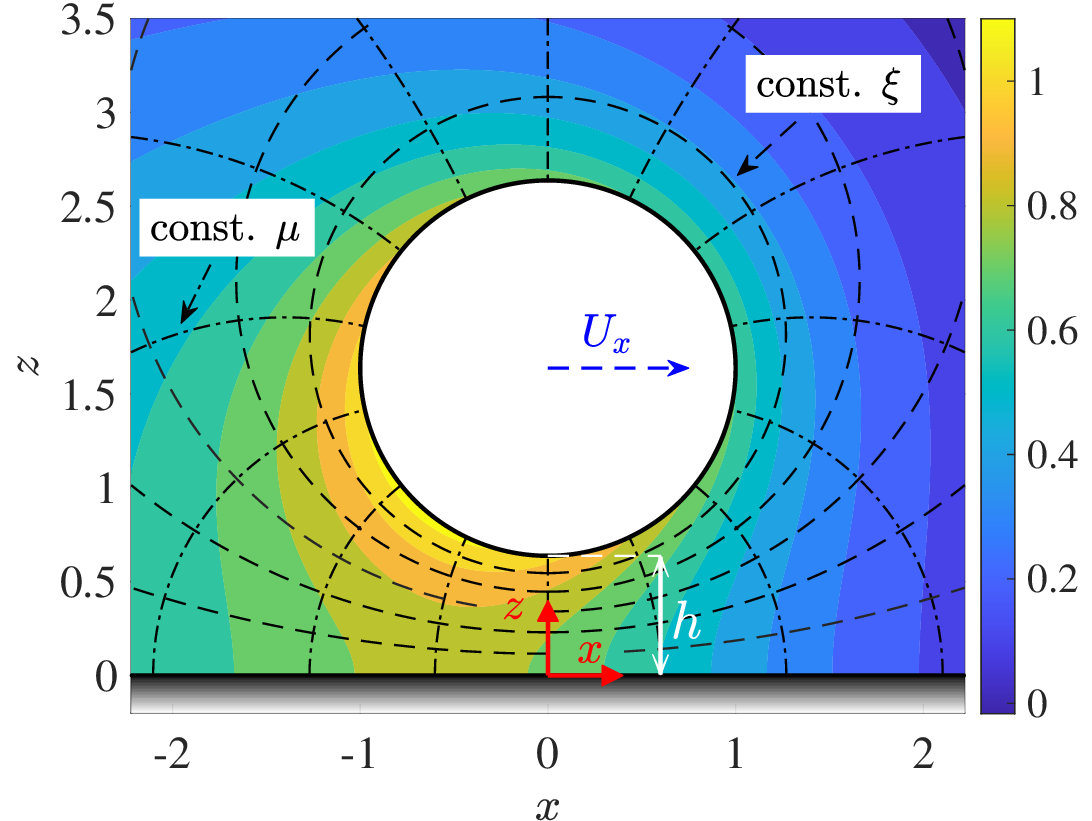}}
\hspace{5mm}
\subfloat[]{\label{bisp_coords_b}\includegraphics[width=5.5cm]{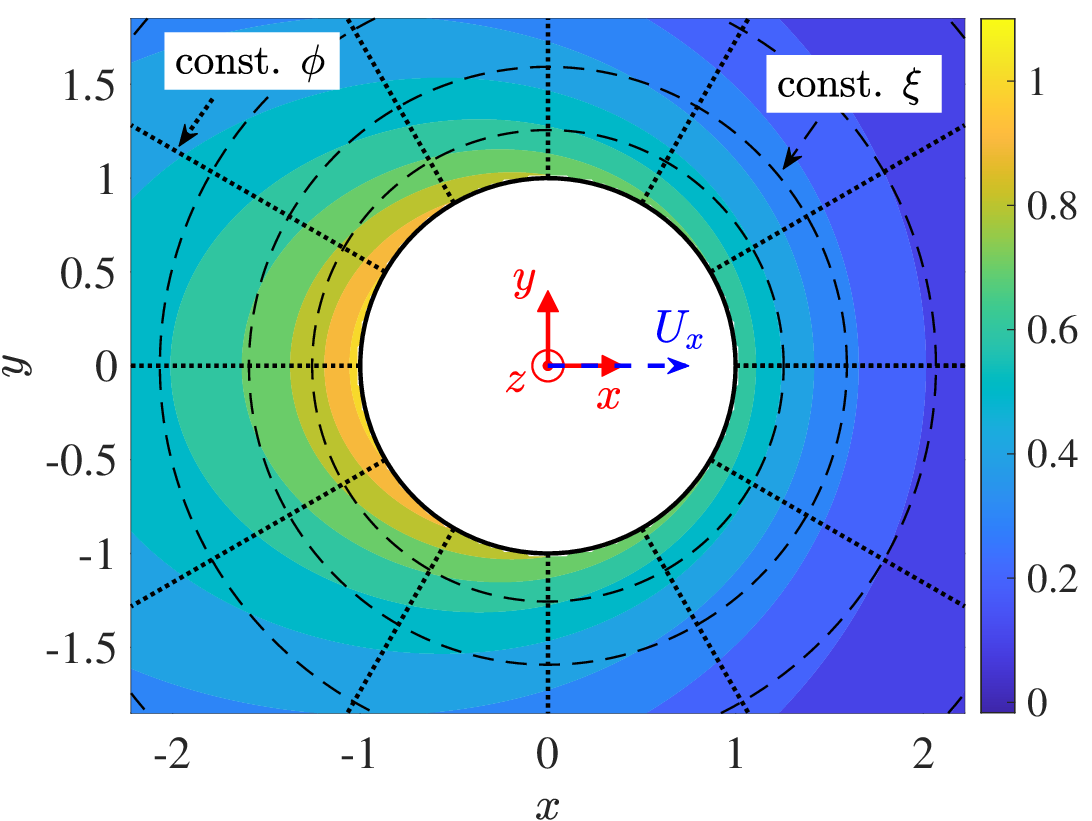}}

\end{center}

\caption{The body-fitted bispherical coordinate $\left( \xi, \mu, \phi \right)$ grid used to solve the chemohydrodynamic problem. The wall is located at $z=0$ (denoted by the grayscale rectangle) and the drop-to-wall separation is denoted $h$. The colormap denotes solute concentration distribution around the drop. Panel~(a) corresponds to the $x-z$ plane $(y=0)$, and shows the $\xi=\textrm{const.}$ and $\mu=\textrm{const.}$ coordinate surfaces. Panel~(b) corresponds to the plane intersecting the drop along its `equator' $(z=h+1)$ and shows $\xi=\textrm{const.}$ and $\phi=\textrm{const.}$ coordinate surfaces. The complete details of this coordinate system and the associated solution methodology can be found in the Supplementary Material.}
\label{bisp_coords}

\end{figure}

We seek steady solutions for the system given by eqns.~\eqref{stokes_oi} to \eqref{force_free}, i.e., when $\partial C/\partial t \equiv 0$ and $h$ is fixed. It is best to solve the system in a frame co-moving with the drop, which changes the advection-diffusion equation to (written now in dimensionless form):
\begin{equation}\label{advDiff_C_dropFrame_ND}
    \left( \mathbf{u}^\textrm{o} - \mathbf{U} \right) \cdot \nabla C = \frac{1}{\Pe} \nabla^2 C.
\end{equation}
Eqn.~\eqref{advDiff_C_dropFrame_ND} is non-linear since $\mathbf{u}^\textrm{o}$ and $\mathbf{u}^\textrm{i}$ are linear functions of $C$, determined by the Stokes equations forced by the Marangoni/slip condition at the drop's surface (eqns.~\eqref{tau_BC_dim} and \eqref{u_BC_dim}). Eqn.~\eqref{advDiff_C_dropFrame_ND}, along with the dimensionless boundary and force-free conditions, is thus solved iteratively in a body-fitted, bispherical coordinate system which allows convenient incorporation of the problem geometry and accurate enforcement of the surface boundary conditions (see Fig.~\ref{bisp_coords}). The flow and solute concentration fields are expanded as a series of non-axisymmetric bispherical harmonic eigenfunctions~\citep{Desai2022}, and substituted into the advection-diffusion equation~\eqref{advDiff_C_dropFrame_ND} and the boundary conditions given in Section~\ref{BCs_drop_surf}. These are then projected onto the orthogonal bases of the bispherical harmonics to reduce the system to a set of coupled, nonlinear ordinary differential equations for the solute concentration eigenfunctions. The equations are then discretized and solved iteratively via a fixed-point iteration technique. The iterations are initialized with flow and concentration $\left( \mathbf{u}^\textrm{o,i}, C \right)$ corresponding to the axisymmetric, hovering state of the drop for a given drop-to-wall separation (say $d$) and physical parameters $\left( \Pe, \lambda, \tilde{m} \right)$. Additionally, nonaxisymmetry is imposed by initializing $\mathbf{U} = 10^{-3}\mathbf{e}_x$. It is verified that for the simulations corresponding to $\Pe<\Pe_\textrm{c}$, $U_x$ converges to $\approx 0$, whereas for $\Pe>\Pe_\textrm{c}$, $U_x$ increases and converges to the steady state values reported in the results. The full details of the numerical method have been included in a Supplementary Material.

An important point to note is that our numerical method does not directly provide the steady-state drop-to-wall separation $h$. Instead, it yields the drop speed $U_x$ and the vertical hydrodynamic force $F^\textrm{h}_z$ for: (i) a \emph{prescribed separation} of the drop above the wall $d$, and, (ii) a prescribed P\'eclet number $\Pe$. In general, $F^\textrm{h}_z = F^\textrm{h}_z \left( d, \Pe \right)$ will not satisfy the vertical force balance of eqn.~\eqref{force_free}, i.e., in general $F^\textrm{h}_z \left( d, \Pe \right) \ne W$. If $F^\textrm{h}_{z,1} \left( d_1, \Pe \right) > W$, then it means that the drop would be expected to get repelled from the wall and could only swim parallel to the wall at a separation $h>d_1$ where the vertical force balance would be satisfied. On the other hand, if $F^\textrm{h}_{z,2} \left( d_2, \Pe \right) < W$, then the drop would settle toward the wall and adjust to a separation $h<d_2$, where again the vertical force balance would be satisfied. Hence, along with the iteration used to solve the advection-diffusion equation, a second iteration is necessary on the separation $d$ (which is an \emph{input} to the numerical method) until the value of $F^\textrm{h}_z$ (which is one of the \emph{outputs} of the numerical method) is ``sufficiently close'' to the value of the external force acting on the drop; in this work, we iterate on $d$ until a value $d_\textrm{c}$ is converged upon, such that,
\begin{equation*}
    \left| \frac{ F^\textrm{h}_z \left( d_\textrm{c}, \Pe \right) - W }{W} \right| < 10^{-2};
\end{equation*}
the steady-state drop-to-wall separation is thus obtained as $h \approx d_\textrm{c}$.

Our numerical method has been validated by reproducing the following known results:
\begin{enumerate}[(\roman*)]
    \item Hydrodynamic drag on a clean passive drop moving near a rigid wall at a specified velocity, for both the wall-normal~\citep{Wacholder1972} and wall-parallel cases~\citep{Keh2001}.
    \item Steady-state swimming speed of a viscous active drop in unbounded fluid, i.e. in the limit $h \gg 1$~\citep{Izri2014, Morozov2019}.
    \item Steady-state swimming speed of a rigid isotropic active particle--i.e. in the limit $\lambda \gg 1$ and $\tilde{m} \gg 1$--moving parallel to a wall~\citep{Desai2022}.
\end{enumerate}
The results of the above computations are shown in Fig.~\ref{validation_figs} in Appendix~\ref{validations}.

\section{Stationary vs. swimming steady states}\label{solnTypes}

\begin{figure}[t]
\begin{center}
\subfloat[]{\label{stat_swim_a}\includegraphics[trim = 80 0 60 0, clip, width=5.5cm]{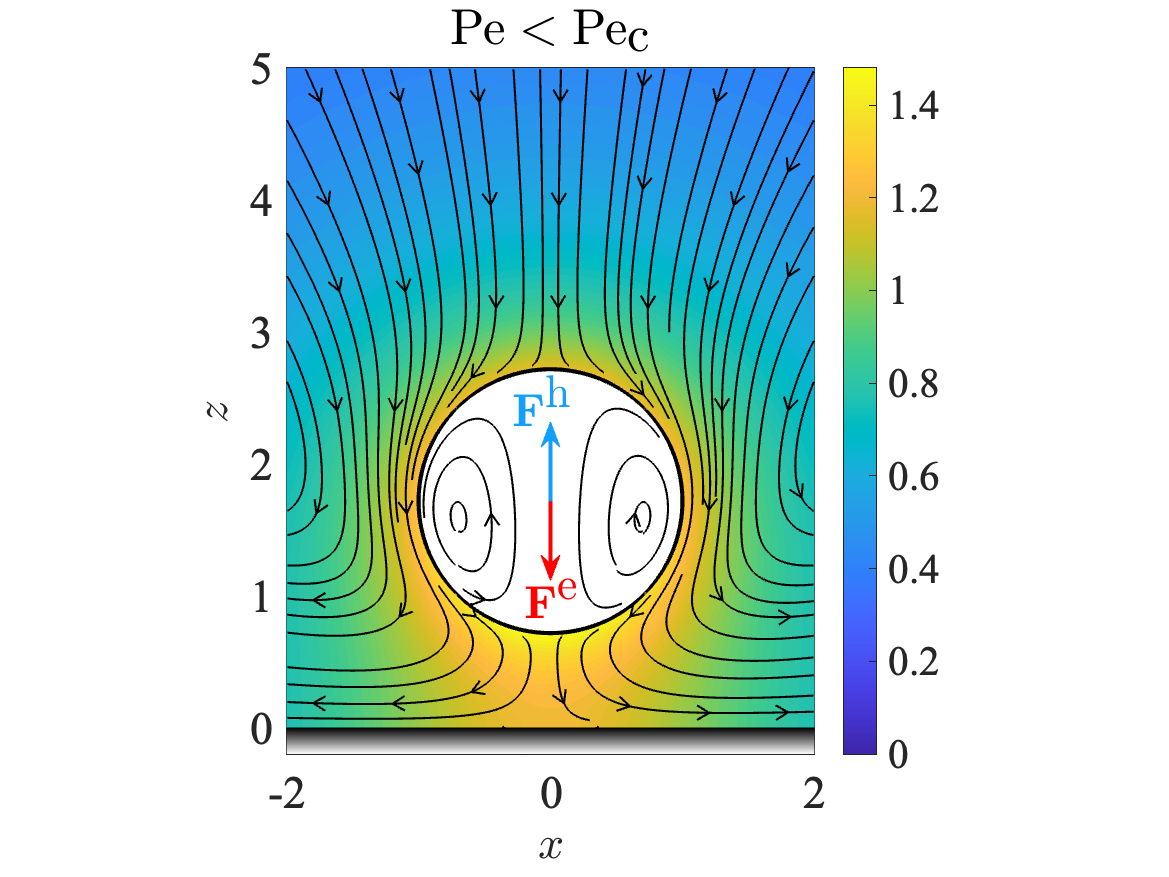}}
\subfloat[]{\label{stat_swim_c}\includegraphics[trim = 80 0 60 0, clip, width=5.5cm]{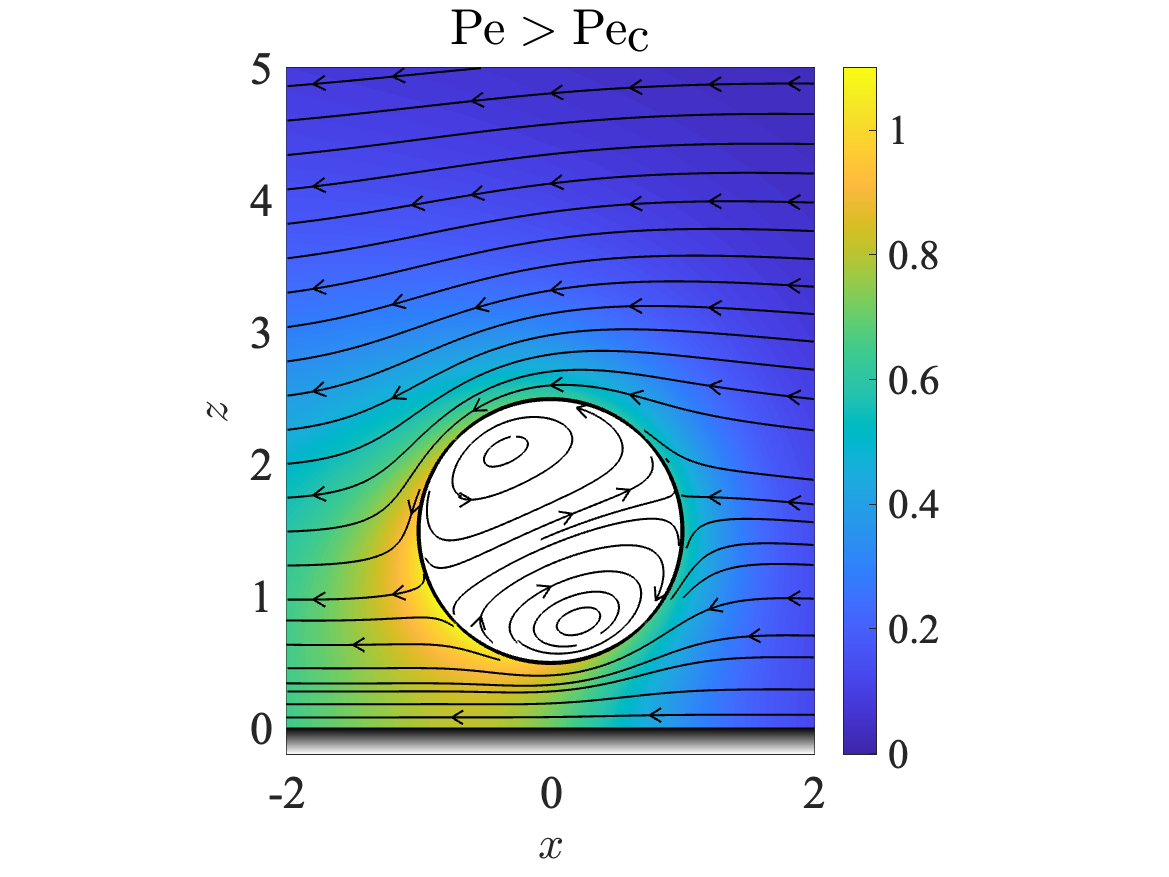}}

\end{center}

\caption{Steady states for an active drop near a wall. The external force on the drop is $\mathbf{F}^\textrm{e}=-W\mathbf{e}_z$. The drop-to-wall separation is determined by a balance between $\mathbf{F}^\textrm{e}$ and the hydrodynamic repulsion, $\mathbf{F}^\textrm{h}$. (a)~A stable stationary state far from $\Pe_\textrm{c}$ $\left( \Pe=1 \right)$; the hovering separation is $h=0.716$. (b)~A swimming state far from $\Pe_\textrm{c}$ $\left( \Pe=10 \right)$; the swimming separation is $h=0.490$. Parameter values are: $W=3, \lambda=1, \tilde{m}=0$. Streamlines are plotted in the frame of reference of the drop and the colors denote dimensionless solute concentration.} 
\label{stat_swim}

\end{figure}

Eqns.~\eqref{stokes_oi} to \eqref{force_free} admit two types of solutions for the parameter range considered. The first is a stationary state where the drop is at rest ($U_x \equiv 0$), as shown in Fig.~\ref{stat_swim_a}. The second is a swimming state, where the drop swims parallel to the wall in a rectilinear motion, as shown in Fig.~\ref{stat_swim_c}.

\subsection{Stationary states}\label{stat_state}
\begin{figure}[t]
\begin{center}

\subfloat[]{\label{stat_states_lambda_a}\includegraphics[width=6.5cm]{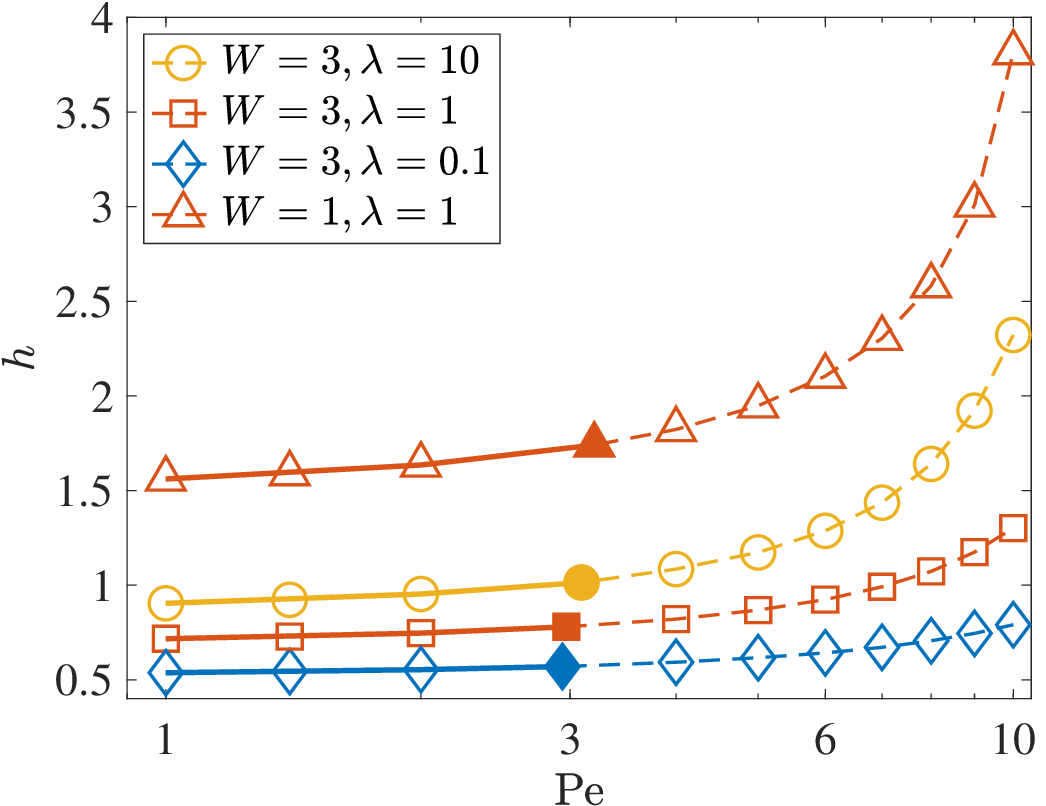}}
\\
\subfloat[]{\label{stat_states_lambda_b}\includegraphics[trim = 100 0 40 0, clip, width=6cm]{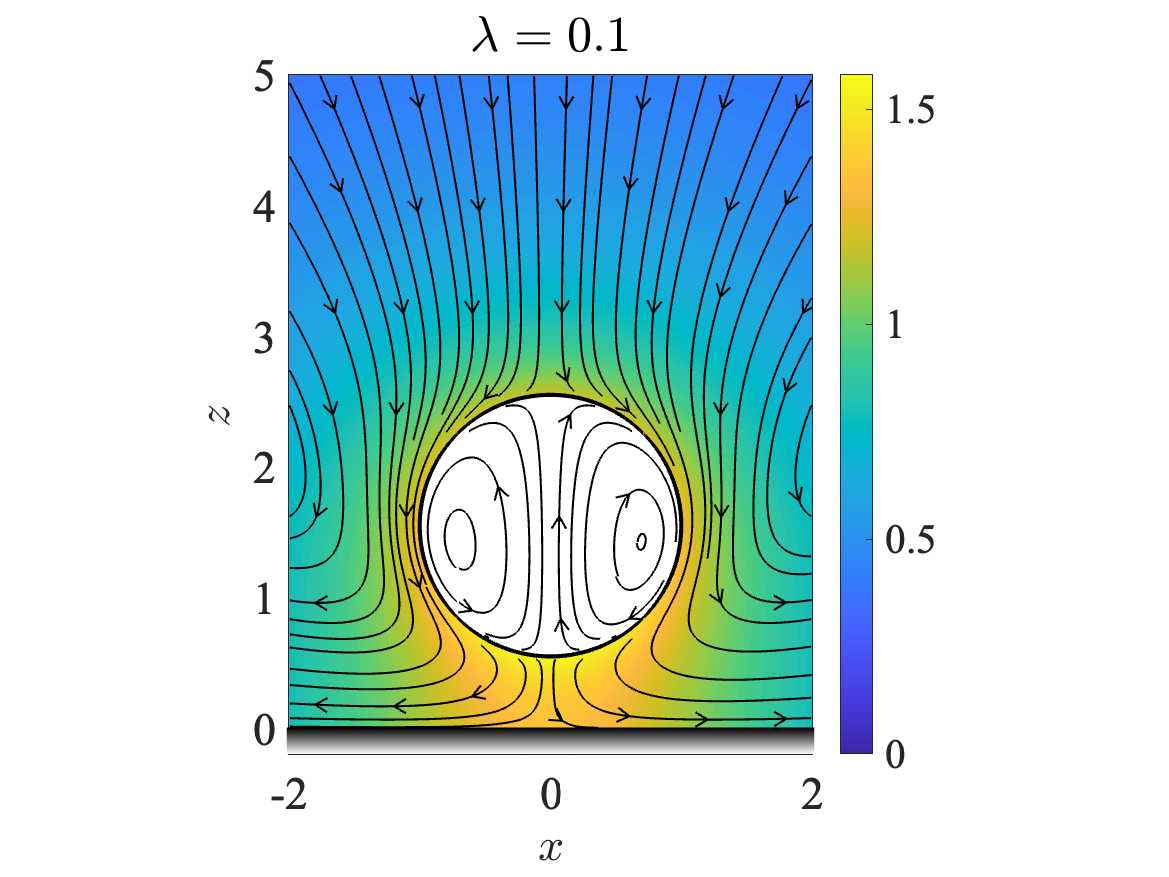}}
\subfloat[]{\label{stat_states_lambda_c}\includegraphics[trim = 100 0 40 0, clip, width=6cm]{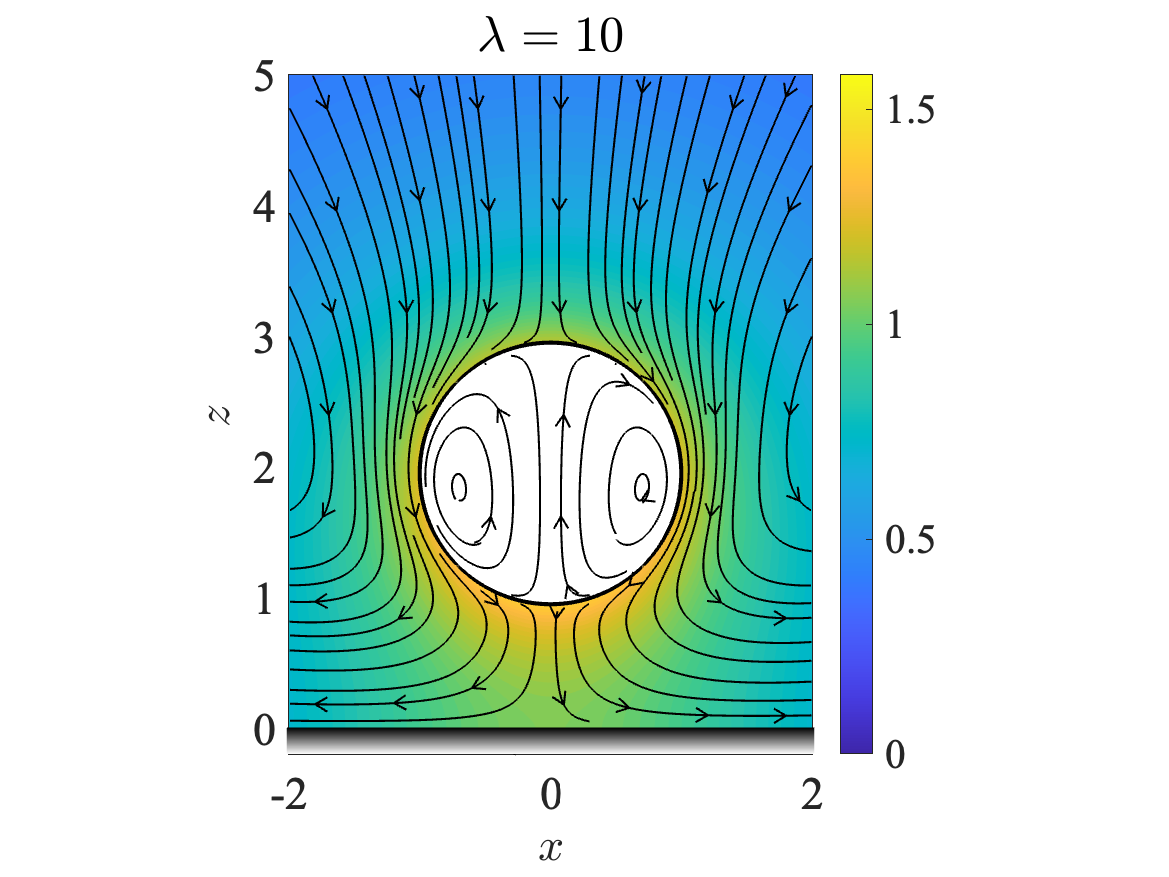}}

\end{center}

\caption{(a) Equilibrium separation $h$ as a function of $\Pe$ for Marangoni-stress-driven drops, i.e., when $\tilde{m} \equiv 0$, for different values of $\left( W, \lambda \right)$. Solid (resp. dashed) lines denote stable (resp. unstable) solutions. Note that the drop is stationary, i.e., $U_x \equiv 0$, in all cases. The filled symbols correspond to the critical P\'eclet number $\Pe_\textrm{c}$ and the corresponding critical separation $h^*$, at which the stationary state destabilizes to the propelling state. (b-c) Flow field and solute concentration around two drops with different viscosity ratio $\lambda=0.1$ (panel (b)) and $\lambda=10$ (panel (c)). For both panels (b) and (c), $\Pe=2$ and the drops experience the same vertical force, $W=3$.}
\label{stat_states_lambda}

\end{figure}

In the stationary state, the wall-induced polarity in the solute concentration around the drop causes axisymmetric fluid flow toward the wall. The drop thus hovers at an equilibrium separation $h$, where the hydrodynamic force balances the external force on the drop. Fig.~\ref{stat_states_lambda} shows $h$ as a function of the P\'eclet number $\Pe$, for stationary drops, i.e. with $U_x \equiv 0$. Intuitively, drops that experience a greater external force $W$ stay closer to the wall. For fixed $W$, an increase in $\Pe$ increases solute advection towards the wall, causing further solute accumulation in the drop-to-wall gap and resulting in increased equilibrium separation due to stronger hydrodynamic repulsion.

The equilibrium separation will also depend on the viscosity ratio $\lambda$ and the mobility ratio $\tilde{m}$. The dependence on $\tilde{m}$ is weak and is shown in Fig.~\ref{stat_states_mc} in the Appendix. The dependence on $\lambda$ is more pronounced and is explored next for the case of drops driven purely by Marangoni stresses, i.e. when $\tilde{m} \equiv 0$. Figs.~\ref{stat_states_lambda_a} to \ref{stat_states_lambda_c} show that a less viscous drop rests closer to the wall than a more viscous drop. To understand this, we compare the hydrodynamic force acting on two drops held stationary at the same separation, say $h_\textrm{fix}$, but having different viscosity ratios, $\lambda=0.1$ and $\lambda=10$. The reciprocal theorem can be used to express this force as a weighted integral of the solute concentration gradient along the drop surface~\citep{Stone1996}:
\begin{equation}\label{reciproc_drop_lam_FHz}
    F^\textrm{h}_z = \int_{\partial V} { \nabla_s C \cdot \left[ \left( 3 \lambda + 2 \right) \mathbf{u}^\perp_\textrm{aux} \right] \,dS },
\end{equation}
where, $\partial V$ denotes the surface of the drop, $\nabla_sC = \left( \boldsymbol{\delta} - \mathbf{nn} \right) \cdot \nabla C$ is the surface gradient of solute concentration, and $\mathbf{u}^\perp_\textrm{aux}$ is the flow field associated with the auxiliary Stokes flow problem. For the purpose of eqn.~\eqref{reciproc_drop_lam_FHz}, this is chosen as the flow around a passive, clean drop (zero activity and mobility) moving normal to the wall at a prescribed velocity $\mathbf{U}^\perp_\textrm{aux}=\mathbf{e}_z$~\citep{Wacholder1972}. Eqn.~\eqref{reciproc_drop_lam_FHz} then states that the contribution of $\nabla_sC$ toward the hydrodynamic force acting on the drop, $F^\textrm{h}_z$, is weighted by the ``influence function'' $\left( 3\lambda + 2 \right) \mathbf{u}^\perp_\textrm{aux}$.

\begin{figure}[t]
\begin{center}

\subfloat[]{\label{analyze_stat_states_a}\includegraphics[width=6cm]{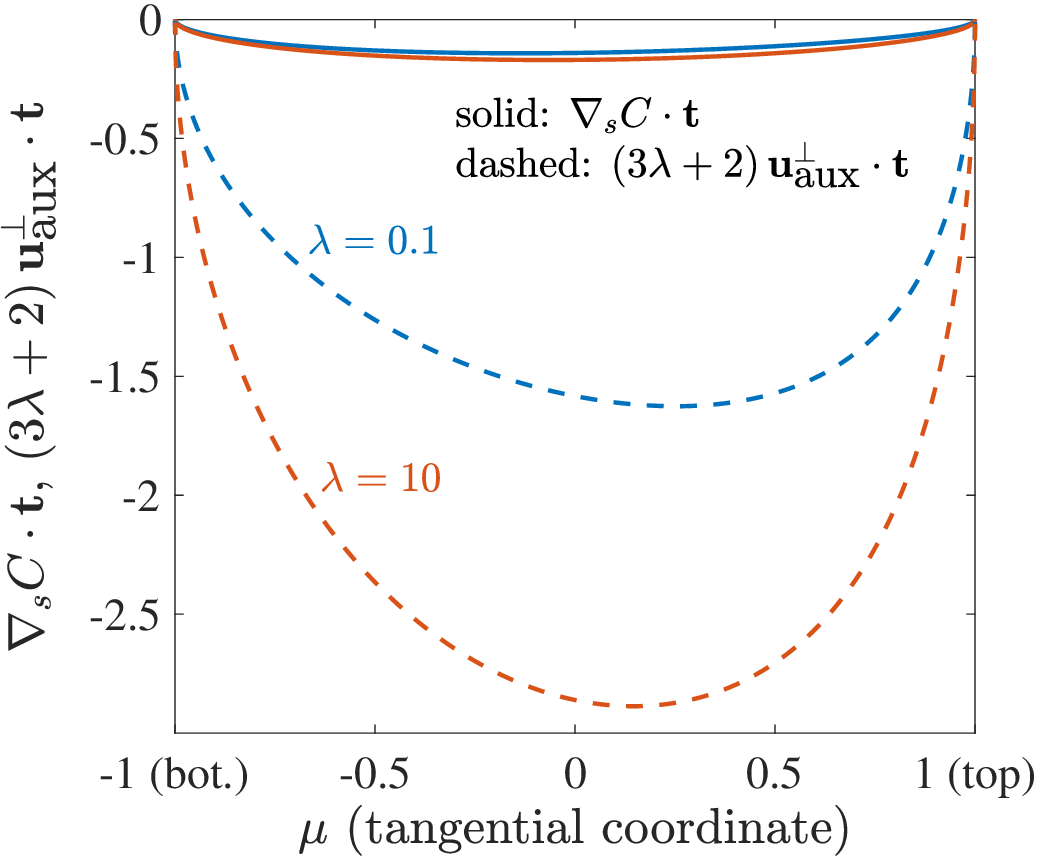}}
\subfloat[]{\label{analyze_stat_states_b}\includegraphics[width=6.5cm]{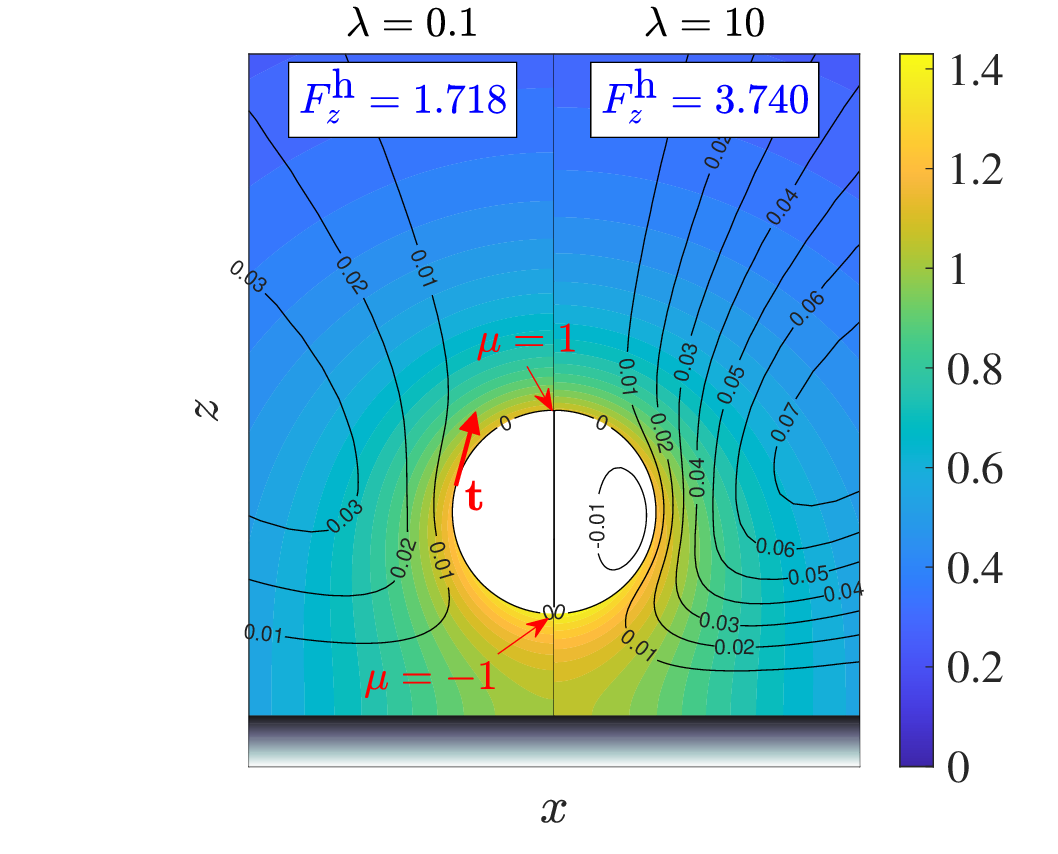}}

\end{center}

\caption{Hydrodynamic force as a function of viscosity ratio. (a)~Tangential component of the (axisymmetric) concentration gradient, $\nabla_sC$, and the influence function $\left( 3\lambda+2 \right)\mathbf{u}^\perp_\textrm{aux}$ along the drop surface; here, $\mathbf{t}$ is the local tangent vector in the polar direction, and $\mu$ is a local tangential coordinate. (b)~Comparison between flow field and solute distribution around the two drops, $\lambda=0.1$ (left) and $\lambda=10$ (right), held fixed at a separation $h_\textrm{fix}=1$, at $\Pe=5$. The lines denote iso-contours of the Stokes streamfunction whose value is given by the overlying numbers. Also shown are the tangent vector $\mathbf{t}$ and the tangential coordinate $\mu$.}
\label{analyze_stat_states}

\end{figure}

Fig.~\ref{analyze_stat_states_a} shows the surface variation of $\nabla_s C$ and $\left( 3\lambda + 2 \right) \mathbf{u}^\perp_\textrm{aux}$ as a function of the tangential bispherical coordinate $\mu$ that increases along the polar direction $\mathbf{t}$ on the drop surface: from -1 at the point closest to the wall, to +1 at the point furthest from the wall (see Figs.~\ref{bisp_coords_a} and  \ref{analyze_stat_states_b}). The solute distribution changes only modestly between $\lambda=0.1$ and $\lambda=10$ but the influence function, $\left(3\lambda+2\right)\mathbf{u}^\perp_\textrm{aux}$, is much larger for $\lambda=10$ than for $\lambda=0.1$ (Fig.~\ref{analyze_stat_states_a}). Mathematically, surface flows engendered by solute concentration gradients exert greater hydrodynamic force on the more viscous drop. Fig.~\ref{analyze_stat_states_b} shows the flow field and the solute concentration around the two drops, plotted side-by-side for comparison. The lines denote iso-contours of the Stokes streamfunction~\citep{HappelBrenner2012}, and they are more concentrated around the drop with $\lambda=10$. Thus, the more viscous drop experiences stronger velocity gradients and a higher hydrodynamic force compared to the less viscous drop, when both are fixed at the same separation, $h_\textrm{fix}$. In Fig.~\ref{analyze_stat_states_b}, $h_\textrm{fix}=1$, with $F^\textrm{h}_z\left(\lambda=0.1\right) \approx 1.72$ and $F^\textrm{h}_z\left(\lambda=10\right) \approx 3.74$. This in turn means that if the external force on the drops is the same, say $W=3$, and the drops are free to `choose' an equilibrium separation based on vertical force balance, then the more viscous drop would hover further from the wall than the separation $h_\textrm{fix}$ shown in Fig.~\ref{analyze_stat_states_b} (and the less viscous drop would hover closer to the wall than $h_\textrm{fix}$).

The stationary states described above exist only for low values of $\Pe$. As $\Pe$ increases above a critical value $\Pe_\textrm{c}$, the stationary state becomes unstable, and the drop transitions into a swimming state. In Fig.~\ref{stat_states_lambda_a}, this $\Pe_\textrm{c}$ is shown with filled symbols. The solutions for $\Pe > \Pe_\textrm{c}$ have been depicted using dashed lines because they are unstable. The corresponding stable solutions--when the drop swims parallel to the wall--are discussed next.

\subsection{Swimming states}\label{swim_state}
The swimming state stems from a linear instability of the stationary state above a critical P\'eclet number $\Pe_\textrm{c}$. This linear instability problem was analyzed in detail in \cite{Desai2021} for a rigid isotropic active particle swimming only due to diffusiophoresis, and those results are generalized here for the case of a viscous active drop. The main conclusions are:
\begin{enumerate}[(\roman*)]
    \item $\Pe_\textrm{c}$ depends weakly on the viscosity ratio $\lambda$ (Fig.~\ref{res_stabAna_a}) and very weakly on the mobility ratio $\tilde{m}$ (see Fig.~\ref{stat_states_mc_b} in Appendix~\ref{h_vs_Pe_mc}).
    
    \item For a prescribed set of parameters $\left( W, \lambda, \tilde{m}, \Pe_\textrm{c} \right)$, there exists a critical separation $h^*$ at which the drop destabilizes and swims along the wall. The filled symbols in Fig.~\ref{stat_states_lambda_a} denote values of the $\left( \Pe_\textrm{c}, h^* \right)$ pair, for different $\left( W, \lambda \right)$.

    \item For small values of the external force, e.g., $W =0.02 \ll 1$, the critical P\'eclet number approaches the critical value in the unbounded case, i.e., $\Pe_\textrm{c} \left( W \ll 1 \right) = 4 - \Delta \Pe$, with $\Delta \Pe \ll 1$, and the corresponding critical separation is expectedly $h^* \gg 1$.
    
    \item Stationary drops hovering closer to the wall destabilize at lower advection strengths: $\Pe_\textrm{c}$ and $h^*$ both decrease with an increase in the magnitude of the external force $W$ as well as with a reduction in the viscosity ratio $\lambda$.
\end{enumerate}
In what follows, we focus on the swimming speed of the drop as a function of its viscosity and mobility ratios.

\begin{figure}[t]
\begin{center}

\subfloat[]{\label{res_stabAna_a}\includegraphics[width=5.5cm]{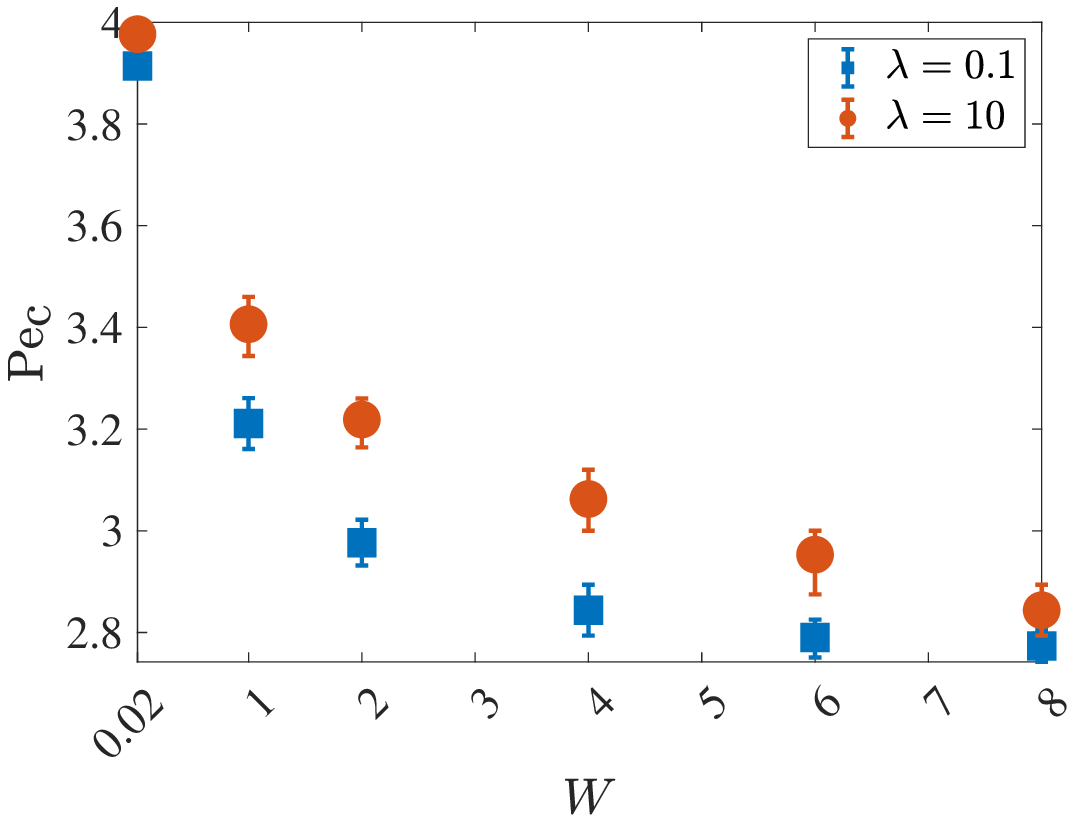}}
\subfloat[]{\label{res_stabAna_b}\includegraphics[width=5.5cm]{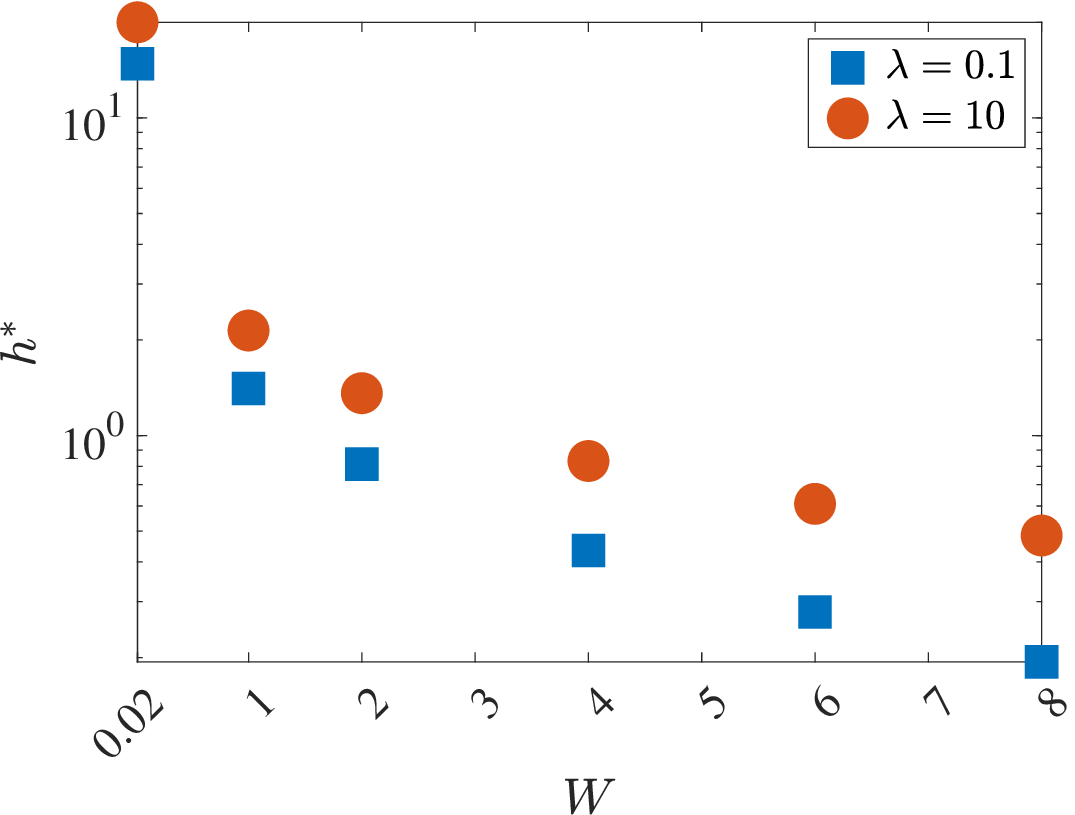}}

\end{center}

\caption{(a) The critical P\'eclet number for self-propulsion, $\Pe_\textrm{c}$, above which a hovering drop (as shown in Fig.~\ref{stat_swim_a}) becomes unstable to longitudinal perturbations and spontaneously begins to swim parallel to the wall (as shown in Fig.~\ref{stat_swim_c}). The variation in $\Pe_\textrm{c}$ is plotted as a function of the external force $W$, for different values of the viscosity ratio and a fixed mobility ratio $\tilde{m} \equiv 0$. The error bars denote uncertainty due to a (minor) dependence of $\Pe_\textrm{c}$ on the number of bispherical eigenfunction modes retained in the numerical method (see Supplementary Material). (b) The critical separation $h^*$ at which the active drop destabilizes (identified by the filled symbols in Fig.~\ref{stat_states_lambda_a}).}
\label{res_stabAna}

\end{figure}

\section{Steady swimming of the viscous active drop}
\subsection{Influence of the viscosity ratio}\label{effect_lambda}

\begin{figure}[t]
\begin{center}

\subfloat[]{\label{res_onlyMar_Ux}\includegraphics[width=6cm]{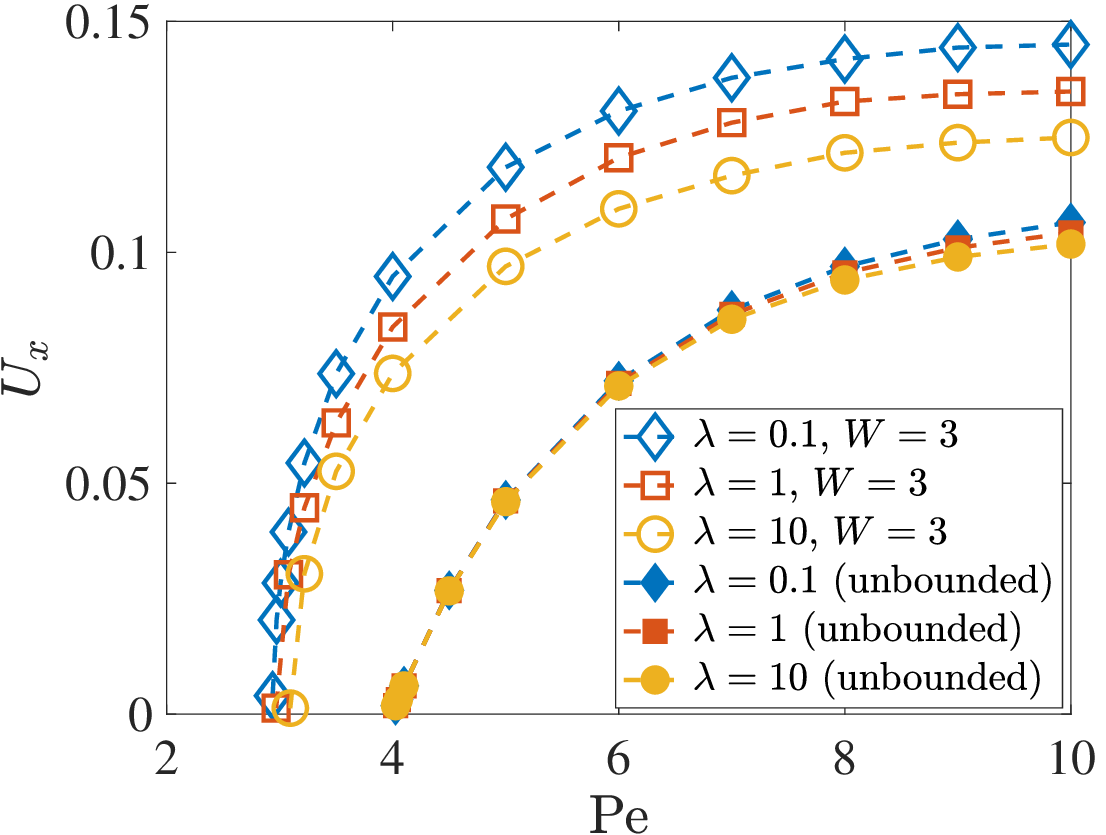}}
\subfloat[]{\label{res_onlyMar_h}\includegraphics[width=6cm]{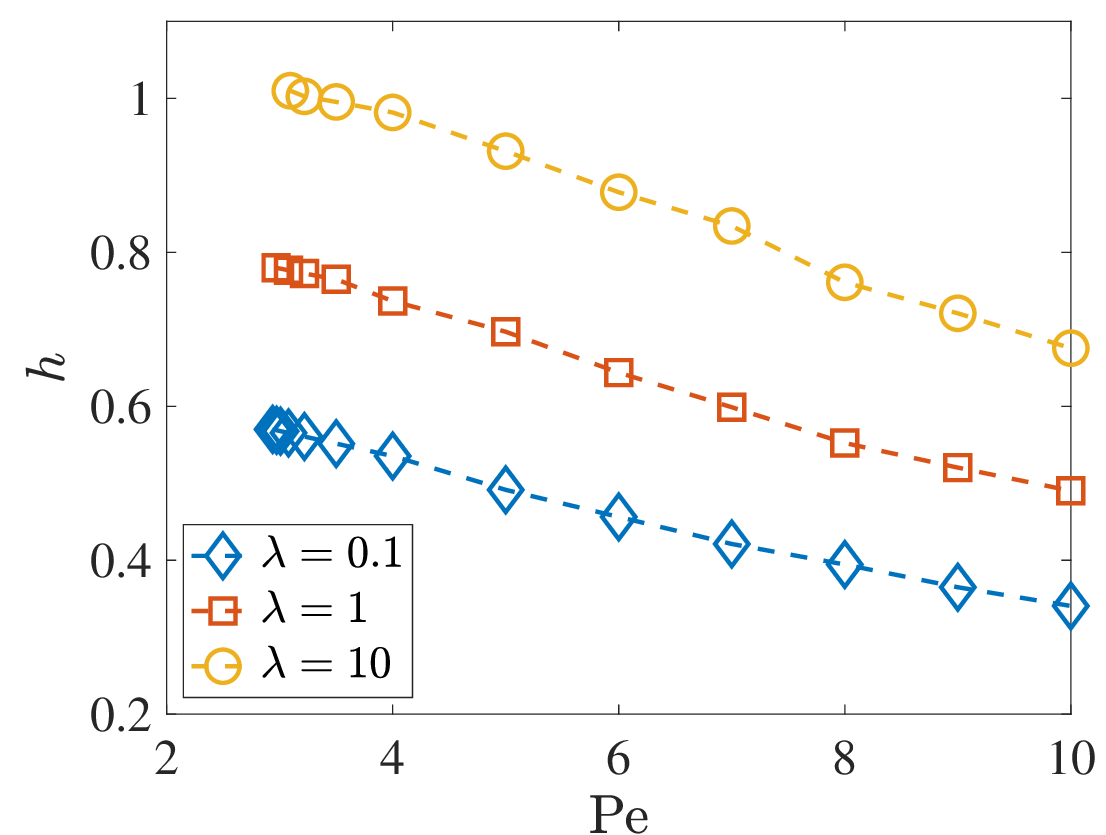}}
\\
\subfloat[]{\label{flow_conc_onlyMar_a}\includegraphics[width=5.5cm]{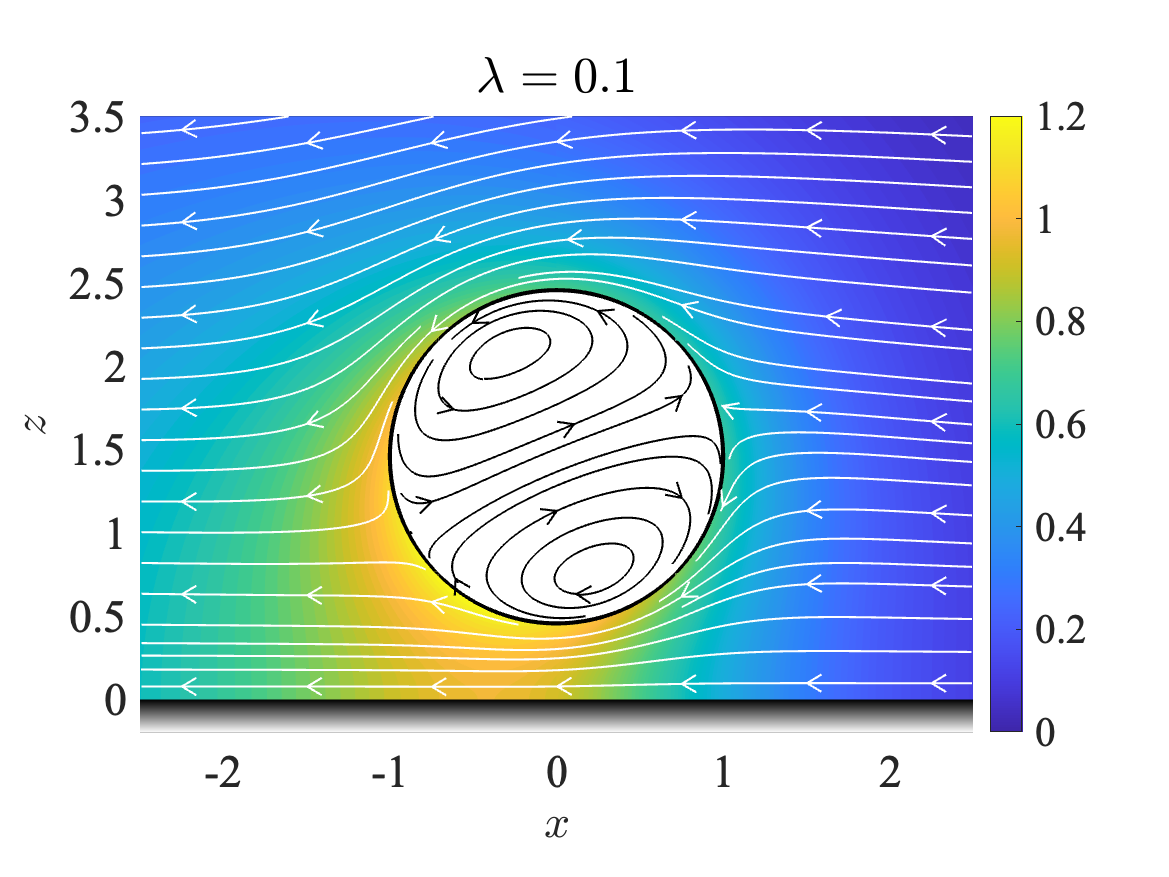}}
\subfloat[]{\label{flow_conc_onlyMar_c}\includegraphics[width=5.5cm]{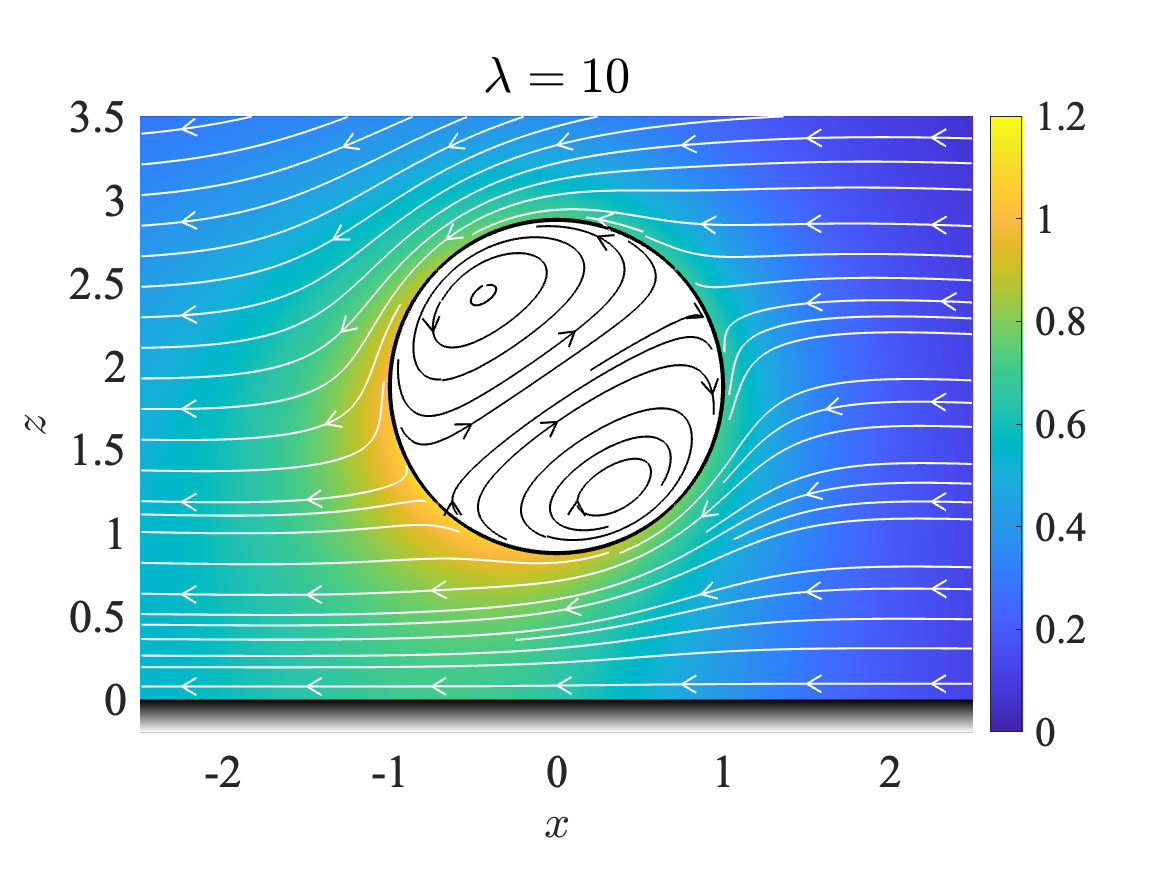}}

\end{center}

\caption{(a) Steady-state swimming speed, $U_x$, and, (b) the corresponding drop-to-wall separation, $h$, for an active drop swimming parallel to a rigid wall and experiencing a wall-normal external force $W=3$. Also shown in panel (a) are the results for an active drop swimming in an unbounded fluid ($h\to\infty$). The mobility ratio, $\tilde{m} \equiv 0$, meaning that the drops propel due to purely Marangoni forcing. (c, d) Solute distribution around the active drop, along with the associated streamlines. The vertical external force in each case is $W=3$, and the P\'eclet number $\Pe = 6$. The panels correspond to different viscosity ratios, (c) $\lambda = 0.1$, (d) $\lambda = 10$. The streamlines are plotted in a reference frame co-moving with the drop.}
\label{res_onlyMar}

\end{figure}

We first consider the motion of an active drop that swims solely due to Marangoni stresses, with the flow field being continuous across the interface (see eqn.~\eqref{tau_BC_dim} and eqn.~\eqref{u_BC_dim} with $\mathcal{M}\equiv0$). In terms of the dimensionless parameters defined in Section~\ref{dl_pars}, this means $\tilde{m}\equiv0$. The results are summarized in Fig.~\ref{res_onlyMar}, with focus on the effect of the drop's viscosity ratio $\lambda$ on its swimming speed $U_x$ (and swimming separation $h$) as a function of the P\'eclet number $\Pe$. We note that for the unbounded case ($h\to\infty$), there exists only a modest difference between the swimming speeds of drops with different viscosity ratios. However, wall proximity introduces significant differences w.r.t. the viscosity ratio. Fig.~\ref{res_onlyMar_Ux} shows that for all $\lambda$, the swimming speed increases rapidly with $\Pe$ for $\Pe \gtrsim \Pe_\textrm{c}$ and saturates for higher $\Pe$. The drop-to-wall separation $h$, in the swimming state, reduces progressively as $\Pe$ increases (Fig.~\ref{res_onlyMar_h}). An increase in $\Pe$ enhances solute advection downstream of the drop's direction of motion (along $-\mathbf{e}_x$ in Figs.~\ref{flow_conc_onlyMar_a}, \ref{flow_conc_onlyMar_c}), which reduces the solute accumulation below the drop. The drop thus settles to a lower separation such that the vertical polarity in solute concentration once again becomes strong enough to withstand the external force $W$. This qualitative behavior is identical to that of a diffusiophoresis-driven rigid active particle~\citep{Desai2022}, except for a fundamental difference: the rigid particle rotates (about the $y$-axis) as it translates but the drop does not. This is expected since a viscous drop cannot, by definition, sustain a hydrodynamic torque.

The swimming speed of an active drop of viscosity ratio $\lambda$ can be quantified by using the reciprocal theorem for Stokes flows:
\begin{equation}\label{reciproc_drop_lam}
    U_x = \int_{\partial V} { \nabla_s C \cdot \left[ \left( 3 \lambda + 2 \right) \mathbf{u}^\parallel_\textrm{aux} \right] \,dS }.
\end{equation}
In this instance, the reciprocal theorem is applied in a reference frame fixed to the drop and the auxiliary problem is the motion of a passive, clean drop moving along the wall under the action of an external force $\mathbf{F}^\parallel_\textrm{aux}=\mathbf{e}_x$. A general derivation of eqn.~\eqref{reciproc_drop_lam} has been included in Appendix~\ref{reciproc_derive}. Like eqn.~\eqref{reciproc_drop_lam_FHz}, eqn.~\eqref{reciproc_drop_lam} states that the contribution of $\nabla_sC$ toward the swimming speed of the drop, $U_x$, is influenced by the function $\left( 3\lambda + 2 \right) \mathbf{u}^\parallel_\textrm{aux}$.

\begin{figure}[t]
\begin{center}

\subfloat[]{\label{infl_plots_Pe6_onlyMar_a}\includegraphics[width=5.5cm]{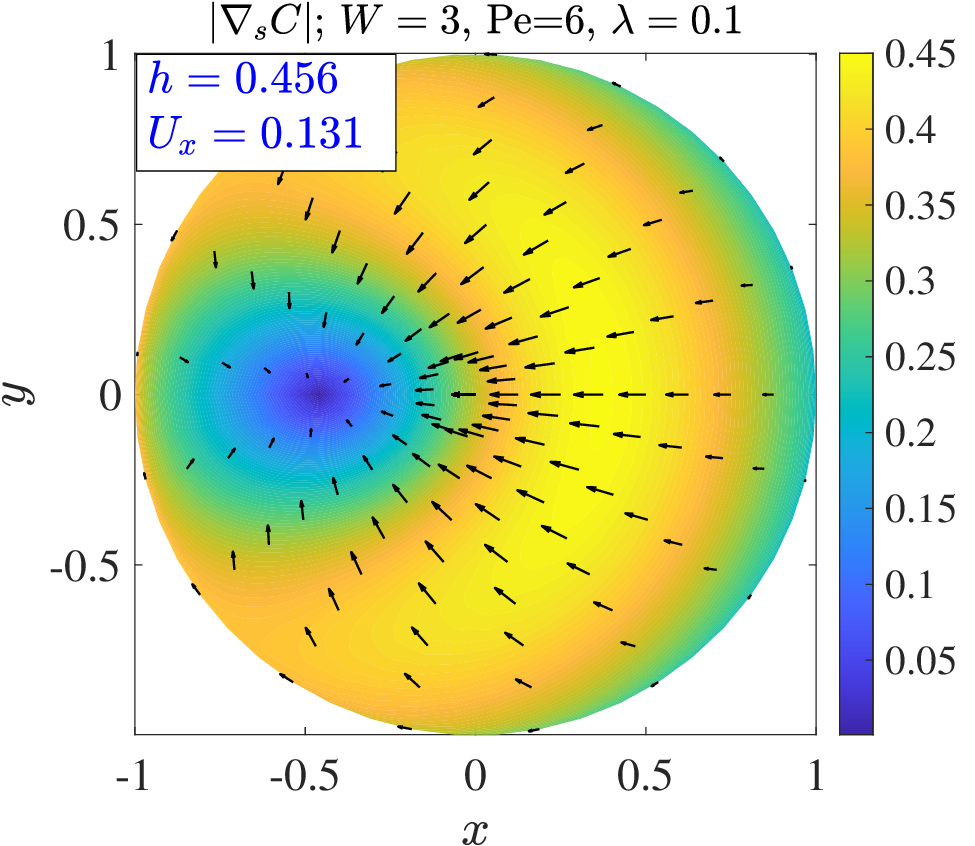}}
\subfloat[]{\label{infl_plots_Pe6_onlyMar_c}\includegraphics[width=5.5cm]{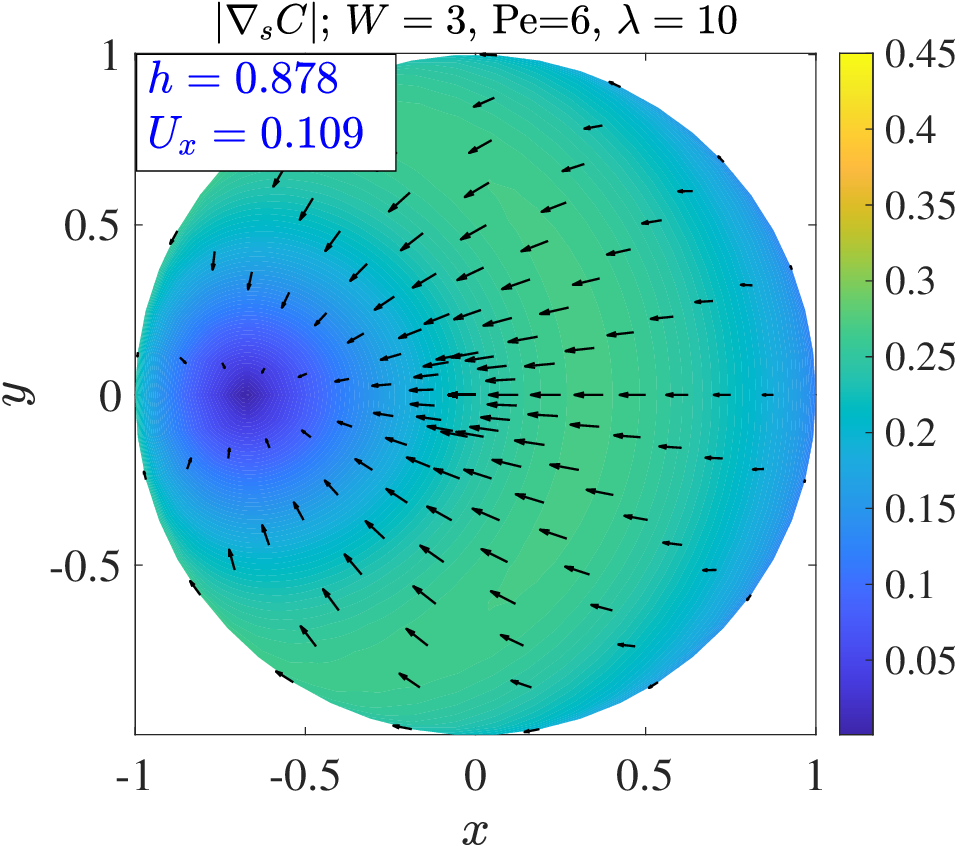}}

\subfloat[]{\label{infl_plots_Pe6_onlyMar_d}\includegraphics[width=5.5cm]{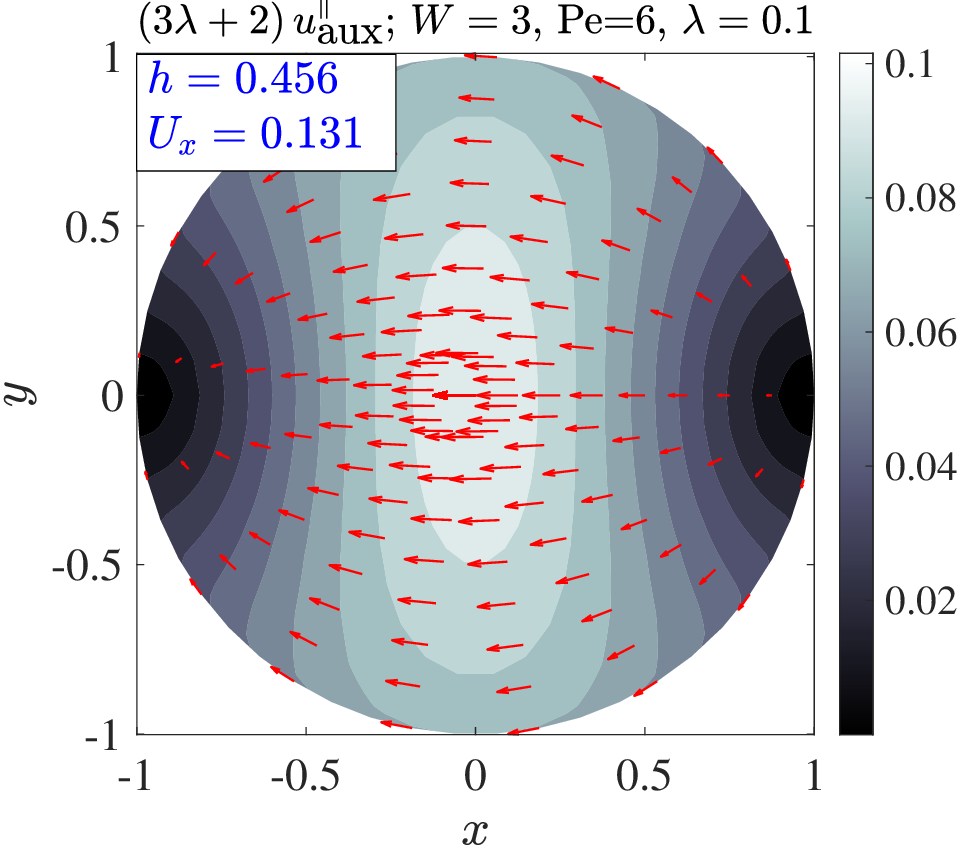}}
\subfloat[]{\label{infl_plots_Pe6_onlyMar_f}\includegraphics[width=5.5cm]{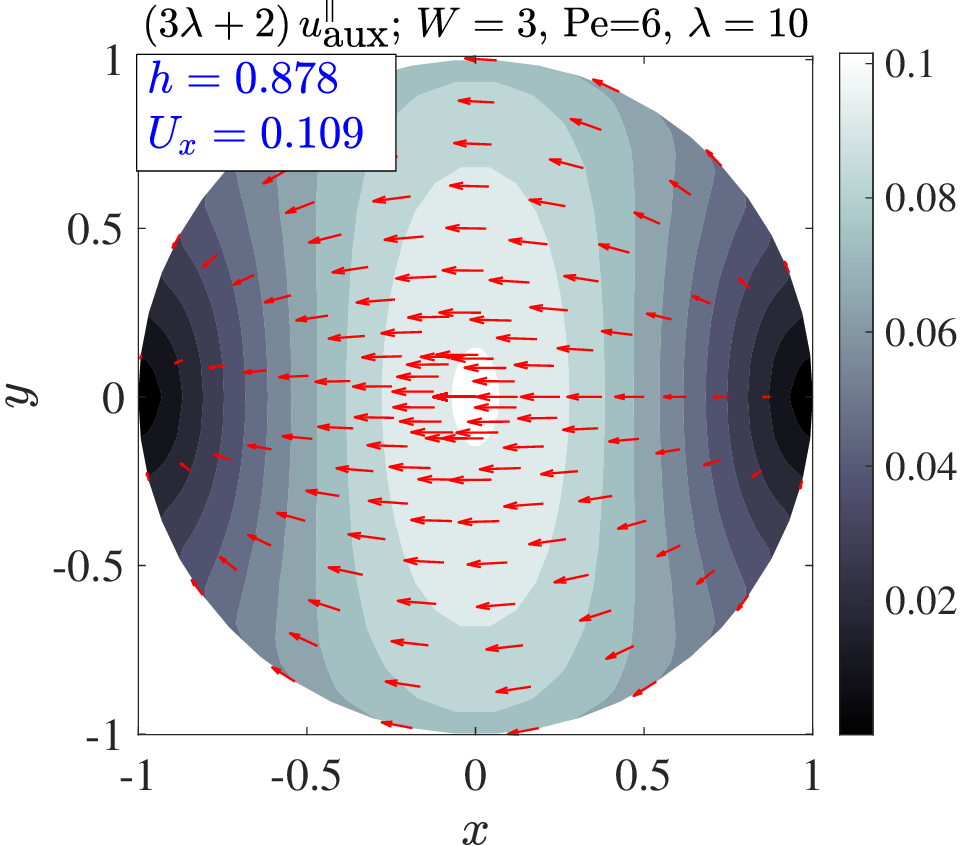}}

\caption{The magnitude of the concentration gradient $\nabla_s C$ (top row) and the influence function $\left( 3\lambda + 2 \right)\mathbf{u}^\parallel_\textrm{aux}$ (bottom row), at the bottom hemisphere of an active drop swimming next to a wall $\left( u^\parallel_\textrm{aux} = \left| \mathbf{u}^\parallel_\textrm{aux} \right| \right)$. $\Pe=6$ in all cases, and two different viscosity ratios are considered: $\lambda=0.1$ for panels (a) and (c), and $\lambda=10$ for panels (b) and (d). The mobility ratio for these figures is identically zero, $\tilde{m} \equiv 0$. The vectors denote concentration gradients and the (tangential) components of the auxiliary flow $\mathbf{u}_\textrm{aux}^\parallel$.}
\label{infl_plots_Pe6_onlyMar}

\end{center}
\end{figure}

Fig.~\ref{infl_plots_Pe6_onlyMar} shows the distribution of $\nabla_sC$ and $\left( 3\lambda + 2 \right) \mathbf{u}^\parallel_\textrm{aux}$ at the bottom hemispherical surface of the drop (the side of the drop facing the wall), for two viscosity ratios. We focus on the bottom hemisphere since that is the region with the highest surface concentration of the solute, as well as the strongest concentration gradients (see, for instance, Fig.~\ref{flow_conc_onlyMar_a}). Since the magnitude and distribution of $\left( 3\lambda + 2 \right) \mathbf{u}^\parallel_\textrm{aux}$ does not change appreciably between $\lambda=0.1$ and $\lambda=10$, the speed $U_x$ then depends mainly on the strength of the driving concentration gradients $\nabla_sC$. The latter is highest for the least viscous drops, since they swim closest to the wall and their proximity restricts the ability of the emitted solute to diffuse out and get advected downstream (compare Figs.~\ref{flow_conc_onlyMar_a} and \ref{flow_conc_onlyMar_c}). Hence, drops with lower viscosities tend to swim faster.

\subsection{Influence of the mobility ratio}\label{effect_mc}

\begin{figure}[t]
\begin{center}

\subfloat[]{\label{res_mixed_Ux}\includegraphics[width=6cm]{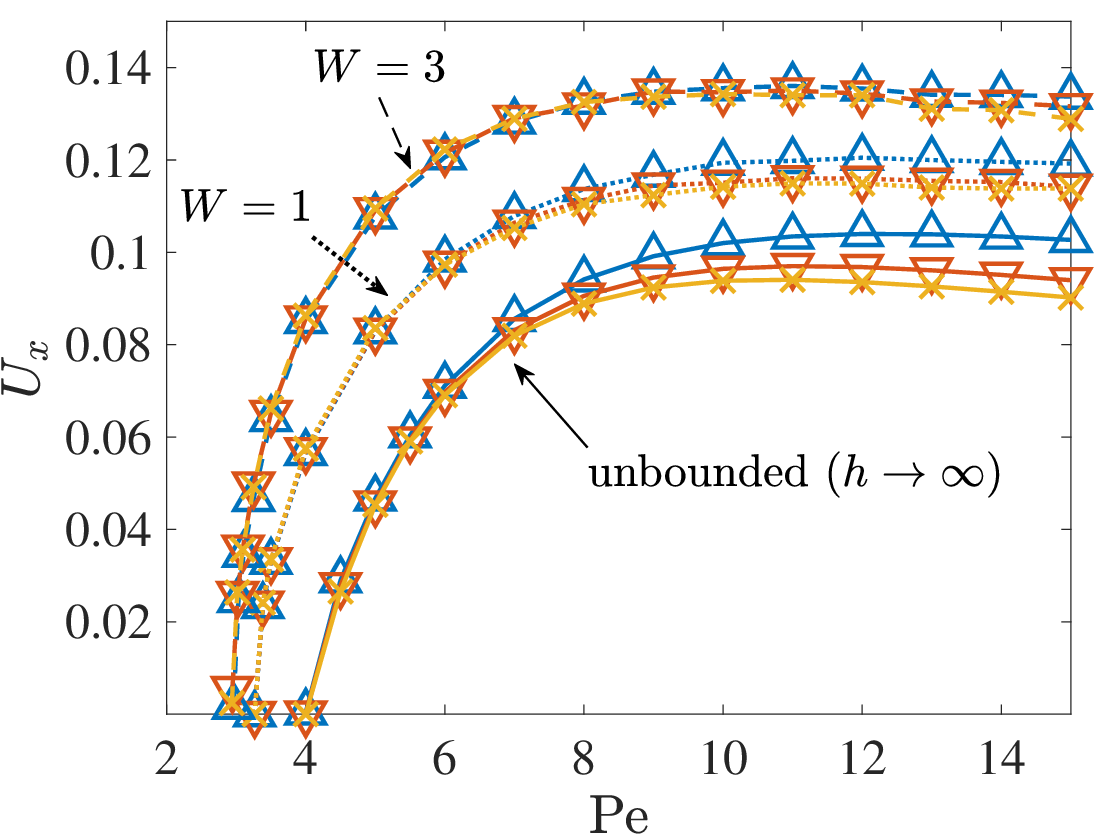}}
\subfloat[]{\label{res_mixed_d}\includegraphics[width=6cm]{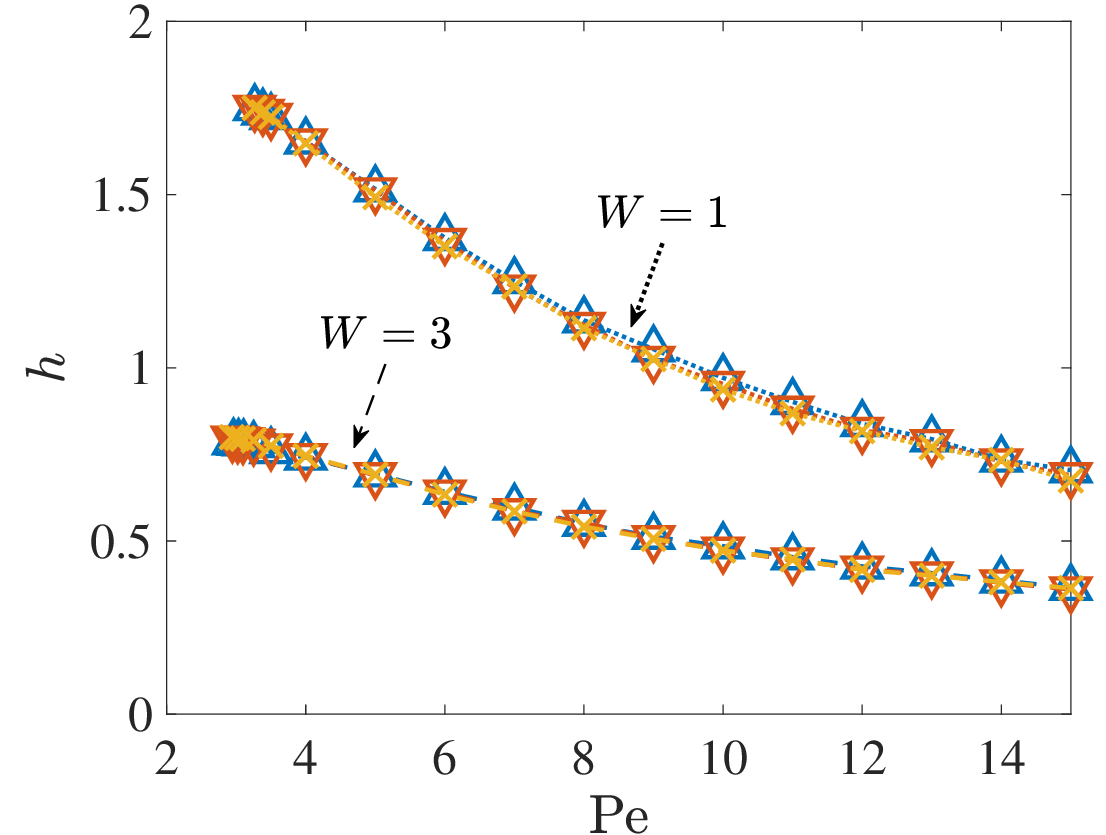}}
\\
\subfloat[]{\label{flow_conc_mixed_a}\includegraphics[width=6cm]{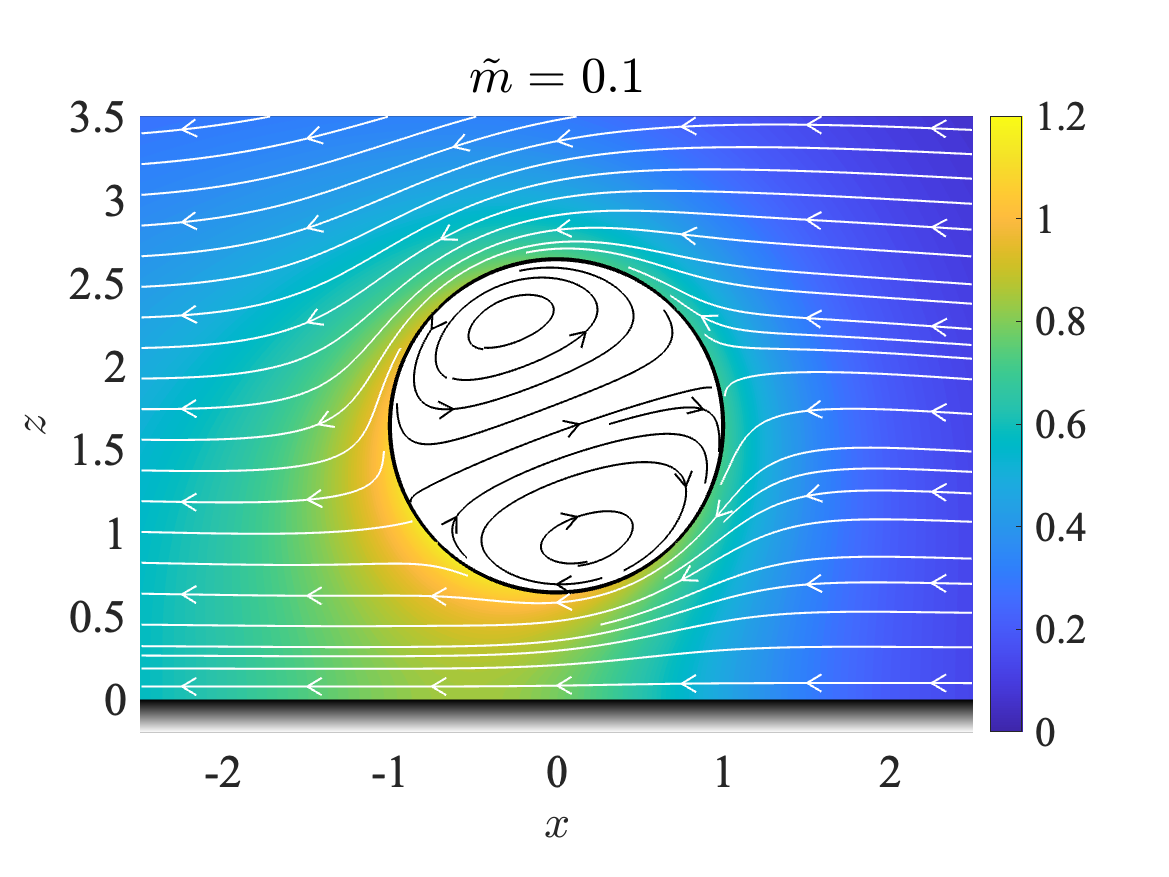}}
\subfloat[]{\label{flow_conc_mixed_c}\includegraphics[width=6cm]{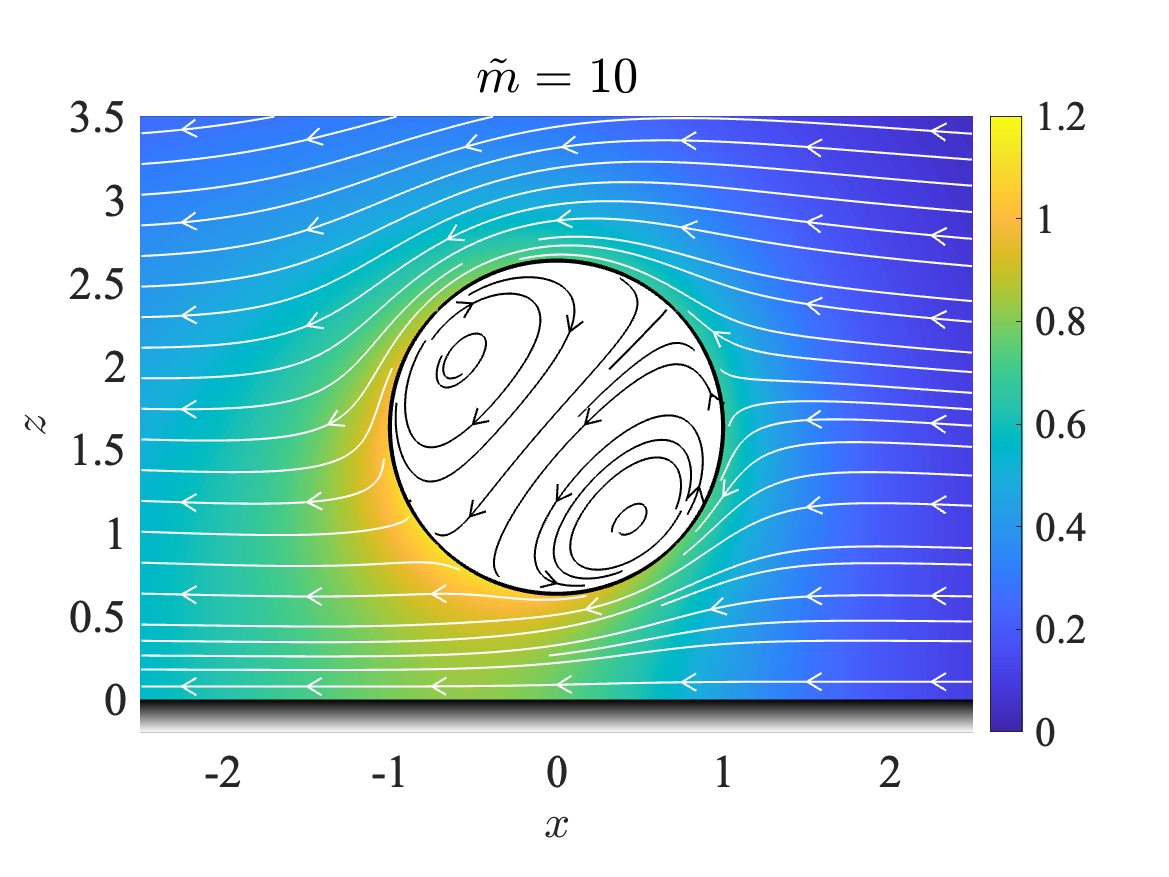}}

\caption{(a) Steady-state swimming speed, $U_x$, and, (b) the corresponding drop-to-wall separation, $h$, for a viscous active drop experiencing wall-normal external force $W = 3$ (dashed lines), $W = 1$ (dotted lines), and in the unbounded case, $h\to\infty$ (solid lines). The results are shown for three mobility ratios (phoretic/Marangoni) $\tilde{m}=0.1$ $(\bigtriangleup)$, $\tilde{m}=1$ $(\bigtriangledown)$ and $\tilde{m}=10$ $(\times)$. The viscosity ratio, $\lambda = 1$, in all cases. (c, d) Solute distribution around the active drop, along with the associated streamlines. The vertical external force in each case is $W=3$, and the P\'eclet number, $\Pe = 6$. The mobility ratios are, (c) $\tilde{m} = 0.1$, (d) $\tilde{m} = 10$. The viscosity ratio in both cases is $\lambda=1$. The streamlines are plotted in a reference frame co-moving with the drop.}
\label{res_mixed}

\end{center}
\end{figure}

We now consider the mobility mechanism wherein surface concentration gradients lead to both Marangoni stresses and diffusiophoretic slip across the interface, i.e $\tilde{m} \ne 0$. Our results are summarized in Fig.~\ref{res_mixed}, which highlights two important points. First, the impact of the mobility ratio on $U_x$ and $h$ reduces as the external force $W$ on the drop increases. In an unbounded fluid, the velocities for different values of $\tilde{m}$ are similar for small $\Pe$, but become reasonably different as $\Pe$ is increased (see Fig.~\ref{res_mixed_Ux} for $h\to\infty$). The difference in swimming speeds for $\Pe \ge 10$ is much smaller for $W=1$, and practically vanishes for $W=3$. Second, drops experiencing larger external force $W$ swim closer to the wall, as expected, and at a higher speed, much like rigid active particles~\citep{Desai2021}.

\begin{figure}[t]
\begin{center}

\subfloat[]{\label{generalize_infl_f}\includegraphics[width=4.5cm]{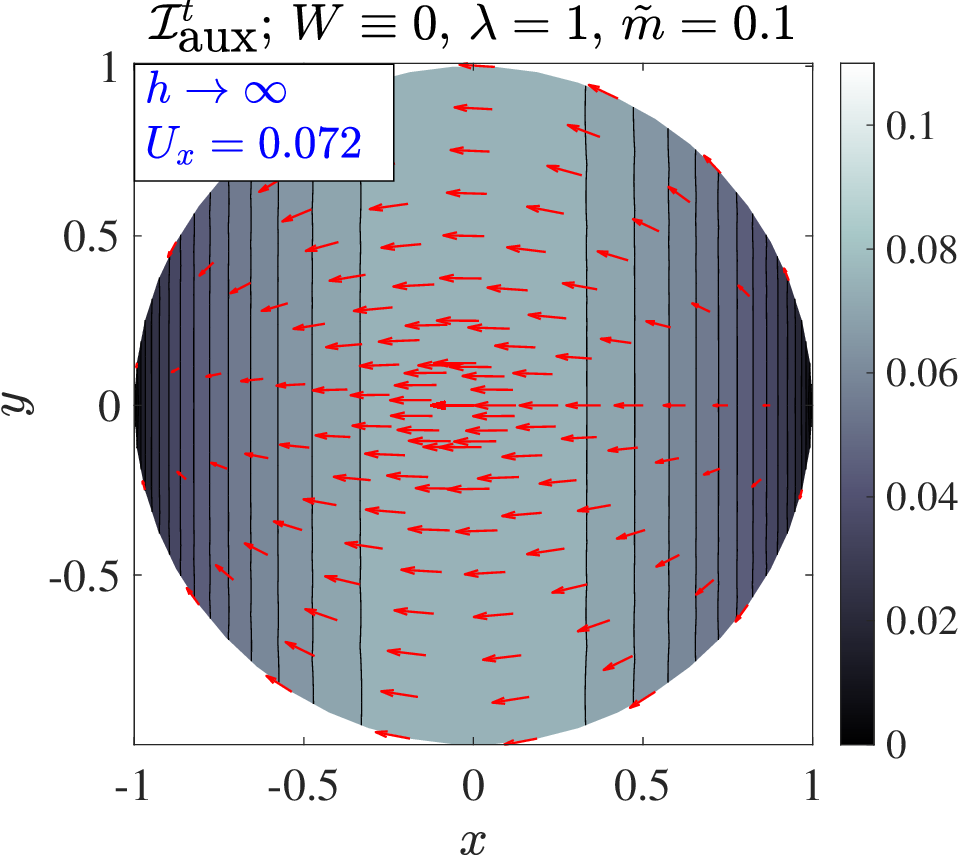}}
\subfloat[]{\label{generalize_infl_c}\includegraphics[width=4.5cm]{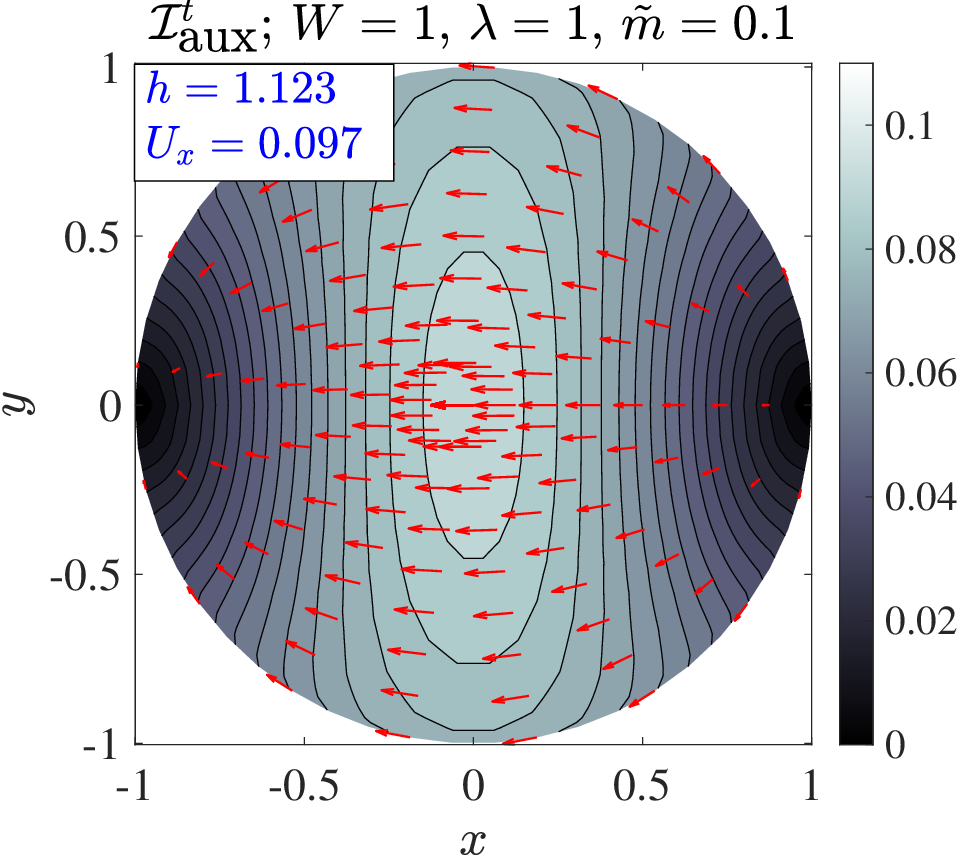}}
\subfloat[]{\label{generalize_infl_d}\includegraphics[width=4.5cm]{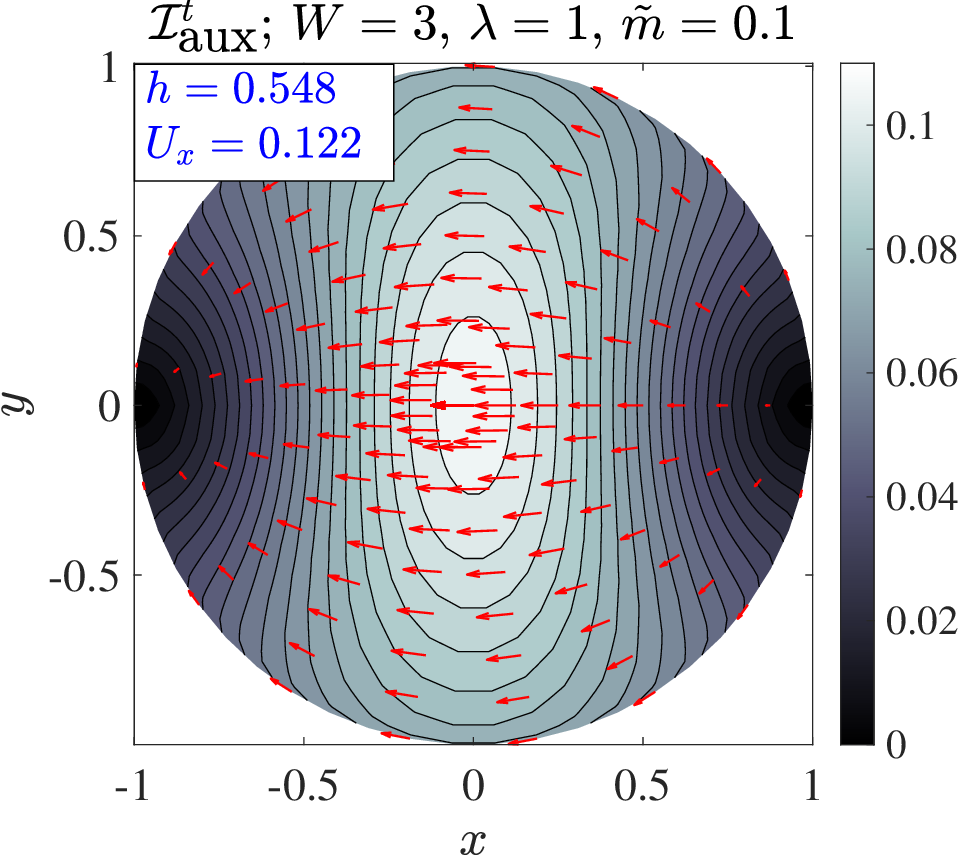}}

\subfloat[]{\label{generalize_infl_e}\includegraphics[width=4.5cm]{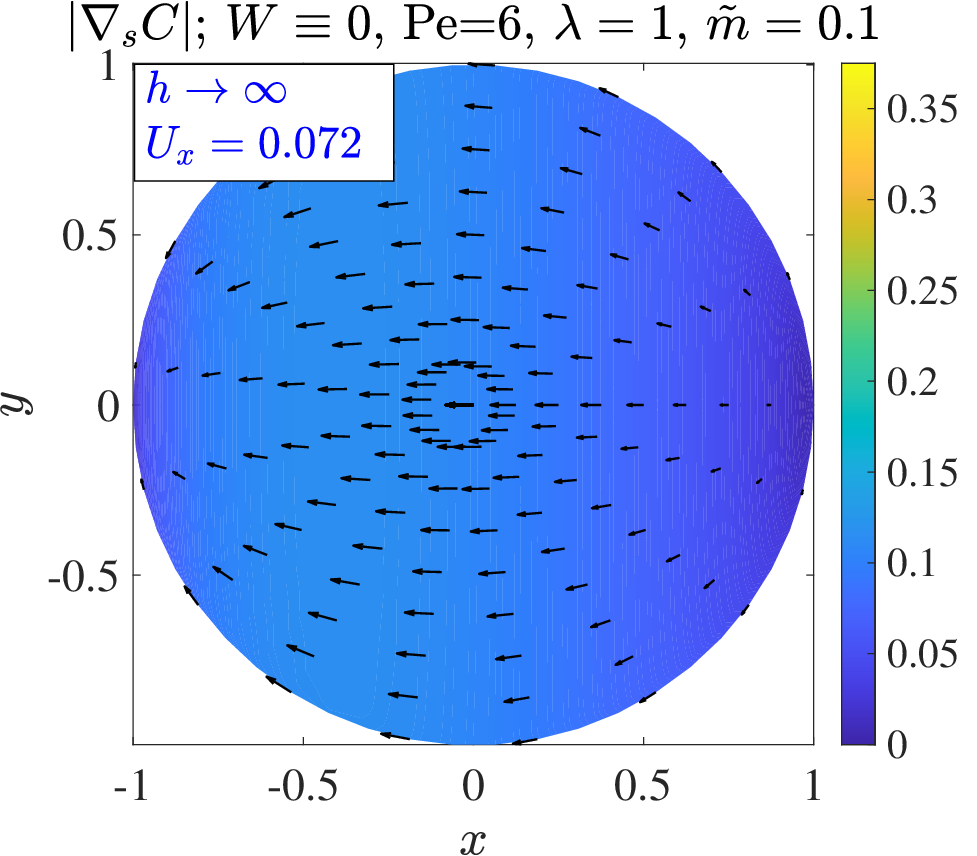}}
\subfloat[]{\label{generalize_infl_a}\includegraphics[width=4.5cm]{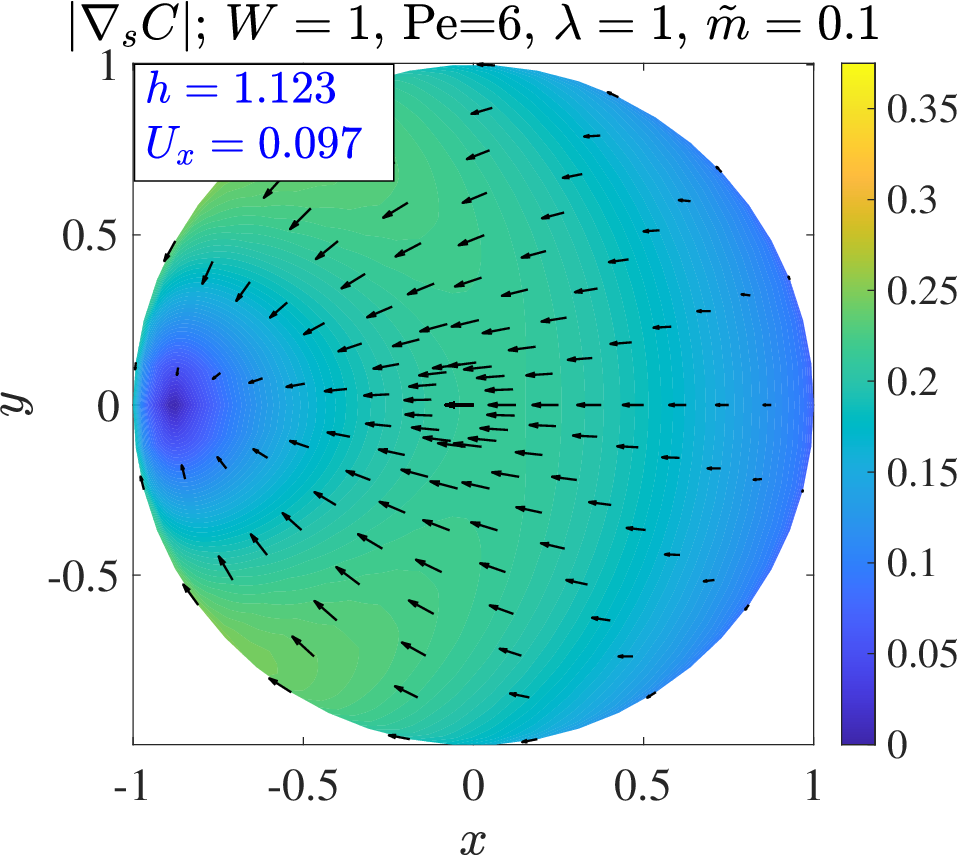}}
\subfloat[]{\label{generalize_infl_b}\includegraphics[width=4.5cm]{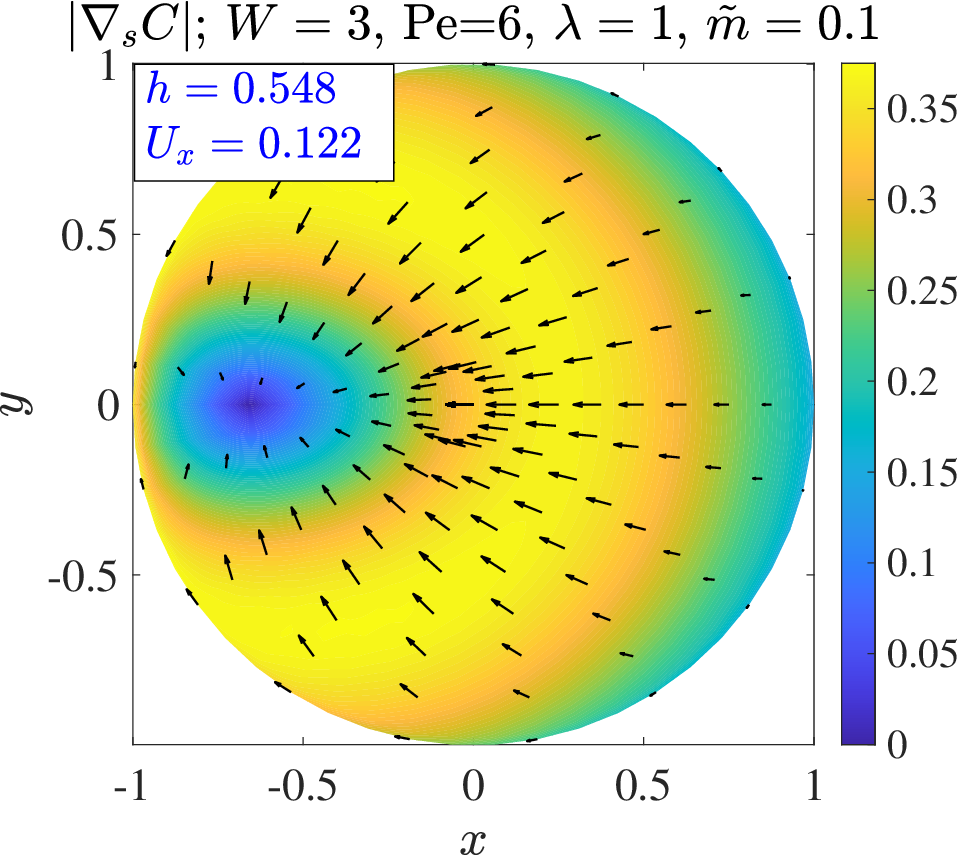}}

\caption{Effect of the wall on the motion of an active drop swimming due to combined diffusiophoretic and Marangoni mobility. (a-c) Distribution of the tangential component of the influence function, $\mathcal{I}^t_\textrm{aux}~=~\left| \left( \boldsymbol{\delta} - \mathbf{nn} \right) \cdot \boldsymbol{\mathcal{I}}_\textrm{aux} \right|$, at drop-to-wall separations corresponding to three different external forces; panel~(a) corresponds to the unbounded case where there is no vertical external force, hence $W\equiv0$. (d-f) Corresponding distribution of the surface concentration gradients, $\left| \nabla_sC \right|$ at $\Pe=6$. The viscosity ratio $\lambda=1$ and the mobility ratio $\tilde{m}=0.1$. The vectors denote local orientations of $\left( \boldsymbol{\delta} - \mathbf{nn} \right) \cdot \boldsymbol{\mathcal{I}}_\textrm{aux}$ and $\nabla_sC$.}
\label{generalize_infl}

\end{center}
\end{figure}

These results can be explained by analyzing the contribution of the concentration gradients toward the swimming speed of the drop. Once again, the Stokes reciprocal theorem yields,
\begin{equation}\label{reciproc_drop_mc}
    U_x = \int_{\partial V} { \nabla_s C \cdot \boldsymbol{\mathcal{I}}_\textrm{aux} \,dS },
\end{equation}
where the auxiliary problem is the same as that considered in Section~\ref{effect_lambda}, but the influence function $\boldsymbol{\mathcal{I}}_\textrm{aux}$ is more general,
\begin{equation}\label{infl_gen}
    \boldsymbol{\mathcal{I}}_\textrm{aux} = \frac{\tilde{m} \left(3 \lambda + 2\right)}{1 + 3 \lambda \tilde{m}} \mathbf{n} \cdot \boldsymbol{\sigma}^\parallel_\textrm{aux} + \frac{3 \lambda + 2}{1 + 3 \lambda \tilde{m}} \mathbf{u}^\parallel_\textrm{aux},
\end{equation}
where, $\mathbf{n} \cdot \boldsymbol{\sigma}^\parallel_\textrm{aux}$ is the traction associated with the auxiliary Stokes flow problem. For the unbounded case, there exists an analytical expression for $\boldsymbol{\mathcal{I}}_\textrm{aux}$, which can be written by using the Hadamard–Rybczynski solution for a clean drop~\citep{Leal2007, Pak2014}:
\begin{subequations}
\begin{align}
   \mathbf{n} \cdot \boldsymbol{\sigma}^\parallel_{\textrm{aux},h\to\infty} &= -\frac{3}{4\pi} \frac{1}{3\lambda+2} \left( \lambda\boldsymbol{\delta} + 2 \mathbf{nn}\right) \cdot \mathbf{e}_x, \\
   \mathbf{u}^\parallel_{\textrm{aux},h\to\infty} &= -\frac{1}{4\pi} \frac{1}{3\lambda+2} \left( \boldsymbol{\delta} - \mathbf{nn}\right) \cdot \mathbf{e}_x,
\end{align}
\end{subequations}
which yields,
\begin{equation}\label{infl_gen_infty}
    \boldsymbol{\mathcal{I}}_{\textrm{aux},h\to\infty} = -\frac{1}{4\pi} \left( \boldsymbol{\delta} - \frac{6\tilde{m}-1}{1+3\lambda\tilde{m}} \mathbf{nn} \right) \cdot \mathbf{e}_x.
\end{equation}
For the wall-bounded cases, $\boldsymbol{\mathcal{I}}_{\textrm{aux}}$ needs to be evaluated numerically. The tangential component of the general influence function, $\mathcal{I}^t_\textrm{aux}= \left| \left( \boldsymbol{\delta} - \mathbf{nn} \right) \cdot \boldsymbol{\mathcal{I}}_\textrm{aux} \right|$, is shown in Figs.~\ref{generalize_infl_f} to \ref{generalize_infl_d} for three cases: $h \to \infty$ (or $W\equiv0$), and $W=1,3$. Note that only this tangential component contributes to the dot product in eqn.~\eqref{reciproc_drop_mc}. An increase in wall proximity causes the region of highest influence (large $\mathcal{I}^t_\textrm{aux}$) to shrink and focus around $\left( x=0,y=0,z=h \right)$, which is the point on the drop that is closest to the wall (see also Fig.~\ref{bisp_coords}). In the absence of a wall, the influence is more ``spread out'', and changes in interfacial flow due to Marangoni stresses and/or diffusiophoresis have a substantial effect on the swimming speed. In the presence of a wall, the sensitivity to $\tilde{m}$ is reduced, since the maximum contribution to self-propulsion comes from a much smaller fraction of the surface.

\begin{figure}[t]
\begin{center}

\subfloat[]{\label{res_drop_vs_part_a}\includegraphics[width=6cm]{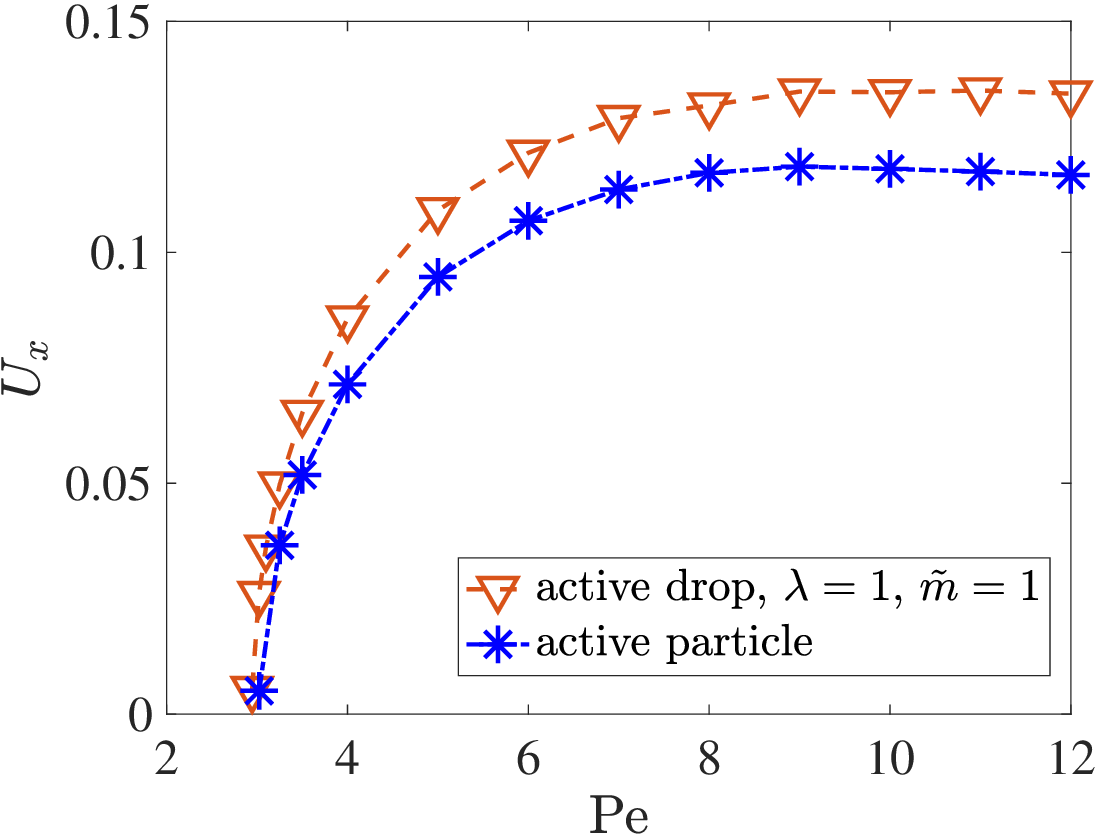}}
\subfloat[]{\label{res_drop_vs_part_b}\includegraphics[width=6cm]{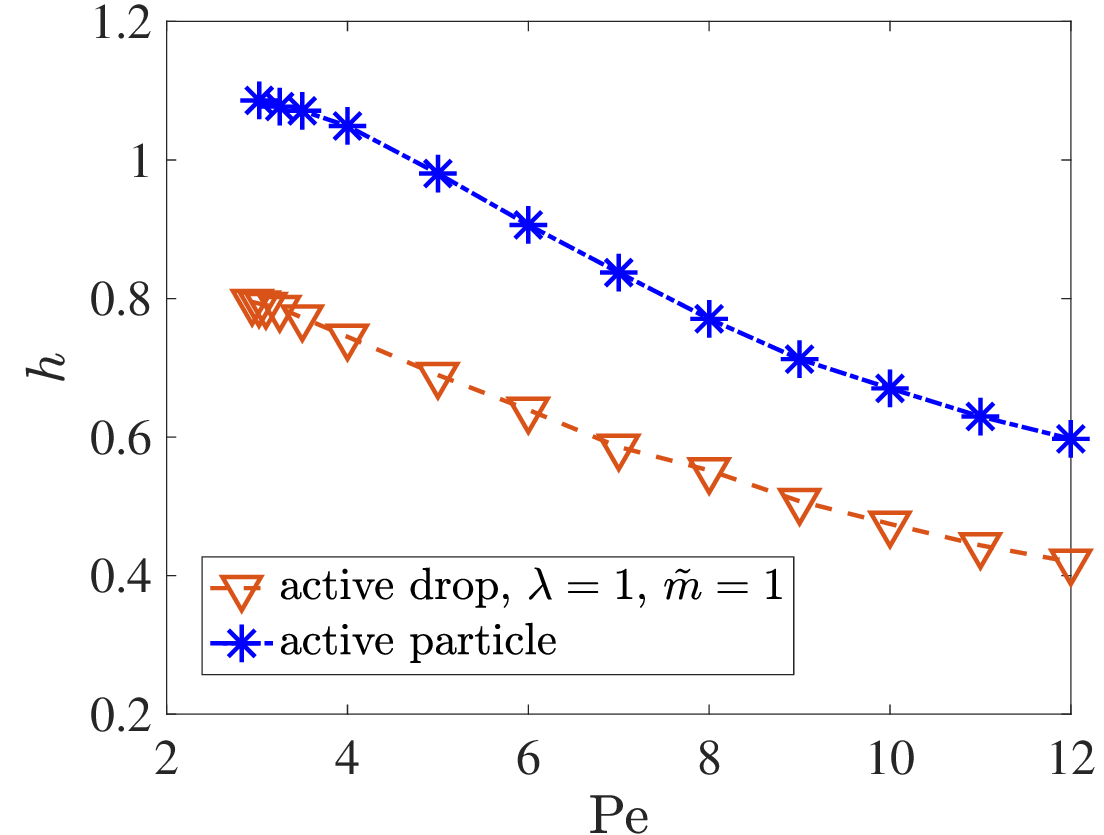}}

\caption{(a) Steady-state swimming speed, $U_x$, and, (b) the corresponding drop-to-wall separation, $h$, for a viscous active drop and a rigid active (phoretic) particle experiencing the same wall-normal external force $W = 3$. For the drop, the viscosity ratio is $\lambda = 1$, and the mobility ratio is $\tilde{m} = 1$.}
\label{res_drop_vs_part}

\end{center}
\end{figure}

The relative magnitude and distribution of $\mathcal{I}^t_\textrm{aux}$ also explains why drops experiencing higher external force $W$ swim faster. Figs.~\ref{generalize_infl_e} to \ref{generalize_infl_b} show the distribution of the concentration gradient $\nabla_sC$ at the wall-facing surface of the drop. Clearly, the gradients are much stronger for the `heavier drop' ($W=3$) and so is their influence on self-propulsion (compare across columns in Fig.~\ref{generalize_infl}). This mechanism, wherein wall proximity enhances the driving interfacial flows and localizes them around the most hydrodynamically efficient/influential regions, is identical to what is seen for rigid phoretic particles~\citep{Desai2022}. Moreover, if the external force is fixed, then the evolution (with $\Pe$) of the swimming speed and separation of a rigid active particle is very similar to that of a fluid active drop, except that the particle speeds are slightly smaller (Figs.~\ref{res_drop_vs_part}). Thus, wall-proximity enhances the swimming speed of fluid active drops in the same qualitative and quantitative manner as it does for isotropic active particles, owing to the same underlying physical mechanisms.

\section{Flow field inside and around the drop}\label{flowFields}

\subsection{Influence of the viscosity ratio}
\begin{figure}[t]
\begin{center}

\subfloat[]{\label{in_flow_lam_a}\includegraphics[width=4.25cm]{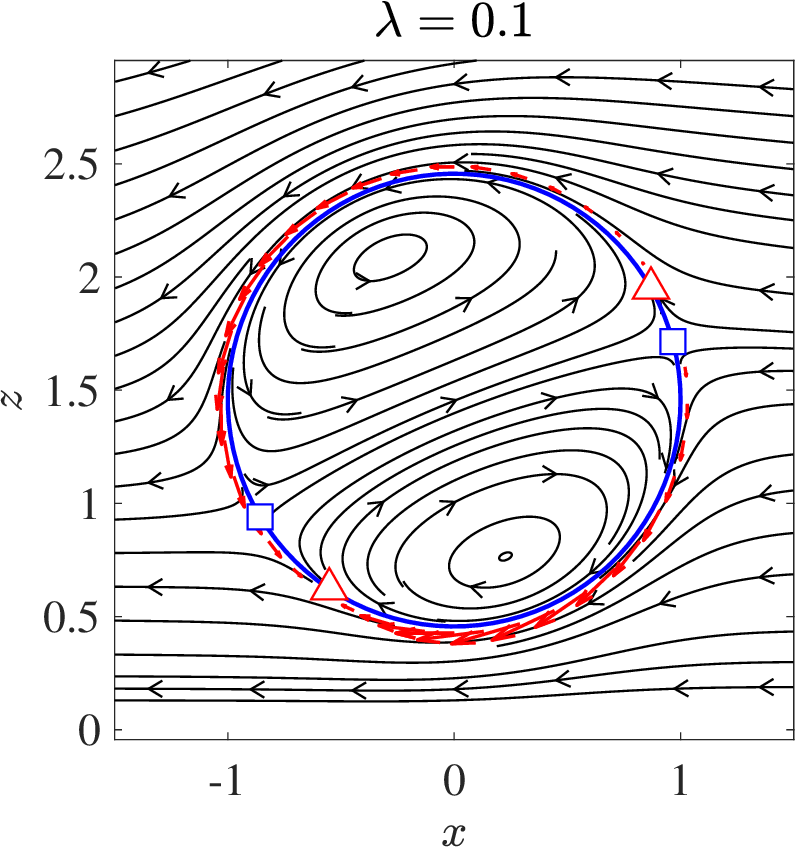}}
\subfloat[]{\label{in_flow_lam_b}\includegraphics[width=4.25cm]{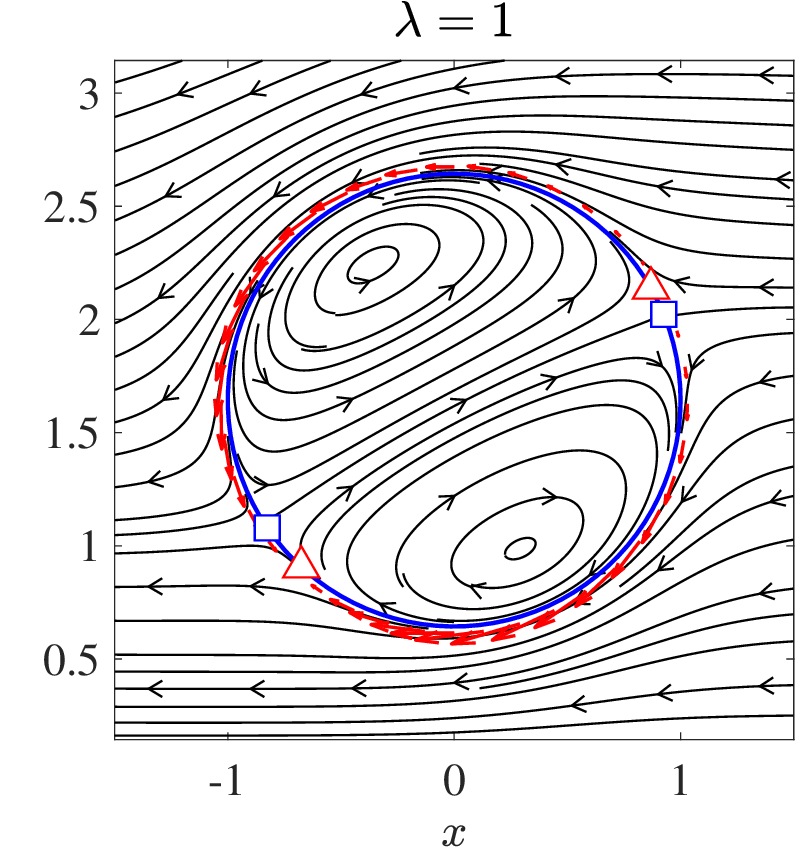}}
\subfloat[]{\label{in_flow_lam_c}\includegraphics[width=4.25cm]{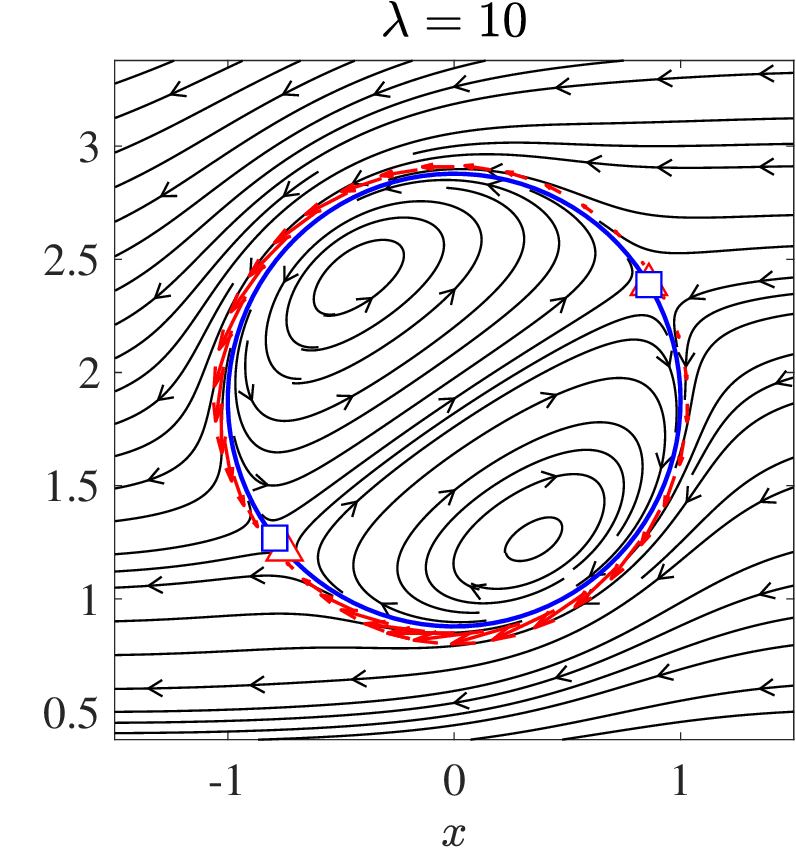}}

\caption{Flow field around and inside an active drop as a function of the viscosity ratio, $\lambda$, shown here on the $y=0$ plane. The red vectors denote the concentration gradients at the drop surface. The white squares and triangles denote surface locations corresponding to zero fluid velocity and solute concentration gradient, respectively. The drops are all swimming along $\mathbf{e}_x$. The other parameters are: $W=3, \Pe=6, \tilde{m}\equiv0$.}
\label{in_flow_lam}

\end{center}
\end{figure}

Even though the drop-to-wall separation changes w.r.t. the viscosity ratio, the structure of the flow, both internal and external, undergoes only a minor change: the stagnation points of the fluid velocity and the surface concentration gradient move closer to each other as $\lambda$ increases from $\lambda=0.1$ to $\lambda=10$ (see Fig.~\ref{in_flow_lam}).

\subsection{Influence of the mobility ratio}
\begin{figure}[t]
\begin{center}

\subfloat[]{\label{rot_in_flow_a}\includegraphics[width=4.25cm]{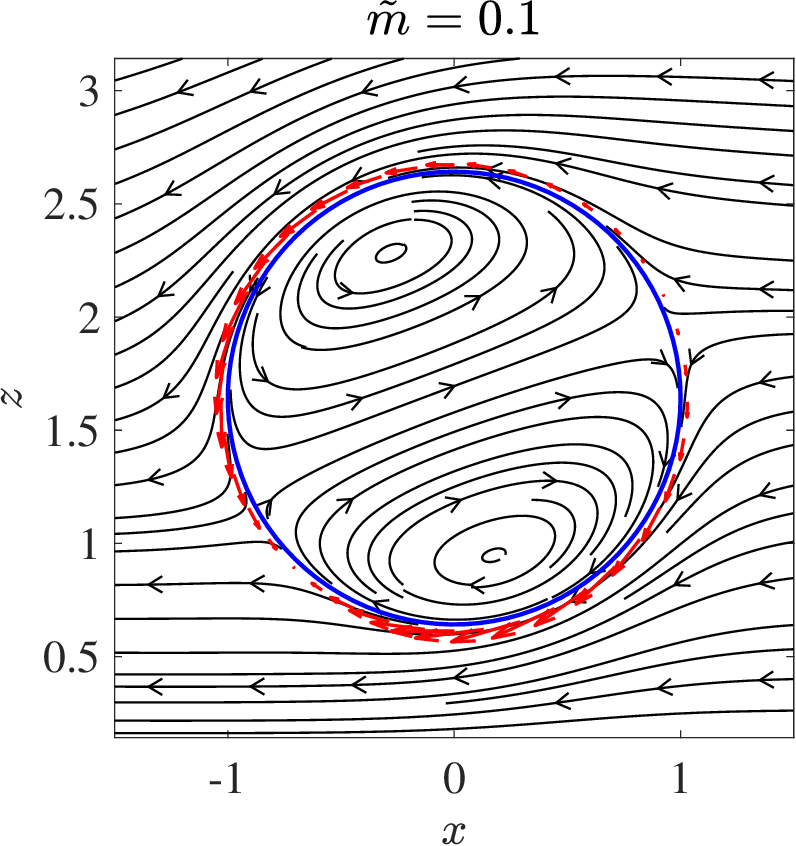}}
\subfloat[]{\label{rot_in_flow_b}\includegraphics[width=4.25cm]{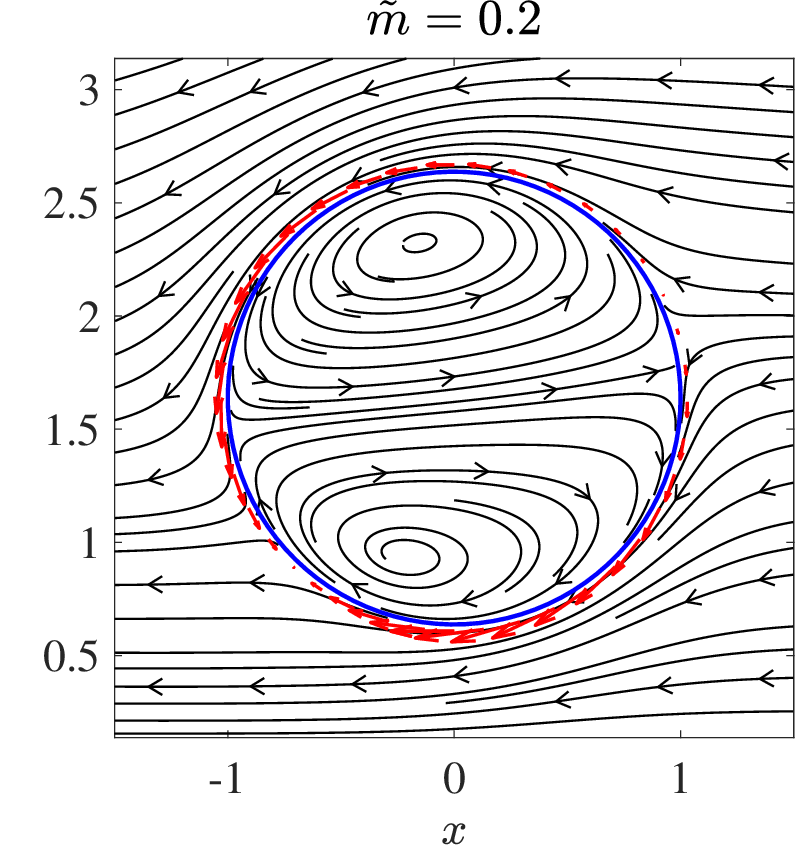}}
\subfloat[]{\label{rot_in_flow_c}\includegraphics[width=4.25cm]{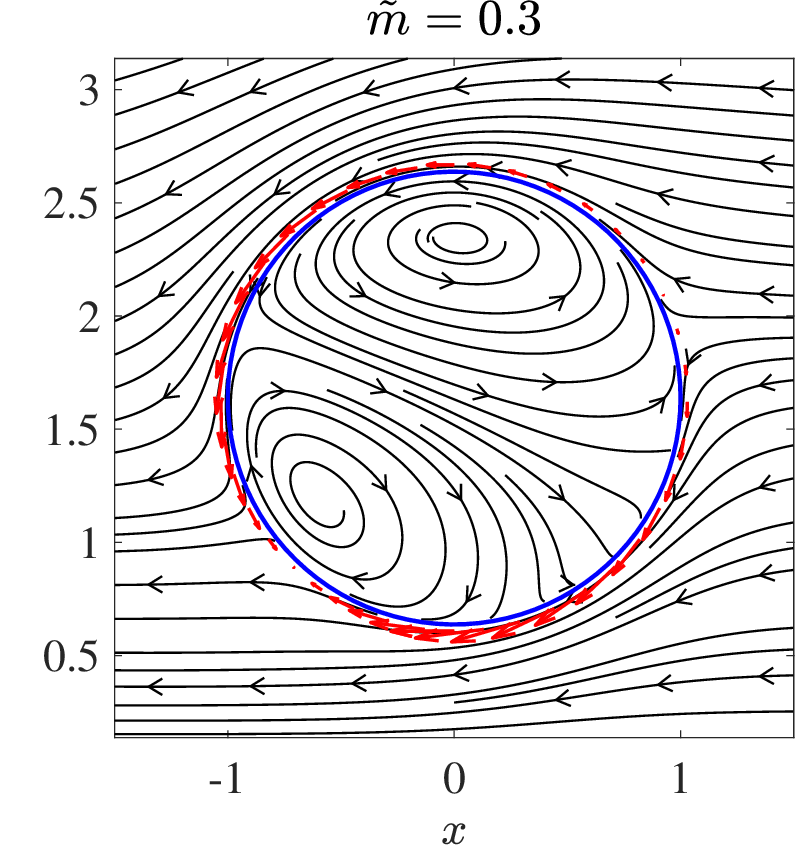}}
\\
\subfloat[]{\label{rot_in_flow_d}\includegraphics[width=4.25cm]{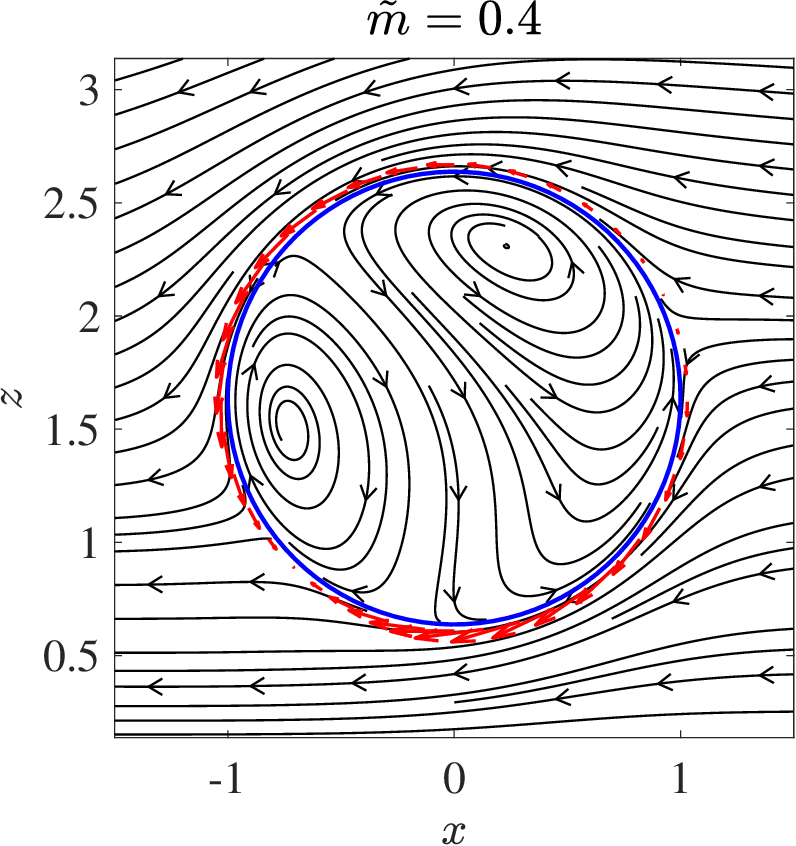}}
\subfloat[]{\label{rot_in_flow_e}\includegraphics[width=4.25cm]{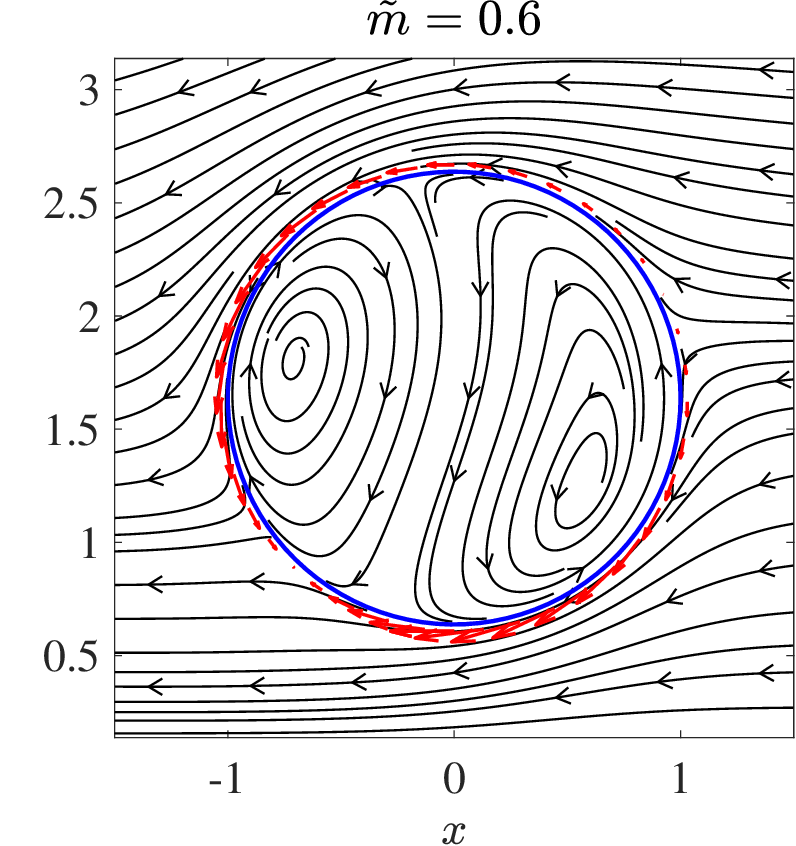}}
\subfloat[]{\label{rot_in_flow_f}\includegraphics[width=4.25cm]{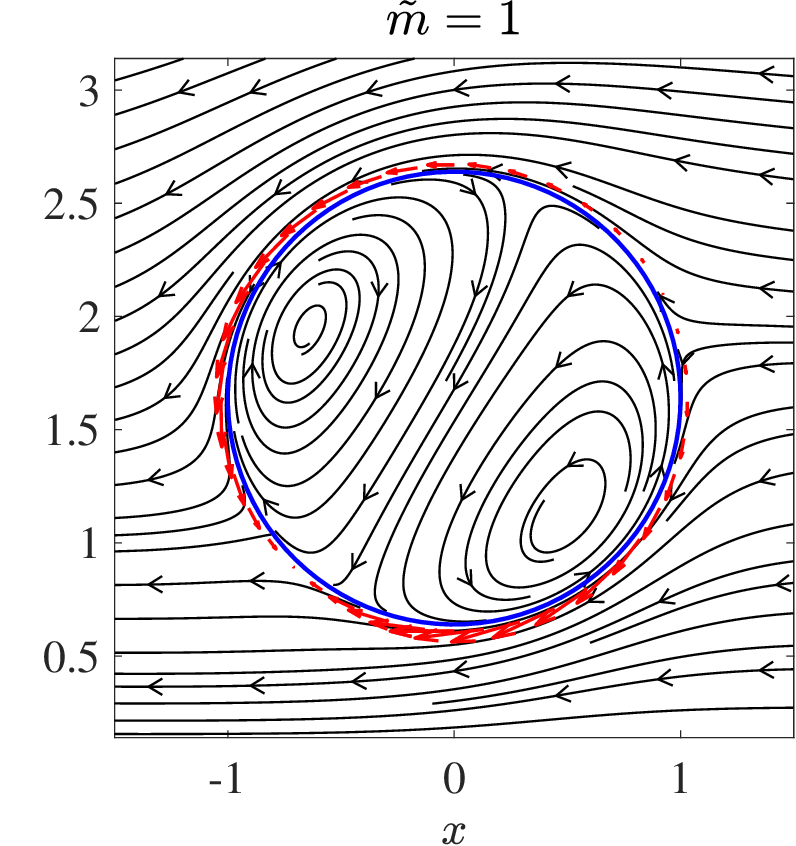}}

\caption{Flow field around and inside an active drop as a function of the mobility ratio, $\tilde{m}$, shown here on the $y=0$ plane. The red vectors denote the concentration gradients at the drop surface. It is apparent that the concentration gradients and the external flow remain essentially unchanged across all mobility ratios (see also Fig.~\ref{analyze_stag_b}). The drops are all swimming along $\mathbf{e}_x$. The other parameters are: $W=3, \Pe=6, \lambda=1$.}
\label{rot_in_flow}

\end{center}
\end{figure}

An increase in the mobility ratio significantly alters the flow field inside the drop, which is reversed w.r.t. the external flow (compare the streamlines inside the drop between Figs.~\ref{flow_conc_mixed_a} and \ref{flow_conc_mixed_c}). This is directly because of the phoretic slip boundary condition (eqn.~\eqref{u_BC_dim}). A dimensionless and scalar form of eqn.~\eqref{u_BC_dim} is:
\begin{equation}\label{u_BC_non_dim}
    \left( \mathbf{u}^\textrm{o} - \mathbf{u}^\textrm{i} \right) \cdot \mathbf{e}_{\theta_s} = \frac{\tilde{m}\left( 3\lambda+2\right)}{\left( 1 + 3\lambda\tilde{m} \right)} \nabla_s C \cdot \mathbf{e}_{\theta_s},
\end{equation}
which has been written for the drop surface intersecting with the $y=0$ plane (i.e. on the surface visible in Figs.~\ref{flow_conc_mixed_a} and \ref{flow_conc_mixed_c}). Due to symmetry, the azimuthal or $\phi$-component of the tangential velocity is identically zero in this plane. In eqn.~\eqref{u_BC_non_dim}, $\theta_s$ is an angular coordinate measured counter-clockwise such that the point $\theta_s=0$ (also $\theta_s=2\pi$) is closest to the wall and $\theta_s=\pi$ is the furthest; $\mathbf{e}_{\theta_s}$ is a tangent unit vector on the drop surface along the direction of increasing $\theta_s$. Figs.~\ref{flow_conc_mixed_a} and \ref{flow_conc_mixed_c} suggest that the flow-field and solute distribution outside the drop remain nearly unchanged between $\tilde{m}=0.1$ and $\tilde{m}=10$. This is quantitatively verified for a range of mobility ratios, in Figs.~\ref{analyze_stag_a} and \ref{analyze_stag_b} in Appendix~\ref{in_flow_mc}. The flow-field inside the drop is shown in Fig.~\ref{rot_in_flow} for drops with increasing mobility ratios. Note that according to Figs.~\ref{res_mixed} and \ref{analyze_stag}, the swimming velocity and external conditions are very similar for all the panels shown and only the inner flow varies. The red vectors in Fig.~\ref{rot_in_flow} denote the concentration gradients at the drop surface in the $y=0$ plane, which have an average clockwise bias. It is then clear, from eqn.~\eqref{u_BC_non_dim}, that for a fixed external flow and concentration gradient, an increase in phoretic mobility rotates the internal flow along $\nabla_sC$; see Figs.~\ref{rot_in_flow_a} to \ref{rot_in_flow_f}. This rotation saturates for large $\tilde{m}$, as reflected in the $\frac{\tilde{m}\left( 3\lambda + 2 \right)}{1 + 3\lambda\tilde{m}}$ factor in the right-hand-side of eqn.~\eqref{u_BC_non_dim}. An interesting flow feature is the location of the stagnation points inside the drop. In the absence of a wall, the flow is axisymmetric around the swimming direction and the stagnation points lie on the swimming axis for all $\tilde{m}$~\citep{Morozov2019}, but this is no longer true in the presence of a wall. One practical implication of the change in internal flow as a function of the mobility ratio is that experimental visualization of the flow--for example, locating the stagnation points--can be used as a means to get insight into the drop’s mobility mechanism.

\section{Discussion}\label{discu}
\subsection{Summary of results}
This study considered the motion of a chemically active fluid drop parallel to a rigid wall due to a combination of diffusiophoretic and Marangoni mobility mechanisms. Importantly, the drop always stays close to the wall due to the action of a vertical external force. It was shown that less viscous drops swim closer to the wall (as compared to more viscous drops) since they require stronger fluid shear to balance the external force (see Figs.~\ref{stat_states_lambda_b}, \ref{stat_states_lambda_c} and Figs.~\ref{flow_conc_onlyMar_a}, \ref{flow_conc_onlyMar_c}). This proximity causes more accumulation of chemical solute between the drop and the wall, whose gradients provide a stronger thrust, causing less viscous drops to swim faster (Fig.~\ref{infl_plots_Pe6_onlyMar}). This is in stark contrast to unbounded flow, where drops with different viscosities swim at much more comparable speeds. The effect of the mobility ratio (strength of interfacial flows caused by diffusiophoresis relative to those caused by Marangoni stresses) was also considered, and it was shown that for sufficiently large external force, the swimming speed was independent of the mobility ratio (Fig.~\ref{res_mixed_Ux}).

\subsection{Similarities and differences between active drops and active particles}
The principle motivation of the present analysis was to determine the extent to which the motion of a viscous active drop may be approximated by a rigid active particle. It was shown that, qualitatively, the two models produce very similar results, with the active drop swimming at slightly larger speeds than an active particle (Fig.~\ref{res_drop_vs_part_a}). More importantly, we concluded that the physical mechanism behind faster swimming of active particles near a wall is equally applicable to active drops. In both cases, a reduction in the separation between the particle/drop and the wall is accompanied by an efficient rearrangement of the driving interfacial flows and/or stresses in the particle-/drop-wall gap. For particles, this is a consequence of strong shear stresses in a related auxiliary Stokes flow problem, while for fluid drops this occurs due to an additional contribution from the auxiliary fluid flow (eqn.~\eqref{infl_gen}). Notably, in close proximity to the wall, the evolution of the swimming speed of the drop does not depend on its mobility, and hence phoretic particle models are especially useful for studying near-wall propulsion or other such near-field interactions between chemically active drops. Of course, one caveat is when drop deformation becomes important, in which case one must indeed work with the more complete viscous active drop model~\citep{Guchhait2025}. Overall, the similarity between our results and those for an active particle are encouraging and support the validity of modeling active drops as isotropic active particles. This is expected to enable mathematically simpler treatment of a broad class of active drop problems, without sacrificing any mechanistic insights underpinning their motion. At the same time, caution should be exercised on the more quantitative aspects of active drop motion since physical properties of the drop itself, mainly its viscosity ratio, can have noticeable consequences for its motion near surfaces.

\subsection{Perspectives for future work}
The present work focuses on steady wall-parallel propulsion and does not explore simultaneous (unsteady) wall-normal motion, which may be of two types: (i) a transient phase followed by steady horizontal swimming, wherein a hovering drop destabilizes (for $\Pe > \Pe_\textrm{c}$), moves normal to the wall and then attains the steady state described in this work, and, (ii) an oscillatory regime, wherein the drop does not attain a fixed swimming height and instead oscillates normal to the wall while also translating horizontally. In principle, these problems can be tackled by adapting our numerical framework to an unsteady, body-fitted bispherical grid, which would require careful consideration of the temporal derivative $\partial C/\partial t$ since points with fixed bispherical coordinates would not be stationary in space as the drop moves normal to the wall~\citep{Lippera2020}.

Another unsteady problem is two-dimensional meandering of the drop at a fixed separation from the wall. The emergence of quasi-2D motion can be investigated by analyzing the stability of our steady solutions to perturbations along the direction orthogonal to both the velocity of the drop and the external force on the drop (i.e., along $\mathbf{e}_y$ in Fig.\ref{bisp_coords}). This can be done by employing more general bispherical harmonic expansions in the azimuthal $\left( \phi \right)$ coordinate. An open question here would be the characterization of ``chaotic'' motion, perhaps at large P\'eclet numbers, as seen in experiments~\citep{Hokmabad2021}, 3D simulations of unbounded isotropic active particles~\citep{Hu2022} and 2D simulations of `active discs'~\citep{Li2022}. However, such an analysis would likely require a different numerical technique since a numerical method relying on truncated sums of bispherical harmonic eigenfunctions becomes impractical at large $\Pe$.

Finally, a limiting case of our numerical method also enables us to solve for the motion of passive drops near a rigid surface, perhaps for the first time in bispherical coordinates. This opens avenues for research in related topics, for example, to study the general motion of passive surfactant-laden drops near boundaries~\citep{Jadhav2021, Jadhav2024}.

\appendix

\section{Parameter values}\label{param_values}
An estimate of the P\'eclet number $\Pe=u_\textrm{ref}R/D$ requires knowledge of the solute diffusivity $D$, and the solute properties $\mathcal{A}$, $\mathcal{K}$ and $\mathcal{M}$ (to obtain $u_\textrm{ref}$ as given in eqn.~\eqref{u_c}). It is important to note that these properties depend drastically on the type of solute being modeled (surfactant monomers, oil-in-water micelles, water-in-oil `reverse' micelles, etc). The diffusivity of this solute can be estimated as $D = k_BT/\left( 6\pi\eta_\textrm{o}\delta \right)$, where $\delta$ is the size of the solute molecule. For solubilizing drops, the activity $\mathcal{A}$ is estimated by equating the rate of change of drop volume to the total volume of micelles `emitted' per unit time, $\mathcal{A} \sim \left( 3/4\pi \right) \kappa/\delta^3$, where $\kappa>0$ is the measurable shrinkage rate of the drop. The surface tension sensitivity $\mathcal{K} \sim k_BT\epsilon$ and the phoretic mobility $\mathcal{M} \sim k_BT\epsilon^2/\left( 2 \eta_\textrm{o} \right)$ are related to the interaction length, $\epsilon$, between the solute and the surface of the drop~\citep{Anderson1989}.

The dimensionless external force (due to gravity) depends on the drop's excess density $\Delta \rho$ and is estimated as $W = \frac{4\pi}{3} \Delta\rho g R^3/\left( \eta_\textrm{o} u_\textrm{ref} R \right)$. The viscosity ratio varies over a large range, depending on the drop and the suspending fluid. As for the mobility ratio $\tilde{m}$, a robust estimate, particularly for highly viscous and/or surfactant-packed drops, seems unavailable in the literature. We thus consider values across two orders of magnitude $\tilde{m} \sim O\left(0.1\right)-O\left(10\right)$, to gauge its effect on the motion of the drop. Finally, the capillary number--ratio of viscous to surface tension forces--is defined as $\textrm{Ca} \sim \eta_\textrm{o}u_\textrm{ref}/\gamma$, where $\gamma$ is the interfacial tension between the fluids inside and outside the drop. Table~\ref{table_values} summarizes the values of these parameters; instead of listing individual estimates for $\left( \mathcal{A}, \mathcal{K}, \mathcal{M} \right)$, we directly provide an estimate for the characteristic velocity scale $u_\textrm{ref}$.

\begin{table}[h]
  \begin{center}
\def~{\hphantom{0}}
  \begin{tabular}{llc}
      Parameter  & Description   &   Typical value (units) \\[3pt]
       $R$   & Radius of the drop & 10-100 ($\mu$m) \\
       $\Delta \rho$ & Excess density of the drop & 0.014-0.067 (g/mL) \\
       $\eta_\textrm{o}$   & Viscosity of the outer fluid & 1-30 (mPa-s) \\
       $\eta_\textrm{i}$   & Viscosity of the inner fluid & 1-30 (mPa-s) \\
       $\gamma$   & Interfacial tension & $\sim O(10)$ (mN-m) \\
       $\kappa$   & Solubilization rate & $\sim O(10^{-2})$ ($\mu$m/s) \\
       $\delta$   & Size of filled micelles & $\sim O(10)$ (nm) \\
       $\epsilon$   & Interaction length & $\sim O(10)$ (nm) \\
       $D = k_BT/\left( 6\pi\eta_\textrm{o}\delta \right)$  & Solute diffusivity & $\sim O\left( 1\right)$ ($\mu$m$^2$/s) \\
       $u_\textrm{ref} = \frac{\mathcal{A}}{D} \frac{R \mathcal{K} + 3 \eta_\textrm{i} \mathcal{M}}{2 \eta_\textrm{o} + 3 \eta_\textrm{i}}$ & Reference velocity & $\sim O(10)$ ($\mu$m/s) \\
       \hline
       $\Pe=u_\textrm{ref}R/D$ & P\'eclet number & $\sim O(10-10^4)$ \\
       $W = \frac{4\pi}{3} \Delta\rho g R^2/\left( \eta_\textrm{o} u_\textrm{ref}\right)$ & Dimensionless external force & $\sim O(0.1-10)$ \\
       $\lambda=\eta_\textrm{i}/\eta_\textrm{o}$ & Viscosity ratio & $\sim O(0.01-10)$ \\
       $\tilde{m}=\eta_\textrm{o}\mathcal{M}/\left(R \mathcal{K} \right)$ & Mobility ratio & $\sim O(0.1-10)$ \\
       $\textrm{Ca}=\eta_\textrm{o}u_\textrm{ref}/\gamma$ & Capillary number & $\sim O(10^{-5})$
  \end{tabular}
  \caption{Typical values of the dimensional (above horizontal line) and the dimensionless (below horizontal line) physico-chemical parameters of the problem (\cite{Izri2014, Moerman2017, Hokmabad2021, Castonguay2023, Singh2025}).}
  \label{table_values}
  \end{center}
\end{table}

\section{Validation of numerical method}\label{validations}
\begin{figure}[t]
\begin{center}
\subfloat[]{\label{validation_figs_a}\includegraphics[width=6cm]{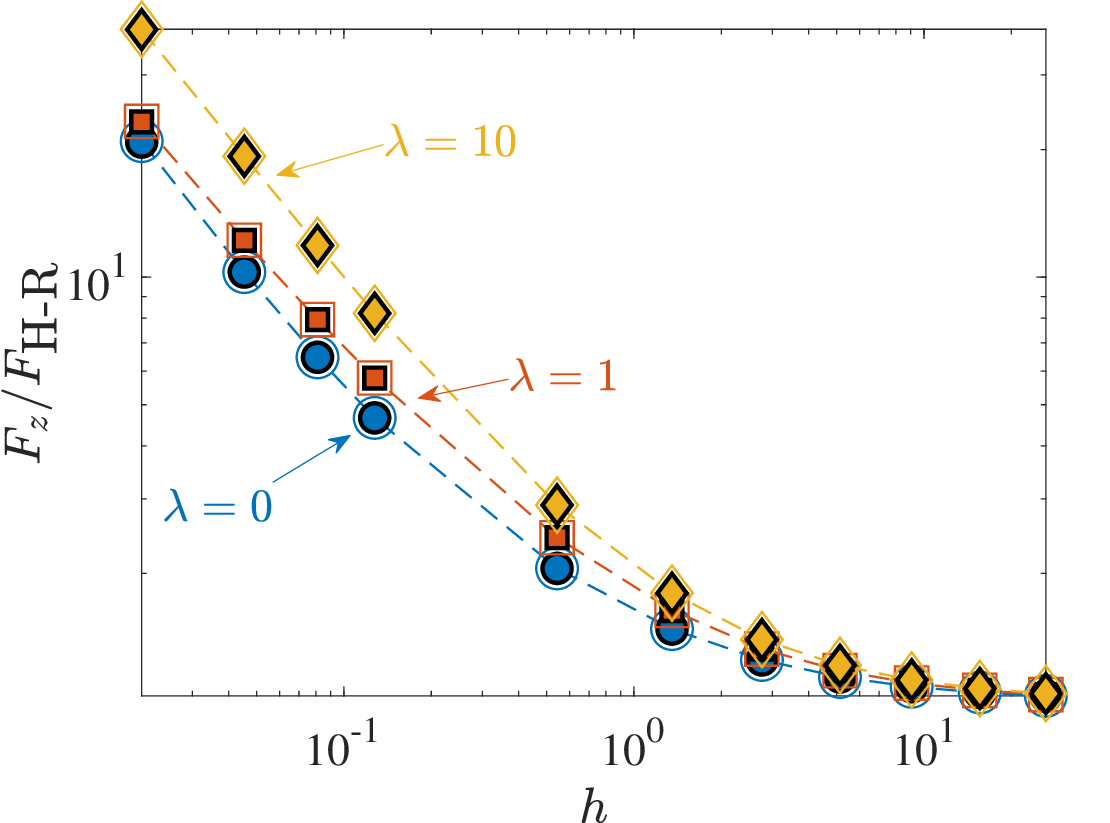}}
\subfloat[]{\label{validation_figs_b}\includegraphics[width=6cm]{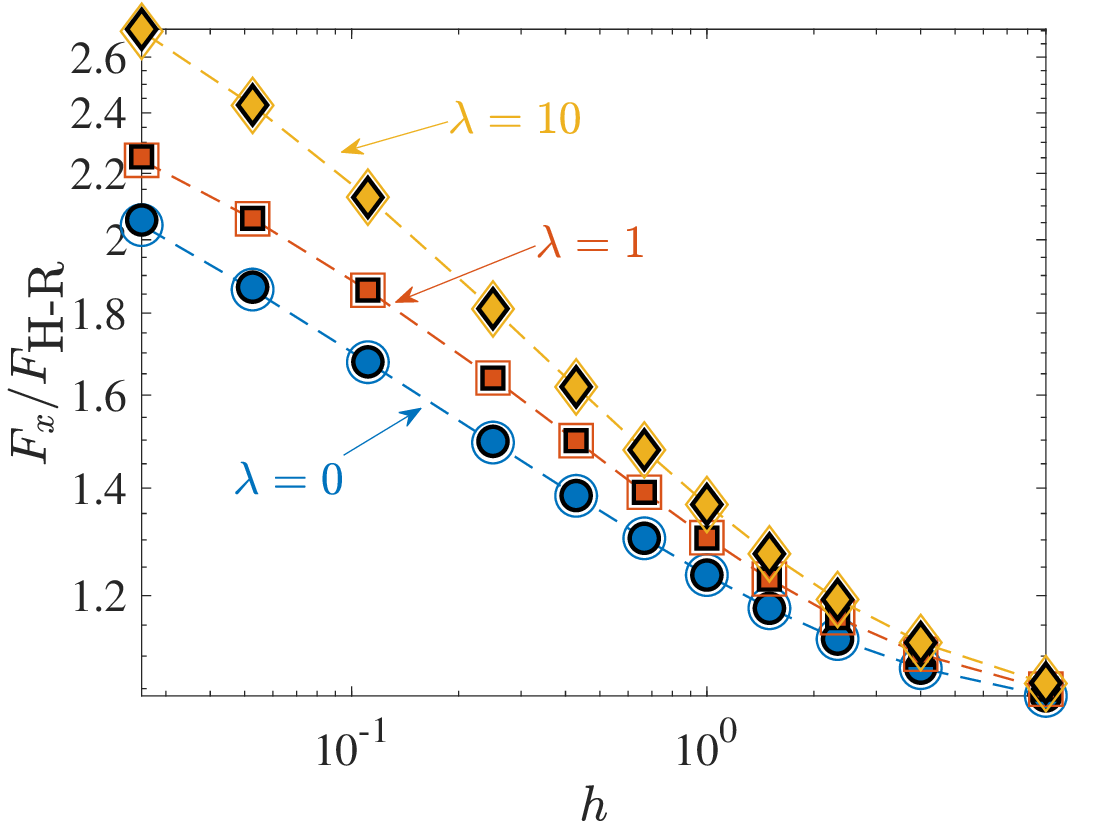}}
\\
\subfloat[]{\label{validation_figs_c}\includegraphics[width=6cm]{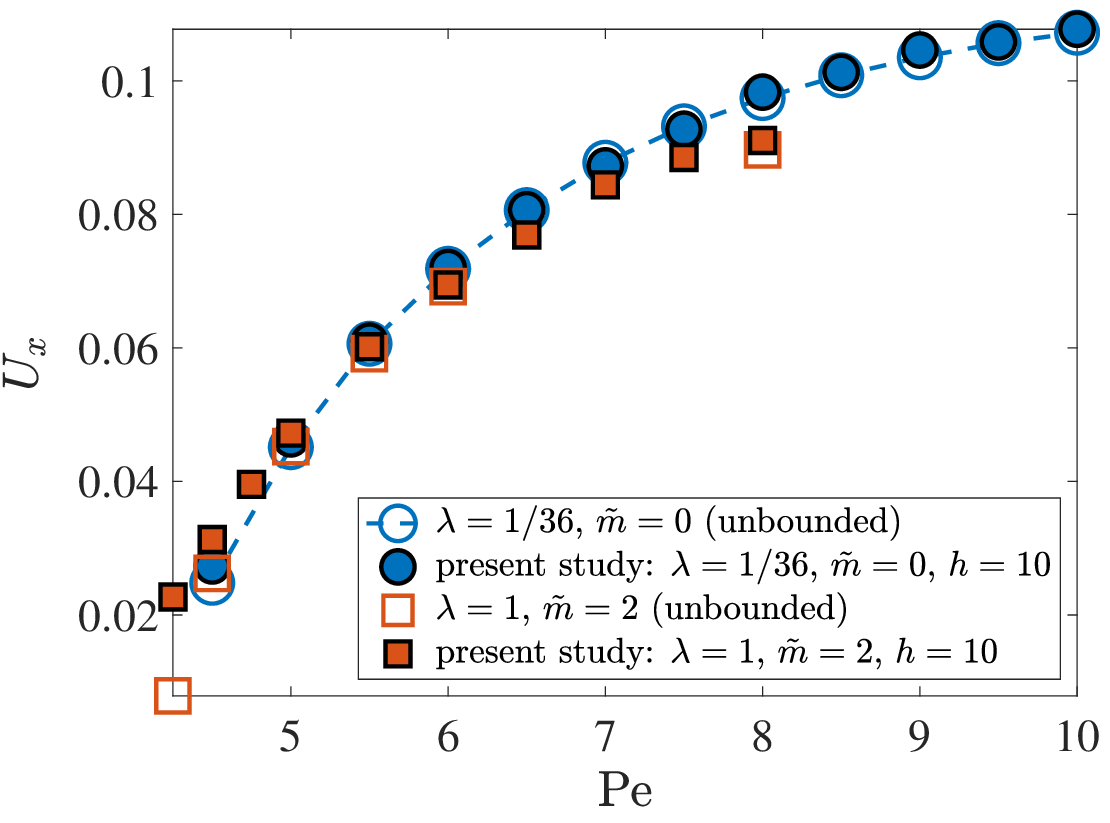}}
\subfloat[]{\label{validation_figs_d}\includegraphics[width=6cm]{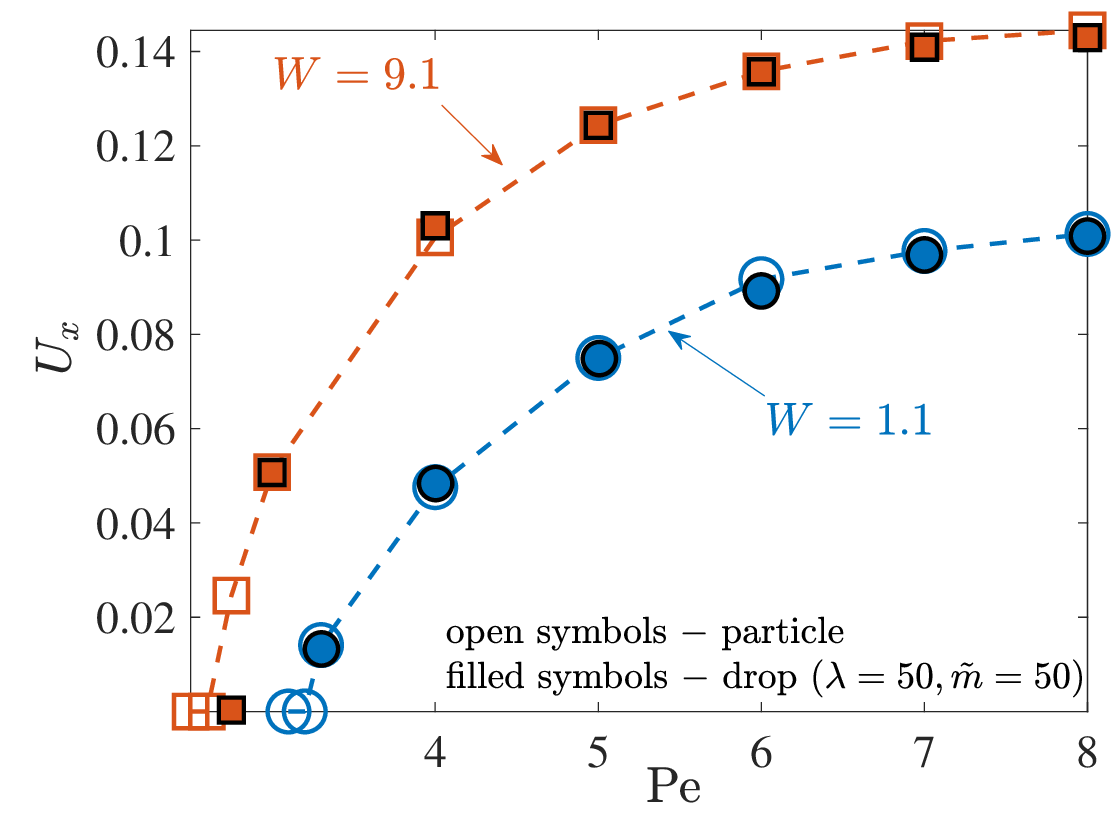}}

\caption{Validation of the numerical method. (a) Normalized hydrodynamic drag on a passive clean drop of viscosity $\lambda$ w.r.t. the external fluid, translating normal to a rigid wall at a separation $h$ and a prescribed speed (see text for the full expression of $F_\textrm{H-R}$). (b) Same as in panel (a), but for a drop translating parallel to a rigid wall. In (a) and (b), the empty symbols denote results from \cite{Wacholder1972} and \cite{Keh2001}, respectively, and the filled symbols denote results from our numerical method adapted for a passive drop. (c)~The steady swimming speed $U_x$ of an unbounded active drop driven solely due to Marangoni stress (empty circles, from \cite{Izri2014}) and one driven due to a combination of Marangoni and diffusiophoretic effects (empty squares, from \cite{Morozov2019}). The filled symbols denote results from our numerical method, for a large separation $h=10\gg1$ from the wall. (d) The swimming speed $U_x$ of an isotropic active particle confined to a rigid wall by an external force $-W \mathbf{e}_z$, and driven by diffusiophoresis (empty circles, from \cite{Desai2022}). The filled symbols denote results from our numerical method in the limit $\lambda,\tilde{m} \gg 1$.}
\label{validation_figs}

\end{center}
\end{figure}

We validate our numerical method against the following benchmark studies:
\begin{enumerate}[(\roman*)]
    \item We first consider the motion of a passive viscous drop normal and parallel to a rigid wall. This problem requires us to set $\mathcal{K}, \mathcal{M} \equiv 0$ in eqns.~\eqref{tau_BC_dim} and \eqref{u_BC_dim}. The velocity of the drop is specified and the normalized hydrodynamic drag $F/F_\textrm{H-R}$ is calculated, where $F_\textrm{H-R} = 6\pi\eta_\textrm{o}RU \frac{\left( 3\lambda + 2 \right)}{\left( 3\lambda + 3 \right)}$ is the Hadamard-Rybczynski drag on an isolated clean drop translating with velocity $U$~\citep{Leal2007}. Our results are compared, for different values of the viscosity ratio $\lambda$ and separation from the wall $h$, to classical results obtained in \cite{Wacholder1972} and \cite{Keh2001} (see Figs.~\ref{validation_figs_a} and \ref{validation_figs_b}).
    
    \item Second, we solve the current problem in the limit $h \gg 1$. Here, the swimming speed of a Marangoni-stress-driven active drop in an unbounded fluid is calculated and compared to the work of \cite{Izri2014}. We also compare our results with \cite{Morozov2019}, who studied the unbounded motion of a viscous active drop propelled by both Marangoni stresses and diffusiophoresis (see Fig.~\ref{validation_figs_c}).
    
    \item Third, we solve our problem for finite values of $h$, but in the limit $\lambda, \tilde{m} \gg 1$, i.e., we solve for the wall-parallel propulsion of an isotropic (rigid) active colloid that swims purely due to diffusiophoretic slip along its surface. Our results are compared to those of \cite{Desai2022} (see Fig.~\ref{validation_figs_d}).
\end{enumerate}
These results are shown in Fig.~\ref{validation_figs} and excellent quantitative agreement is noted in all cases. This validates our numerical method, particularly the implementation of the tangential stress and velocity boundary conditions across the drop's interface, and the iterative solution to the advection-diffusion equation~\eqref{advDiff_C_dropFrame_ND}. Note that Fig.~\ref{validation_figs_d} compares the velocity of a highly viscous and phoretically driven active drop to that of an isotropic, active phoretic particle (i.e., the limit $\lambda, \tilde{m} \to \infty$). The agreement is very good, except for $W=9.1$, when $\Pe \approx 2.75$, where it looks like the critical $\Pe$ for self-propulsion is slightly larger for the drop than for the particle. This could be because the large but finite values of $\lambda$ and $\tilde{m}$ for the active drop computations prevent us from realizing the true active particle limit, specifically for strong confinement.

\section{Equilibrium separation in stationary state, as a function of the mobility ratio}\label{h_vs_Pe_mc}

\begin{figure}[t]
\begin{center}

\subfloat[]{\label{stat_states_mc_a}\includegraphics[width=6.5cm]{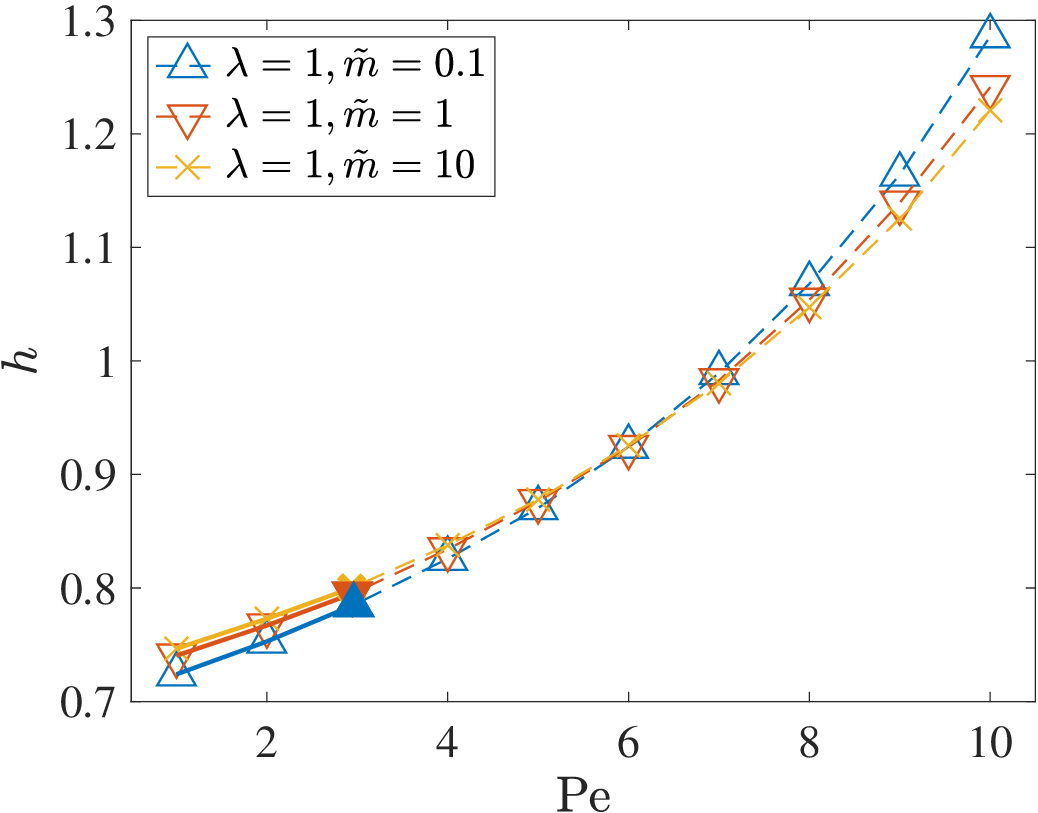}}
\\
\subfloat[]{\label{stat_states_mc_b}\includegraphics[width=5.5cm]{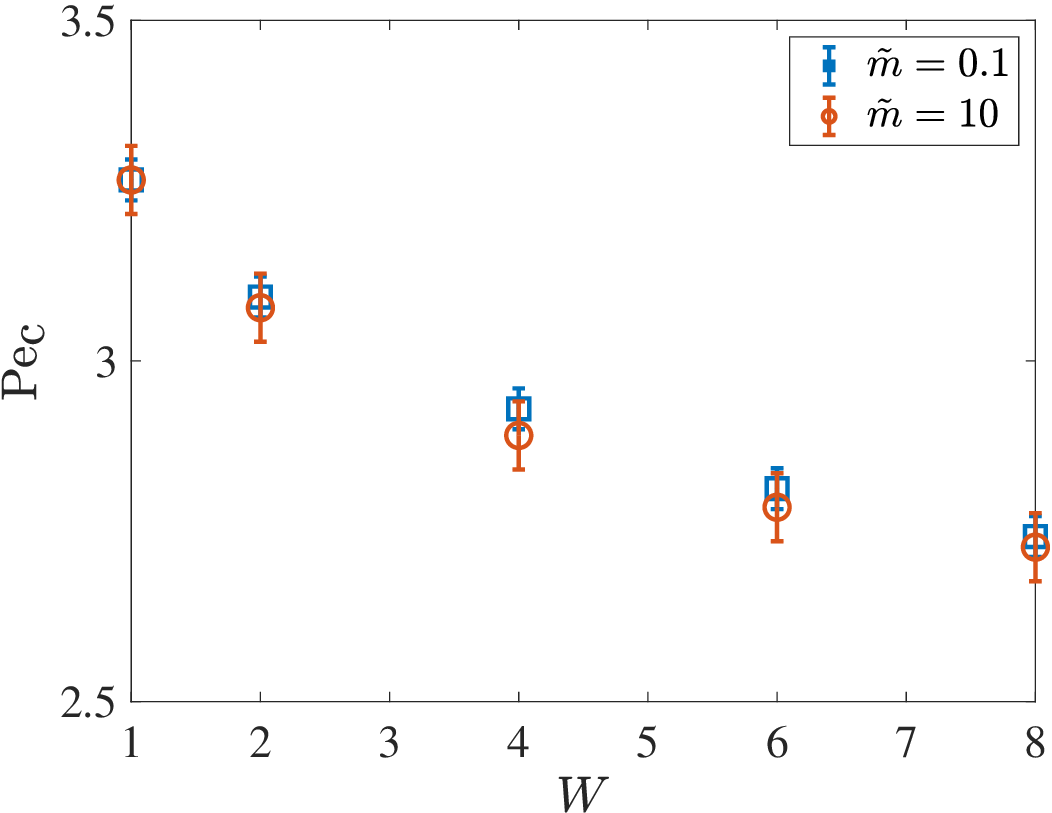}}
\hspace{5mm}
\subfloat[]{\label{stat_states_mc_c}\includegraphics[width=5.5cm]{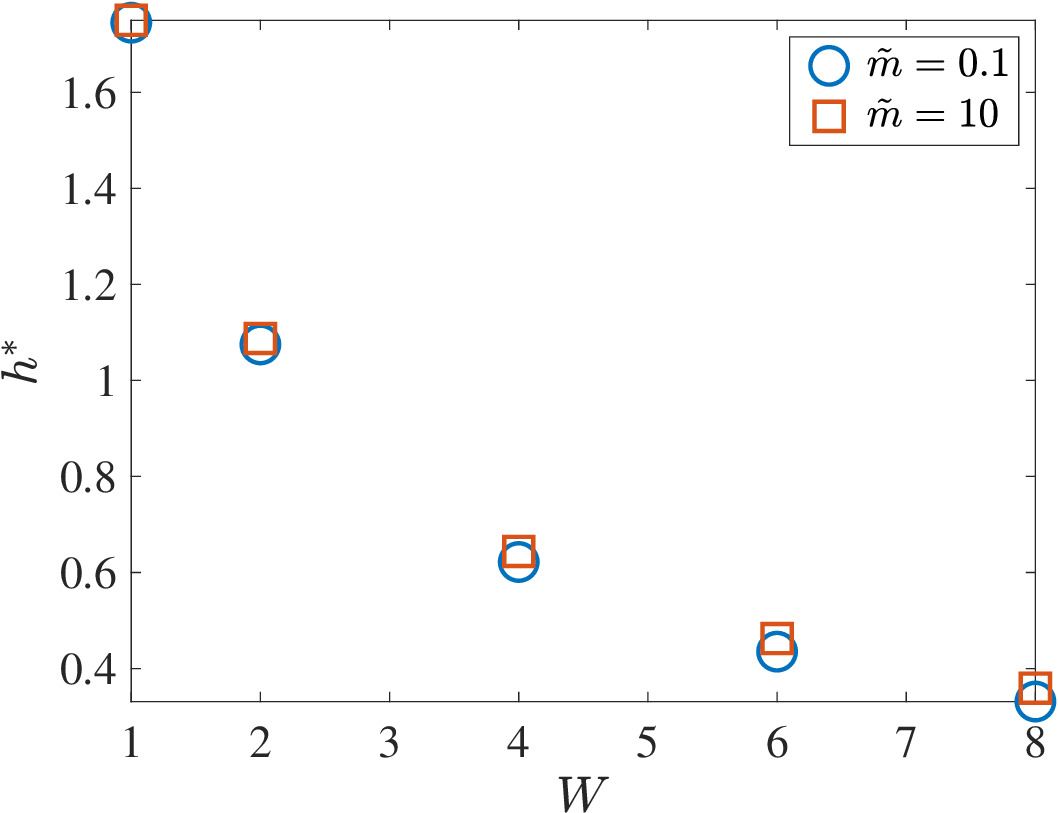}}

\end{center}

\caption{(a) Equilibrium separation $h$ as a function of $\Pe$ for active drops driven by a combination of Marangoni and diffusiophoretic effects, i.e., when $\tilde{m} \ne 0$. The results are shown here for $W=3$. Solid (resp. dashed) lines denote stable (resp. unstable) solutions. Note that the drop is stationary, i.e., $U_x \equiv 0$, in all cases. (b) The critical P\'eclet number for self-propulsion, $\Pe_\textrm{c}$, above which the stationary hovering solution becomes unstable and the drop begins to spontaneously swim parallel to the wall. The variation in $\Pe_\textrm{c}$ is plotted as a function of the external force $W$, for two different values of the mobility ratio and a fixed viscosity ratio $\lambda = 1$. (c) The critical separation $h^*$ at which the stationary active drop destabilizes and begins to swim; these have been shown in panel (a) by the filled/thick symbols.}
\label{stat_states_mc}

\end{figure}

Fig.~\ref{stat_states_mc_a} shows the dependence of the equilibrium height $h$ as a function of $\Pe$ for stationary active drops with different mobility ratios, all experiencing a constant external force $W=3$. The weak dependence on $\tilde{m}$ is apparent, importantly for lower values of $\Pe$ which correspond to stable equilibrium solutions (solid lines in Fig.~\ref{stat_states_mc_a}). The critical P\'eclet number $\Pe_\textrm{c}$, and the corresponding critical separation $h^*$, have been highlighted in Fig.~\ref{stat_states_mc_a} by the filled/thick symbols. In Figs.~\ref{stat_states_mc_b} and \ref{stat_states_mc_c} we plot $\Pe_\textrm{c}$ and $h^*$ as a function of $W$, for $\tilde{m}=0.1$ and $\tilde{m}=10$.

\section{Reciprocal theorem for calculation of the swimming speed of the drop}\label{reciproc_derive}
We present here the derivation of the integral expressions--eqns.~\eqref{reciproc_drop_lam} and \eqref{reciproc_drop_mc}--for the swimming speed $U_x$, using the reciprocal theorem. We denote the variables of the auxiliary Stokes flow problem with a hat and the variables of our active drop problem (the primary problem) without a hat. Importantly, we consider all flow fields in the drop frame, for both the primary and the auxiliary problem. The reciprocal theorem for the fluid outside the drop reads:
\begin{align}\label{reciproc_gen_o}
    &\int_{S_\textrm{p}} {\bar{\mathbf{n}} \cdot \boldsymbol{\sigma}^\textrm{o} \cdot \hat{\mathbf{u}}^\textrm{o}\;dS}\;+\;\int_{S_\infty} {\bar{\mathbf{n}} \cdot \boldsymbol{\sigma}^\textrm{o} \cdot \hat{\mathbf{u}}^\textrm{o}\;dS}\;+\;\int_{S_\textrm{w}} {\bar{\mathbf{n}} \cdot \boldsymbol{\sigma}^\textrm{o} \cdot \hat{\mathbf{u}}^\textrm{o}\;dS} = \\ \nonumber
    &\int_{S_\textrm{p}} {\bar{\mathbf{n}} \cdot \hat{\boldsymbol{\sigma}}^\textrm{o} \cdot \mathbf{u}^\textrm{o}\;dS}\;+\;\int_{S_\infty} {\bar{\mathbf{n}} \cdot \hat{\boldsymbol{\sigma}}^\textrm{o} \cdot \mathbf{u}^\textrm{o}\;dS}\;+\;\int_{S_\textrm{w}} {\bar{\mathbf{n}} \cdot \hat{\boldsymbol{\sigma}}^\textrm{o} \cdot \mathbf{u}^\textrm{o}\;dS},
\end{align}
where $S_\textrm{p}$, $S_\textrm{w}$ and $S_\infty$ denote the drop surface, the wall and a surface at infinity, respectively. Here, $\bar{\mathbf{n}}$ is the normal pointing away from the fluid at all surfaces. Let us first focus on the $S_\infty$ and $S_\textrm{w}$ terms. Since we are considering flows in the moving drop frame, at $r\to\infty$ and at $S_\textrm{w}$, we have,
\begin{subequations}
    \begin{align}
    \mathbf{u}^\textrm{o} &= -\mathbf{U}, \\
    \hat{\mathbf{u}}^\textrm{o} &= -\hat{\mathbf{U}}.
\end{align}
\end{subequations}
The second and third integrals on both sides can thus be written together as,
\begin{equation}
    \int_{S_\infty} {\bar{\mathbf{n}} \cdot \hat{\boldsymbol{\sigma}}^\textrm{o} \cdot \mathbf{u}^\textrm{o}\;dS}\;+\;\int_{S_\textrm{w}} {\bar{\mathbf{n}} \cdot \hat{\boldsymbol{\sigma}}^\textrm{o} \cdot \mathbf{u}^\textrm{o}\;dS} = -\mathbf{U} \cdot \int_{S_\infty+S_\textrm{w}} {\bar{\mathbf{n}} \cdot \hat{\boldsymbol{\sigma}}^\textrm{o}\;dS}.
\end{equation}
Now,
\begin{equation*}
    \nabla \cdot \hat{\boldsymbol{\sigma}}^\textrm{o} = 0,
\end{equation*}
which can be integrated over the fluid volume, and after application of the divergence theorem, this yields:
\begin{equation}
    \int_{S_\textrm{p}} {\bar{\mathbf{n}} \cdot \hat{\boldsymbol{\sigma}}^\textrm{o}\;dS}\;+\;\int_{S_\infty} {\bar{\mathbf{n}} \cdot \hat{\boldsymbol{\sigma}}^\textrm{o}\;dS}\;+\;\int_{S_\textrm{w}} {\bar{\mathbf{n}} \cdot \hat{\boldsymbol{\sigma}}^\textrm{o}\;dS} = 0,
\end{equation}
or,
\begin{subequations}
\begin{align}
    \int_{S_\infty+S_\textrm{w}} {\bar{\mathbf{n}} \cdot \hat{\boldsymbol{\sigma}}^\textrm{o}\;dS} &= -\int_{S_\textrm{p}} {\bar{\mathbf{n}} \cdot \hat{\boldsymbol{\sigma}}^\textrm{o}\;dS}, \\ \nonumber
    &= \hat{\mathbf{F}}^\textrm{H}, \\ \nonumber
    &=-\hat{\mathbf{F}}^\textrm{ext}.
\end{align}    
\end{subequations}
Therefore, the reciprocal theorem eqn.~\eqref{reciproc_gen_o} simplifies to:
\begin{equation}\label{reciproc_gen_Fext}
    \int_{S_\textrm{p}} {\bar{\mathbf{n}} \cdot \boldsymbol{\sigma}^\textrm{o} \cdot \hat{\mathbf{u}}^\textrm{o}\;dS}\;+\;\underbrace{\mathbf{F}^\textrm{ext} \cdot \hat{\mathbf{U}}}_{=0} = \int_{S_\textrm{p}} {\bar{\mathbf{n}} \cdot \hat{\boldsymbol{\sigma}}^\textrm{o} \cdot \mathbf{u}^\textrm{o}\;dS}\;+\;\hat{\mathbf{F}}^\textrm{ext} \cdot \mathbf{U}.
\end{equation}
Note that in eqn.~\eqref{reciproc_gen_Fext}, $\mathbf{F}^\textrm{ext}=-W\mathbf{e}_z$ and we have chosen the auxiliary problem where a clean drop is translating parallel to the wall, hence $\hat{\mathbf{U}} \cdot \mathbf{e}_z = 0$; this means the second term on the left-hand-side vanishes. Now, the reciprocal theorem inside the drop reads,
\begin{equation}\label{reciproc_gen_i}
    -\int_{S_\textrm{p}} {\bar{\mathbf{n}} \cdot \boldsymbol{\sigma}^\textrm{i} \cdot \hat{\mathbf{u}}^\textrm{i}\;dS} = -\int_{S_\textrm{p}} {\bar{\mathbf{n}} \cdot \hat{\boldsymbol{\sigma}}^\textrm{i} \cdot \mathbf{u}^\textrm{i}\;dS}.
\end{equation}
Multiplying eqn.~\eqref{reciproc_gen_i} by the viscosity ratio $\lambda$, and adding to eqn.~\eqref{reciproc_gen_Fext}, we obtain:
\begin{equation}\label{reciproc_pre_final}
    \int_{S_\textrm{p}} {\left( \bar{\mathbf{n}} \cdot \boldsymbol{\sigma}^\textrm{o} - \lambda\bar{\mathbf{n}} \cdot \boldsymbol{\sigma}^\textrm{i} \right) \cdot \hat{\mathbf{u}}^\textrm{o}\;dS} = \int_{S_\textrm{p}} {\left( \bar{\mathbf{n}} \cdot \hat{\boldsymbol{\sigma}}^\textrm{o}\cdot \mathbf{u}^\textrm{o} - \lambda\bar{\mathbf{n}} \cdot \hat{\boldsymbol{\sigma}}^\textrm{i}\cdot \mathbf{u}^\textrm{i} \right)\;dS}\;+\;\hat{\mathbf{F}}^\textrm{ext} \cdot \mathbf{U}.
\end{equation}
In eqn.~\eqref{reciproc_pre_final}, we have used the fact that the fluid velocity is continuous across the interface for the auxiliary problem. Importantly, only the tangential components of $\bar{\mathbf{n}} \cdot \boldsymbol{\sigma}^\textrm{i,o}$ and $\bar{\mathbf{n}} \cdot \hat{\boldsymbol{\sigma}}^\textrm{i,o}$ will contribute to the dot product, since the normal component of the fluid velocity is zero in the drop frame. For the auxiliary problem,
\begin{equation}\label{BCs_reciproc_aux}
    \left( \bar{\mathbf{n}} \cdot \hat{\boldsymbol{\sigma}}^\textrm{o} - \lambda\bar{\mathbf{n}} \cdot \hat{\boldsymbol{\sigma}}^\textrm{i} \right) \cdot \left( \boldsymbol{\delta} - \bar{\mathbf{n}}\bar{\mathbf{n}} \right) = \mathbf{0},
\end{equation}
whereas, for the primary problem, the dimensionless form of the boundary conditions~\eqref{tau_BC_dim} and \eqref{u_BC_dim} yields, respectively,
\begin{subequations}\label{BCs_reciproc_prim}
\begin{align}
        \left( \bar{\mathbf{n}} \cdot \boldsymbol{\sigma}^\textrm{o} - \lambda\bar{\mathbf{n}} \cdot \boldsymbol{\sigma}^\textrm{i} \right) \cdot \left( \boldsymbol{\delta} - \bar{\mathbf{n}}\bar{\mathbf{n}} \right) &= \frac{3\lambda + 2}{1+3\lambda\tilde{m}}\nabla_sC, \\
        \left( \mathbf{u}^\textrm{o} - \mathbf{u}^\textrm{i} \right) \cdot \left( \boldsymbol{\delta} - \bar{\mathbf{n}}\bar{\mathbf{n}} \right) &= \tilde{m}\frac{3\lambda + 2}{1+3\lambda\tilde{m}}\nabla_sC.
\end{align}
\end{subequations}
Note that in eqn.~\eqref{BCs_reciproc_prim}, $\bar{\mathbf{n}}$ is the normal pointing out of the fluid and into the drop (compared with eqn.~\eqref{tau_BC_dim}, $\bar{\mathbf{n}}=-\mathbf{n}$). Substituting eqns.~\eqref{BCs_reciproc_aux} and \eqref{BCs_reciproc_prim} into \eqref{reciproc_pre_final}, and choosing $\hat{\mathbf{F}}^\textrm{ext}=\mathbf{e}_x$ yields eqn.~\eqref{reciproc_drop_mc}. Eqn.~\eqref{reciproc_drop_lam} is easily recovered in the limit $\tilde{m} \equiv 0$.

\section{Interfacial flows and concentration gradients as a function of mobility ratio}\label{in_flow_mc}
\begin{figure}[t]
\begin{center}

\subfloat[]{\label{analyze_stag_a}\includegraphics[width=6cm]{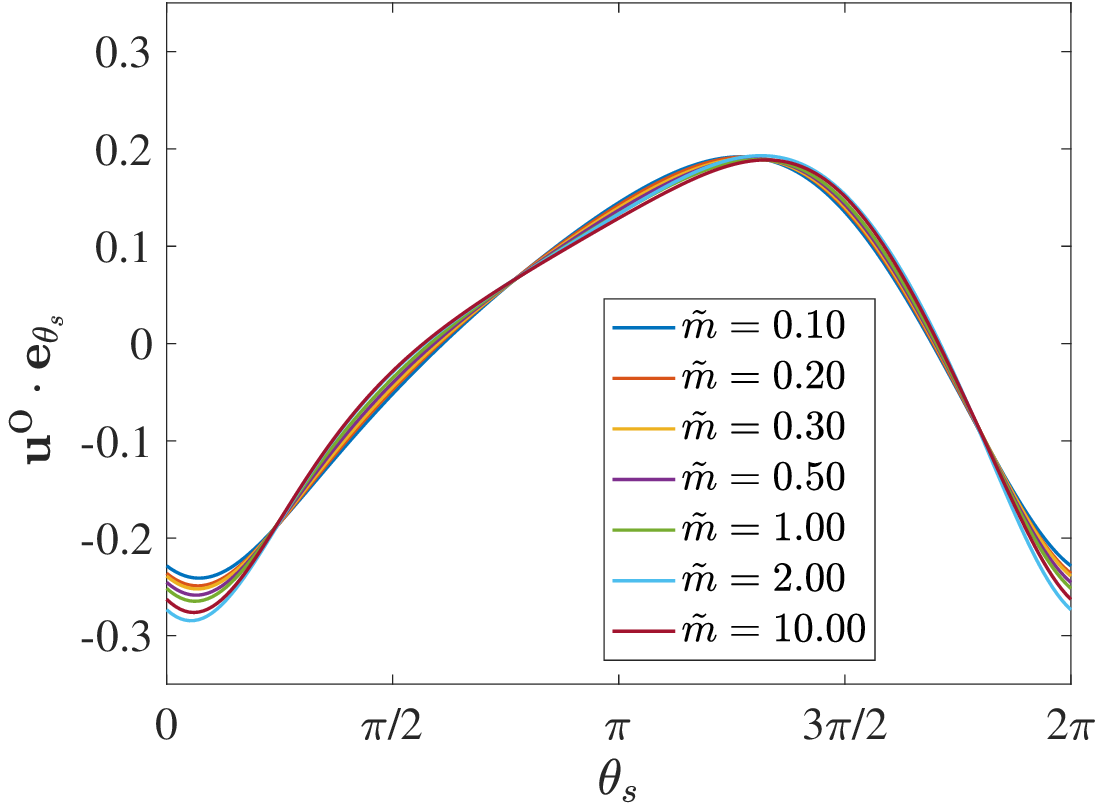}}
\subfloat[]{\label{analyze_stag_b}\includegraphics[width=6cm]{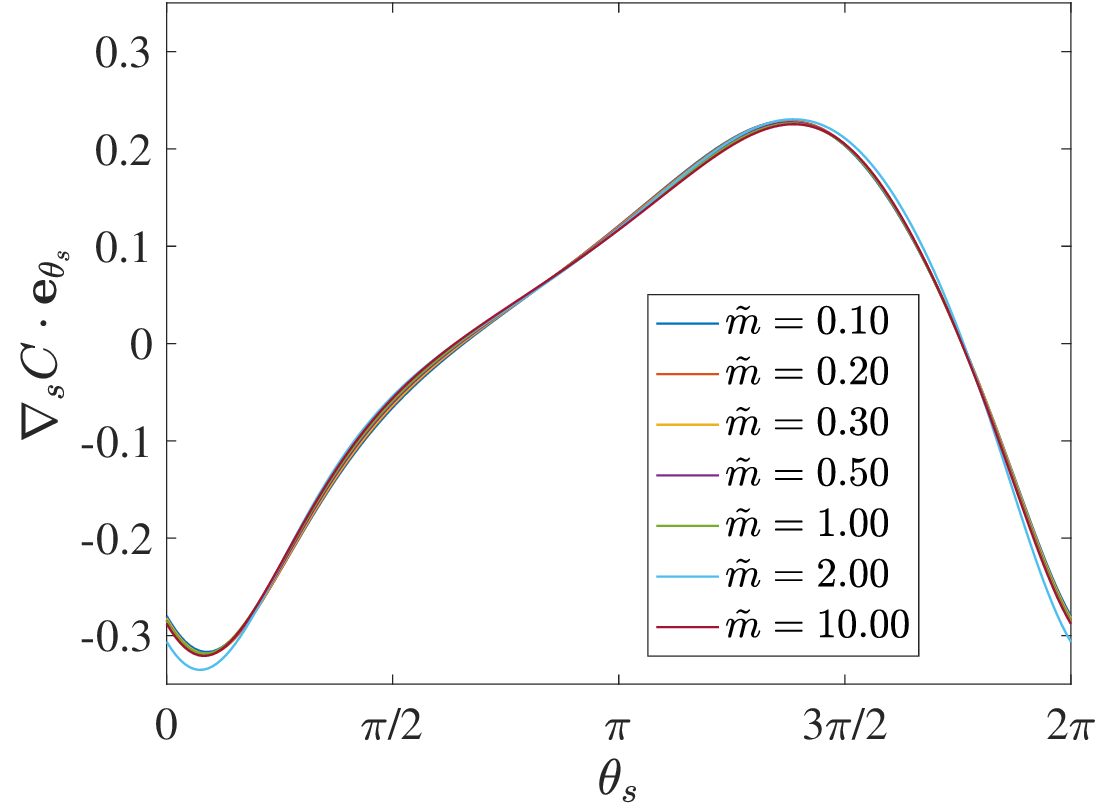}}
\\
\subfloat[]{\label{analyze_stag_c}\includegraphics[width=6cm]{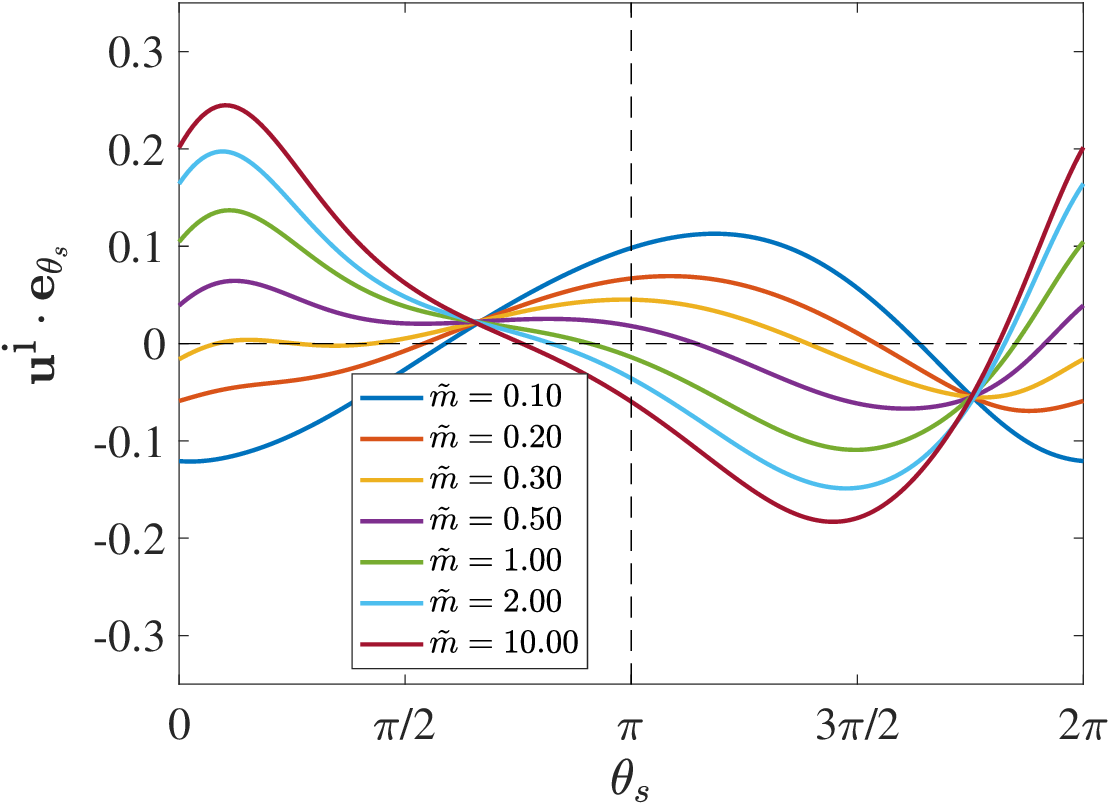}}

\caption{Distribution of (a) the fluid velocity, and, (b) the solute concentration gradients at the outer surface of the drop, as function of the mobility ratio, $\tilde{m}$. (c) Distribution of the fluid velocity at the inner surface of the drop, as function of the mobility ratio, $\tilde{m}$. All quantities have been plotted along the curve defined by the intersection of the drop surface and the $y=0$ plane. The velocities have been plotted in the frame of reference of the drop. The other parameters are: $W=3,\lambda=1,\Pe=6$.}
\label{analyze_stag}

\end{center}
\end{figure}

Fig.~\ref{analyze_stag} shows the fluid velocities and concentration gradients at the drop surface, for eqn.~\eqref{u_BC_non_dim}. The velocity is shown both outside (Fig.~\ref{analyze_stag_a}) and inside (Fig.~\ref{analyze_stag_c}) the drop. Since $\theta_s$ increases in the counter-clockwise direction and $\nabla_sC$ is more negative than positive, this means that the concentration gradients have a clockwise rotating effect on the internal flow (Fig.~\ref{analyze_stag_b}). This results in a reversal of the internal flow as shown in Fig.~\ref{rot_in_flow}, and is also apparent in the change in sign of $\mathbf{u}^\textrm{i} \cdot \mathbf{e}_{\theta_s}$ (compared to $\mathbf{u}^\textrm{o} \cdot \mathbf{e}_{\theta_s}$), as $\tilde{m}$ increases from $\tilde{m}=0.20$ to $\tilde{m}=0.50$.

\bibliographystyle{unsrt}
\bibliography{jfm}

@book{Leal2007,
  title={Advanced Transport Phenomena: fluid mechanics and convective transport processes},
  author={Leal, L. G.},
  year={2007},
  publisher={Cambridge University Press}
}

@book{HappelBrenner2012,
  title={Low Reynolds number hydrodynamics: with special applications to particulate media},
  author={Happel, J. and Brenner, H.},
  year={2012},
  publisher={Springer Science \& Business Media}
}

@article{Desai2021,
  doi = {10.1103/physrevfluids.6.114103},
  url = {https://doi.org/10.1103/physrevfluids.6.114103},
  year = {2021},
  month = nov,
  publisher = {American Physical Society ({APS})},
  volume = {6},
  number = {11},
  author = {N. Desai and S. Michelin},
  title = {Instability and self-propulsion of active droplets along a wall},
  journal = {Phys. Rev. Fluids}
}

@article{Keh2001,
  doi = {10.1016/s0009-2509(01)00323-2},
  url = {https://doi.org/10.1016/s0009-2509(01)00323-2},
  year = {2001},
  month = dec,
  publisher = {Elsevier {BV}},
  volume = {56},
  number = {24},
  pages = {6863--6871},
  author = {J. K. Huan and P. Y. Chen},
  title = {Slow motion of a droplet between two parallel plane walls},
  journal = {Chem. Eng. Sci.}
}

@article{Wacholder1972,
  doi = {10.1016/0009-2509(72)85043-7},
  url = {https://doi.org/10.1016/0009-2509(72)85043-7},
  year = {1972},
  month = oct,
  publisher = {Elsevier {BV}},
  volume = {27},
  number = {10},
  pages = {1817--1828},
  author = {E. Wacholder and D. Weihs},
  title = {Slow motion of a fluid sphere in the vicinity of another sphere or a plane boundary},
  journal = {Chem. Eng. Sci.}
}

@article{Izri2014,
  doi = {10.1103/physrevlett.113.248302},
  url = {https://doi.org/10.1103/physrevlett.113.248302},
  year = {2014},
  month = dec,
  publisher = {American Physical Society ({APS})},
  volume = {113},
  number = {24},
  author = {Z. Izri and M. N. van der Linden and S. Michelin and O. Dauchot},
  title = {Self-Propulsion of Pure Water Droplets by Spontaneous Marangoni-Stress-Driven Motion},
  journal = {Phys. Rev. Lett.}
}

@article{Desai2022,
  doi = {10.1103/physrevfluids.7.100501},
  url = {https://doi.org/10.1103/physrevfluids.7.100501},
  year = {2022},
  month = oct,
  publisher = {American Physical Society ({APS})},
  volume = {7},
  number = {10},
  author = {N. Desai and S. Michelin},
  title = {Steady state propulsion of isotropic active colloids along a wall},
  journal = {Phys. Rev. Fluids}
}

@article{Anderson1989,
  title = {Colloid Transport by Interfacial Forces},
  volume = {21},
  ISSN = {1545-4479},
  url = {http://dx.doi.org/10.1146/annurev.fl.21.010189.000425},
  DOI = {10.1146/annurev.fl.21.010189.000425},
  number = {1},
  journal = {Ann. Rev. Fluid Mech.},
  publisher = {Annual Reviews},
  author = {J. L. Anderson},
  year = {1989},
  month = jan,
  pages = {61–99}
}

@article{Maass2016,
  title = {Swimming Droplets},
  volume = {7},
  ISSN = {1947-5462},
  url = {http://dx.doi.org/10.1146/annurev-conmatphys-031115-011517},
  DOI = {10.1146/annurev-conmatphys-031115-011517},
  number = {1},
  journal = {Ann. Rev. Cond. Matt. Phys.},
  publisher = {Annual Reviews},
  author = {Maass,  C. C. and Kr\"{u}ger,  C. and Herminghaus,  S. and Bahr,  C.},
  year = {2016},
  month = mar,
  pages = {171–193}
}

@article{Morozov2019,
  title = {Nonlinear dynamics of a chemically-active drop: From steady to chaotic self-propulsion},
  volume = {150},
  ISSN = {1089-7690},
  url = {http://dx.doi.org/10.1063/1.5080539},
  DOI = {10.1063/1.5080539},
  number = {4},
  journal = {J. Chem. Phys.},
  publisher = {AIP Publishing},
  author = {Morozov,  M. and Michelin,  S.},
  year = {2019},
  month = jan 
}

@article{Stone1996,
  doi = {10.1103/physrevlett.77.4102},
  year = {1996},
  month = nov,
  publisher = {American Physical Society },
  volume = {77},
  number = {19},
  pages = {4102--4104},
  author = {H. A. Stone and A. D. T. Samuel},
  title = {Propulsion of Microorganisms by Surface Distortions},
  journal = {Phys. Rev. Lett.}
}

@article{Pak2014,
  title = {Viscous Marangoni migration of a drop in a Poiseuille flow at low surface Péclet numbers},
  volume = {753},
  ISSN = {1469-7645},
  url = {http://dx.doi.org/10.1017/jfm.2014.380},
  DOI = {10.1017/jfm.2014.380},
  journal = {J. Fluid Mech.},
  publisher = {Cambridge University Press (CUP)},
  author = {Pak,  O. S. and Feng,  J. and Stone,  H. A.},
  year = {2014},
  month = jul,
  pages = {535–552}
}

@article{Picella2021,
  title = {Confined self-propulsion of an isotropic active colloid},
  volume = {933},
  ISSN = {1469-7645},
  url = {http://dx.doi.org/10.1017/jfm.2021.1081},
  DOI = {10.1017/jfm.2021.1081},
  journal = {J. Fluid Mech.},
  publisher = {Cambridge University Press (CUP)},
  author = {Picella,  F. and Michelin,  S.},
  year = {2021},
  month = dec 
}

@article{Brandao2024,
  title = {Isotropically active particle closely fitting in a cylindrical channel: spontaneous motion at small Péclet numbers},
  volume = {980},
  ISSN = {1469-7645},
  url = {http://dx.doi.org/10.1017/jfm.2023.1082},
  DOI = {10.1017/jfm.2023.1082},
  journal = {J. Fluid Mech.},
  publisher = {Cambridge University Press (CUP)},
  author = {Brandão,  R.},
  year = {2024},
  month = jan 
}

@article{Michelin2023,
  title = {Self-Propulsion of Chemically Active Droplets},
  volume = {55},
  ISSN = {1545-4479},
  url = {http://dx.doi.org/10.1146/annurev-fluid-120720-012204},
  DOI = {10.1146/annurev-fluid-120720-012204},
  number = {1},
  journal = {Ann. Rev. Fluid Mech.},
  publisher = {Annual Reviews},
  author = {Michelin,  S.},
  year = {2023},
  month = jan,
  pages = {77–101}
}

@article{Herminghaus2014,
  title = {Interfacial mechanisms in active emulsions},
  volume = {10},
  ISSN = {1744-6848},
  url = {http://dx.doi.org/10.1039/C4SM00550C},
  DOI = {10.1039/c4sm00550c},
  number = {36},
  journal = {Soft Matter},
  publisher = {Royal Society of Chemistry (RSC)},
  author = {Herminghaus,  S. and Maass,  C. C. and Kr\"{u}ger,  C. and Thutupalli,  S. and Goehring,  L. and Bahr,  C.},
  year = {2014},
  pages = {7008–7022}
}

@article{Ebbens2016,
  title = {Active colloids: Progress and challenges towards realising autonomous applications},
  volume = {21},
  ISSN = {1359-0294},
  url = {http://dx.doi.org/10.1016/j.cocis.2015.10.003},
  DOI = {10.1016/j.cocis.2015.10.003},
  journal = {Curr. Opin. Colloid \& Interface Sci.},
  publisher = {Elsevier BV},
  author = {Ebbens,  S.J.},
  year = {2016},
  month = feb,
  pages = {14–23}
}

@article{Jin2017,
  title = {Chemotaxis and autochemotaxis of self-propelling droplet swimmers},
  volume = {114},
  ISSN = {1091-6490},
  url = {http://dx.doi.org/10.1073/pnas.1619783114},
  DOI = {10.1073/pnas.1619783114},
  number = {20},
  journal = {Proc. Natl. Acad. Sci. U.S.A.},
  publisher = {Proceedings of the National Academy of Sciences},
  author = {Jin,  C. and Kr\"{u}ger,  C. and Maass,  C. C.},
  year = {2017},
  month = may,
  pages = {5089–5094}
}

@article{Dey2022,
  title = {Oscillatory rheotaxis of artificial swimmers in microchannels},
  volume = {13},
  ISSN = {2041-1723},
  url = {http://dx.doi.org/10.1038/s41467-022-30611-1},
  DOI = {10.1038/s41467-022-30611-1},
  number = {1},
  journal = {Nat. Commun.},
  publisher = {Springer Science and Business Media LLC},
  author = {Dey,  R. and Buness,  C. M. and Hokmabad,  B. V. and Jin,  C. and Maass,  C. C.},
  year = {2022},
  month = may 
}

@article{Meredith2020,
  title = {Predator–prey interactions between droplets driven by non-reciprocal oil exchange},
  volume = {12},
  ISSN = {1755-4349},
  url = {http://dx.doi.org/10.1038/s41557-020-00575-0},
  DOI = {10.1038/s41557-020-00575-0},
  number = {12},
  journal = {Nat. Chem.},
  publisher = {Springer Science and Business Media LLC},
  author = {Meredith,  C. H. and Moerman,  P. G. and Groenewold,  J. and Chiu,  Y. and Kegel,  W. K. and van Blaaderen,  A. and Zarzar,  L. D.},
  year = {2020},
  month = nov,
  pages = {1136–1142}
}

@article{Ban2014,
  title = {Self-propelled droplets for extracting rare-earth metal ions},
  volume = {10},
  ISSN = {1744-6848},
  url = {http://dx.doi.org/10.1039/c4sm01001a},
  DOI = {10.1039/c4sm01001a},
  number = {33},
  journal = {Soft Matter},
  publisher = {Royal Society of Chemistry (RSC)},
  author = {Ban,  T. and Tani,  K. and Nakata,  H. and Okano,  Y.},
  year = {2014},
  pages = {6316–6320}
}

@article{Birrer2022,
  title = {We the droplets: A constitutional approach to active and self-propelled emulsions},
  volume = {61},
  ISSN = {1359-0294},
  url = {http://dx.doi.org/10.1016/j.cocis.2022.101623},
  DOI = {10.1016/j.cocis.2022.101623},
  journal = {Curr. Opin. Colloid \& Interface Sci.},
  publisher = {Elsevier BV},
  author = {Birrer,  S. and Cheon,  S. I. and Zarzar,  L. D.},
  year = {2022},
  month = oct,
  pages = {101623}
}

@article{Dwivedi2022,
  title = {Self-propelled swimming droplets},
  volume = {61},
  ISSN = {1359-0294},
  url = {http://dx.doi.org/10.1016/j.cocis.2022.101614},
  DOI = {10.1016/j.cocis.2022.101614},
  journal = {Curr. Opin. Colloid \& Interface Sci.},
  publisher = {Elsevier BV},
  author = {Dwivedi,  P. and Pillai,  D. and Mangal,  R.},
  year = {2022},
  month = oct,
  pages = {101614}
}

@article{Li2018,
  title = {Spatiotemporal control of cargo delivery performed by programmable self-propelled Janus droplets},
  volume = {1},
  ISSN = {2399-3650},
  url = {http://dx.doi.org/10.1038/s42005-018-0025-4},
  DOI = {10.1038/s42005-018-0025-4},
  number = {1},
  journal = {Commun. Phys.},
  publisher = {Springer Science and Business Media LLC},
  author = {Li,  M. and Brinkmann,  M. and Pagonabarraga,  I. and Seemann,  R. and Fleury,  J.},
  year = {2018},
  month = jun 
}

@article{Thutupalli2011,
  title = {Swarming behavior of simple model squirmers},
  volume = {13},
  ISSN = {1367-2630},
  url = {http://dx.doi.org/10.1088/1367-2630/13/7/073021},
  DOI = {10.1088/1367-2630/13/7/073021},
  number = {7},
  journal = {New J. Phys.},
  publisher = {IOP Publishing},
  author = {Thutupalli,  S. and Seemann,  R. and Herminghaus,  S.},
  year = {2011},
  month = jul,
  pages = {073021}
}

@article{Moran2017,
  title = {Phoretic Self-Propulsion},
  volume = {49},
  ISSN = {1545-4479},
  url = {http://dx.doi.org/10.1146/annurev-fluid-122414-034456},
  DOI = {10.1146/annurev-fluid-122414-034456},
  number = {1},
  journal = {Ann. Rev. Fluid Mech.},
  publisher = {Annual Reviews},
  author = {Moran,  J. L. and Posner,  J. D.},
  year = {2017},
  month = jan,
  pages = {511–540}
}

@article{Michelin2013,
  title = {Spontaneous autophoretic motion of isotropic particles},
  volume = {25},
  ISSN = {1089-7666},
  url = {http://dx.doi.org/10.1063/1.4810749},
  DOI = {10.1063/1.4810749},
  number = {6},
  journal = {Phys. Fluids},
  publisher = {AIP Publishing},
  author = {Michelin,  S. and Lauga,  E. and Bartolo,  D.},
  year = {2013},
  month = jun 
}

@article{Moerman2017,
  title = {Solute-mediated interactions between active droplets},
  volume = {96},
  ISSN = {2470-0053},
  url = {http://dx.doi.org/10.1103/PhysRevE.96.032607},
  DOI = {10.1103/physreve.96.032607},
  number = {3},
  journal = {Phys. Rev. E},
  publisher = {American Physical Society (APS)},
  author = {Moerman,  P. G. and Moyses,  H. W. and van der Wee,  E. B. and Grier,  D. G. and van Blaaderen,  A. and Kegel,  W. K. and Groenewold,  J. and Brujic,  J.},
  year = {2017},
  month = sep 
}

@article{Chen2025,
  title = {Evolving Motility of Active Droplets Is Captured by a Self-Repelling Random Walk Model},
  volume = {134},
  ISSN = {1079-7114},
  url = {http://dx.doi.org/10.1103/PhysRevLett.134.018301},
  DOI = {10.1103/physrevlett.134.018301},
  number = {1},
  journal = {Phys. Rev. Lett.},
  publisher = {American Physical Society (APS)},
  author = {Chen,  W. and Izzet,  A. and Zakine,  R. and Clément,  E. and Vanden-Eijnden,  E. and Brujic,  J.},
  year = {2025},
  month = jan 
}

@article{Hu2022,
  title = {Spontaneous locomotion of phoretic particles in three dimensions},
  volume = {7},
  ISSN = {2469-990X},
  url = {http://dx.doi.org/10.1103/PhysRevFluids.7.034003},
  DOI = {10.1103/physrevfluids.7.034003},
  number = {3},
  journal = {Phys. Rev. Fluids},
  publisher = {American Physical Society (APS)},
  author = {Hu,  W. and Lin,  T. and Rafai,  S. and Misbah,  C.},
  year = {2022},
  month = mar 
}

@article{Hokmabad2021,
  title = {Emergence of Bimodal Motility in Active Droplets},
  volume = {11},
  ISSN = {2160-3308},
  url = {http://dx.doi.org/10.1103/PhysRevX.11.011043},
  DOI = {10.1103/physrevx.11.011043},
  number = {1},
  journal = {Phys. Rev. X},
  publisher = {American Physical Society (APS)},
  author = {Hokmabad,  B. V. and Dey,  R. and Jalaal,  M. and Mohanty,  D. and Almukambetova,  M. and Baldwin,  K. A. and Lohse,  D. and Maass,  C. C.},
  year = {2021},
  month = mar 
}

@article{Jadhav2021,
  title = {Effect of surfactant on the settling of a drop towards a wall},
  volume = {912},
  ISSN = {1469-7645},
  url = {http://dx.doi.org/10.1017/jfm.2020.1101},
  DOI = {10.1017/jfm.2020.1101},
  journal = {J. Fluid Mech.},
  publisher = {Cambridge University Press (CUP)},
  author = {Jadhav,  S. N. and Ghosh,  U.},
  year = {2021},
  month = feb 
}

@article{Jadhav2024,
  title = {Coupled bulk and interfacial transport of surfactants governs the settling of a drop towards a wall},
  volume = {987},
  ISSN = {1469-7645},
  url = {http://dx.doi.org/10.1017/jfm.2024.406},
  DOI = {10.1017/jfm.2024.406},
  journal = {J. Fluid Mech.},
  publisher = {Cambridge University Press (CUP)},
  author = {Jadhav,  S. N. and Mandal,  S. and Ghosh,  U.},
  year = {2024},
  month = may 
}

@article{Singh2025,
  title = {Active droplets in microfluidic confinement},
  volume = {37},
  ISSN = {1089-7666},
  url = {http://dx.doi.org/10.1063/5.0293093},
  DOI = {10.1063/5.0293093},
  number = {10},
  journal = {Phys. Fluids},
  publisher = {AIP Publishing},
  author = {Singh,  S. and Mondal,  P. K. and Mandal,  S.},
  year = {2025},
  month = Oct 
}

@article{Guchhait2025,
  title = {Flow fields around active droplets squeezing through tight confinements},
  volume = {10},
  ISSN = {2469-990X},
  url = {http://dx.doi.org/10.1103/PhysRevFluids.10.044202},
  DOI = {10.1103/physrevfluids.10.044202},
  number = {4},
  journal = {Phys. Rev. Fluids},
  publisher = {American Physical Society (APS)},
  author = {Guchhait,  S. and Sontakke,  S. S. and Mandal,  S. and Dey,  R.},
  year = {2025},
  month = Apr 
}

@article{Gou2025,
  title = {A computational framework for simulating suspensions of phoretic particles},
  volume = {1016},
  ISSN = {1469-7645},
  url = {http://dx.doi.org/10.1017/jfm.2025.10282},
  DOI = {10.1017/jfm.2025.10282},
  journal = {J. Fluid Mech.},
  publisher = {Cambridge University Press (CUP)},
  author = {Gou,  Z. and Misbah,  C. and Farutin,  A.},
  year = {2020},
  month = Jul 
}

@article{Castonguay2023,
  title = {Gravitational settling of active droplets},
  volume = {107},
  ISSN = {2470-0053},
  url = {http://dx.doi.org/10.1103/PhysRevE.107.024608},
  DOI = {10.1103/physreve.107.024608},
  number = {2},
  journal = {Physical Review E},
  publisher = {American Physical Society (APS)},
  author = {Castonguay,  A. C. and Kailasham,  R. and Wentworth,  C. M. and Meredith,  C. H. and Khair,  A. S. and Zarzar,  L. D.},
  year = {2023},
  month = Feb 
}

@article{Lippera2020,
  title = {Collisions and rebounds of chemically active droplets},
  volume = {886},
  ISSN = {1469-7645},
  url = {http://dx.doi.org/10.1017/jfm.2019.1055},
  DOI = {10.1017/jfm.2019.1055},
  journal = {J. Fluid Mech.},
  publisher = {Cambridge University Press (CUP)},
  author = {Lippera,  K. and Morozov,  M. and Benzaquen,  M. and Michelin,  S.},
  year = {2020},
  month = Jan 
}

@article{Li2022,
  title = {Swimming dynamics of a self-propelled droplet},
  volume = {934},
  ISSN = {1469-7645},
  url = {http://dx.doi.org/10.1017/jfm.2021.1154},
  DOI = {10.1017/jfm.2021.1154},
  journal = {J. Fluid Mech.},
  publisher = {Cambridge University Press (CUP)},
  author = {Li, G.},
  year = {2022},
  month = Jan 
}

\end{document}